\documentclass[iop,apj,twoside,onecolumn,english]{aastex62}

\usepackage{graphicx}
\usepackage{longtable}
\usepackage{amssymb}
\usepackage{multirow}
\usepackage{rotating}

\usepackage{url}
\usepackage{color}
\definecolor{purple}{rgb}{.9,0,.1}
\usepackage{natbib}
\begin{document}

\journalinfo{submitted to ApJ 20180727}
\journalinfo{revised version submitted to ApJ 20181124}
\journalinfo{final version submitted to ApJ 20190127}
\journalinfo{accepted by ApJ 20190128}

\newcommand{\kms}{km s$^{-1}$}
\newcommand{\msun}{M$_{\sun}$}
\newcommand{\rsun}{R$_{\sun}$}
\newcommand{\lsun}{L$_{\sun}$}
\newcommand{\jg}{iPTF\,14jg} 

\title
{PTF 14jg: The Remarkable Outburst and Post-Burst Evolution of a Previously Anonymous Galactic Star}

\author{Lynne A. Hillenbrand}
\affiliation{Department of Astronomy, California Institute of Technology, Pasadena CA 91125}
\author[0000-0001-9515-478X]{Adam A. Miller}
\affiliation{Department of Astronomy, California Institute of Technology, Pasadena CA 91125}
\affiliation{Center for Interdisciplinary Exploration and Research in Astrophysics (CIERA), 
Northwestern University, 2145 Sheridan Road, Evanston, IL 60208}
\author[0000-0003-2251-0602]{John M. Carpenter},
\affiliation{Department of Astronomy, California Institute of Technology, Pasadena CA 91125}
\affiliation{Joint ALMA Observatory, Avenida Alonso de Córdova 3107, Vitacura, Santiago, Chile}
\author{Mansi M. Kasliwal},
\affiliation{Department of Astronomy, California Institute of Technology, Pasadena CA 91125}
\author{Howard Isaacson} 
\affiliation{University of California at Berkeley, Berkeley, CA}
\author{Sumin Tang},
\affiliation{Department of Astronomy, California Institute of Technology, Pasadena CA 91125}
\affiliation{Kavli Institute for Theoretical Physics, University of California, Santa Barbara, CA 93106}
\author{Vishal Joshi},
\affiliation{Astronomy and Astrophysics Division, Physical Research Laboratory, Navrangpura, Ahmedabad, Gujarat 380 009, India},
\author{D.P.K. Banerjee},  \affiliation{Astronomy and Astrophysics Division, Physical Research Laboratory, Navrangpura, Ahmedabad, Gujarat 380 009, India},
\author[0000-0002-0077-2305]{Roc Cutri}, 
\affiliation{Infrared Processing and Analysis Center, California Institute of Technology, Pasadena CA 91125}

\begin{abstract}
We report the outbursting source \jg, 
which prior to the onset of its late 2013 eruption, was a faint, unstudied, 
and virtually uncatalogued star.  
The salient features of the \jg\ outburst are:
{\sl (i)} projected location near the W4 \ion{H}{2} region  and radial velocity consistent with physical association;  
{\sl (ii)} a lightcurve that underwent a $\sim$6-7 mag optical (R-band) 
through mid-infrared (L-band) brightening on a few month time scale, 
that peaked and then faded by $\sim$3 mag, but plateaued still $>$3.5 mag 
above quiescence by $\sim$8 months post-peak, 
lasting to at least four years after eruption;
{\sl (iii)} strong outflow signatures, with velocities reaching -530 \kms; 
{\sl (iv)} a low gravity and broad ($\sim$100-150 \kms\ FWHM) optical absorption
line spectrum that systematically changes its spectral type with wavelength; 
{\sl (v)} lithium; and {\sl (vi)} ultraviolet and infrared excess. 
We tentatively identify the outburst
as exhibiting characteristics of a young star FU Ori event.  
However, the burst would be unusually hot, 
with an absorption spectrum exhibiting high-excitation ($\sim$11,000--15,000 K)
lines in the optical, and no evidence of CO in the near-infrared,
in addition to exhibiting an unusual lightcurve. 
We thus also consider alternative scenarios including 
various forms of novae, nuclear burning instabilities, massive star events, 
and mergers -- finding them all inferior to the atypically hot FU Ori 
star classification.
The source eventually may be interpreted as a new category 
of young star outburst 
with larger amplitude and shorter rise time than most FU Ori-like events.
Continued monitoring of the lightcurve and spectral evolution will reveal its true nature.

\end{abstract}


\section{Introduction}
The W3 / W4 / W5 complex in the Perseus Spiral Arm is one of the more 
dynamic regions of ongoing star formation in the Galaxy 
\citep[see review by][and references therein]{megeath2008}. 
It is large and massive, spanning $\sim$200 x 50 pc$^2$ with
several times $10^5\ M_\odot$ in molecular gas. 
In addition to several rich star clusters containing tens of thousands 
of stars, it features ionized gas localized in \ion{H}{2} regions, and 
neutral gas that has been shaped into shells and chimneys.  Further 
evidence of the interaction of massive stars with the local interstellar medium
includes numerous supernova remnants, large wind-blown bubbles,
cometary and fragmented clouds, and sculpted dust pillars. 

The object of interest identified here, \jg, is located at 
02:40:30.14 +60:52:45.5 (l = 135.81524,  b = 0.76533).    
In the plane of the sky, the position is between two of the sequence of three 
large \ion{H}{2} regions, just east of the ``swept up shell" in W4 
(see \ion{H}{1} and dust maps in Terebey et al. 2003, their Figures 2 and 3),
and $\sim$1.1 deg from the center of the massive cluster IC 1805.  
The position does not appear to be associated with significant molecular gas emission 
(e.g. Carpenter, Heyer, Snell 2000, their Figures 1 and 2; LaGrois \& Joncas 2009, their Figure 9), 
nor with any of the identified infrared clusters.  
Many surveys targeting the stellar population of the \ion{H}{2} regions
have missed covering this area.  Thus, there is no SIMBAD counterpart to 
\jg, meaning no previous literature. There is also no corresponding object
in Vizier, meaning no documented photometric detection at any wavelength.
Evidence below, primarily radial velocity information, associates \jg\ with
the Perseus Arm.  
Based on parallaxes of masing sources in W3, \cite{xu2006} 
derive a distance of 1.95 kpc, which is commonly adopted as the distance to the overall complex.

In this paper, we present the substantial optical brightening of \jg\ that occurred
over a time period of several months in 2013-2014, and our multi-wavelength 
photometric and spectroscopic follow-up observations during the subsequent several years. 
We draw analogies between this source and the rarely populated FU Ori class
\citep{herbig1977,HK1996} of young stellar objects.  

FU Ori stars 
are interpreted as young low mass stars undergoing episodes of rapid accretion 
at rates that are three to four orders of magnitude larger than the quiescent-state 
T Tauri accretion rates, typically $\sim 10^{-8}$ M$_\odot$/yr 
\citep[e.g.][]{gullbring1998}.  
The occurrence rate and duty cycle of FU Ori events is crucial to establish
given our poor understanding of accretion histories during star formation 
and pre-main sequence evolution.
However, the known FU Ori population is small. 
The class is an empirically diverse set of objects that are associated together
based mainly on the definitive spectral type change with wavelength,
strong outflow signatures, infrared excess, and an observed or suspected 
large amplitude, long duration photometric outburst
\cite{RA2010}. Furthermore,  
relatively few FU Ori stars were observed as their outbursts occurred,
and even fewer were well-studied before their outbursts.  
Thus our knowledge of: the progenitors, the range in lightcurve rise shapes,
and the early outburst spectroscopic characteristics is severely limited.

\S 2 reports our detection of \jg\ from quiescence to outburst and \S 3 discusses archival information
at the outburst position.  \S4 presents the follow-up photometry and spectroscopy
we collected, 
\S5 our analysis of the outburst colors and spectral energy distribution,
\S6 our analysis of the multi-year lightcurve, 
and \S7 our analysis of the complex picture presented in the spectroscopy.  
A synthesis of the observational evidence in the context of the FU Ori interpretation 
is presented in \S 8, while \S 9 considers other possible interpretations 
of the collected evidence regarding \jg.  \S10 contains our discussion 
and \S11 a short conclusion.

\begin{figure}
\includegraphics[width=0.25\textwidth]{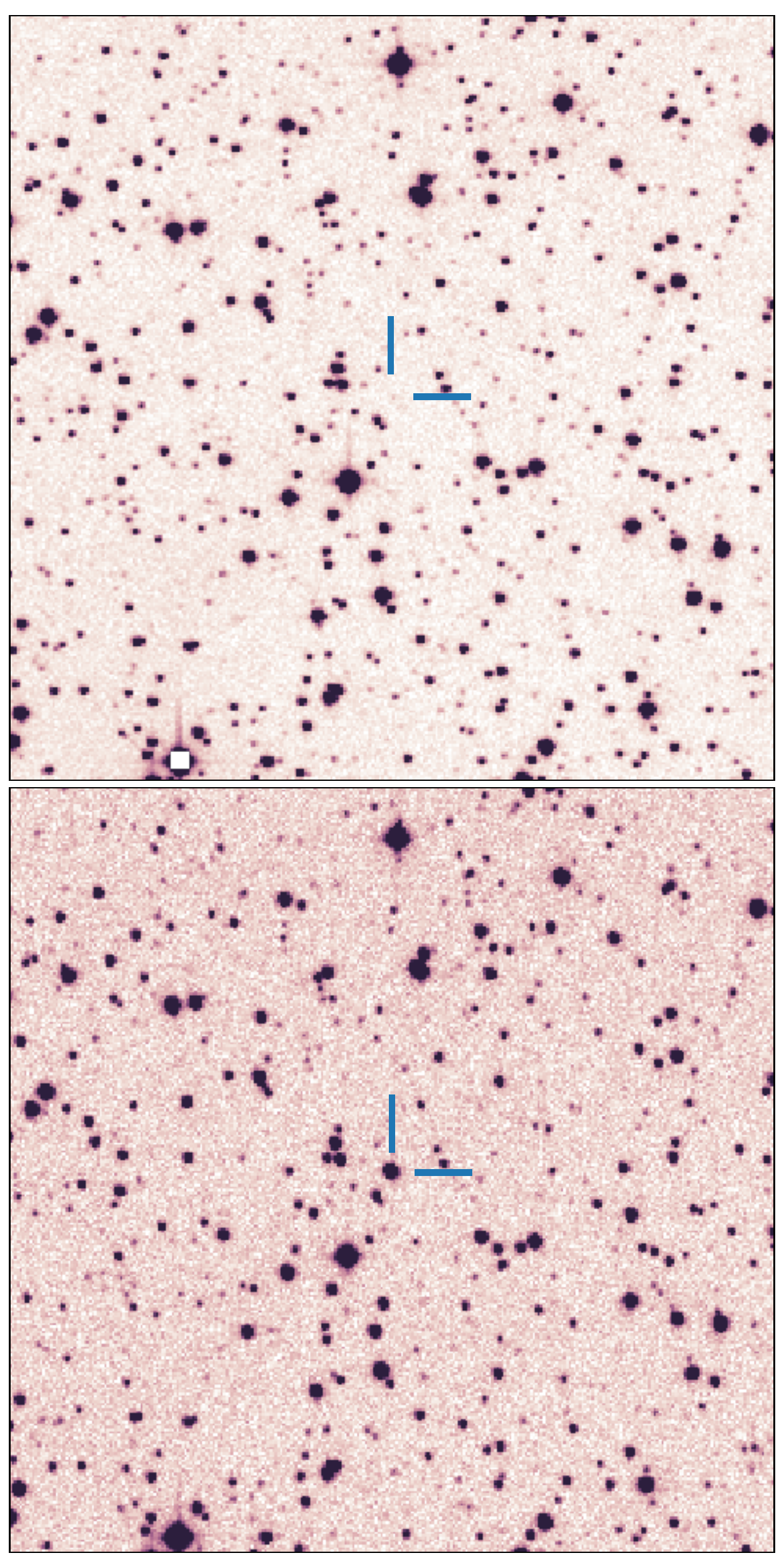}
\includegraphics[width=0.75\textwidth]{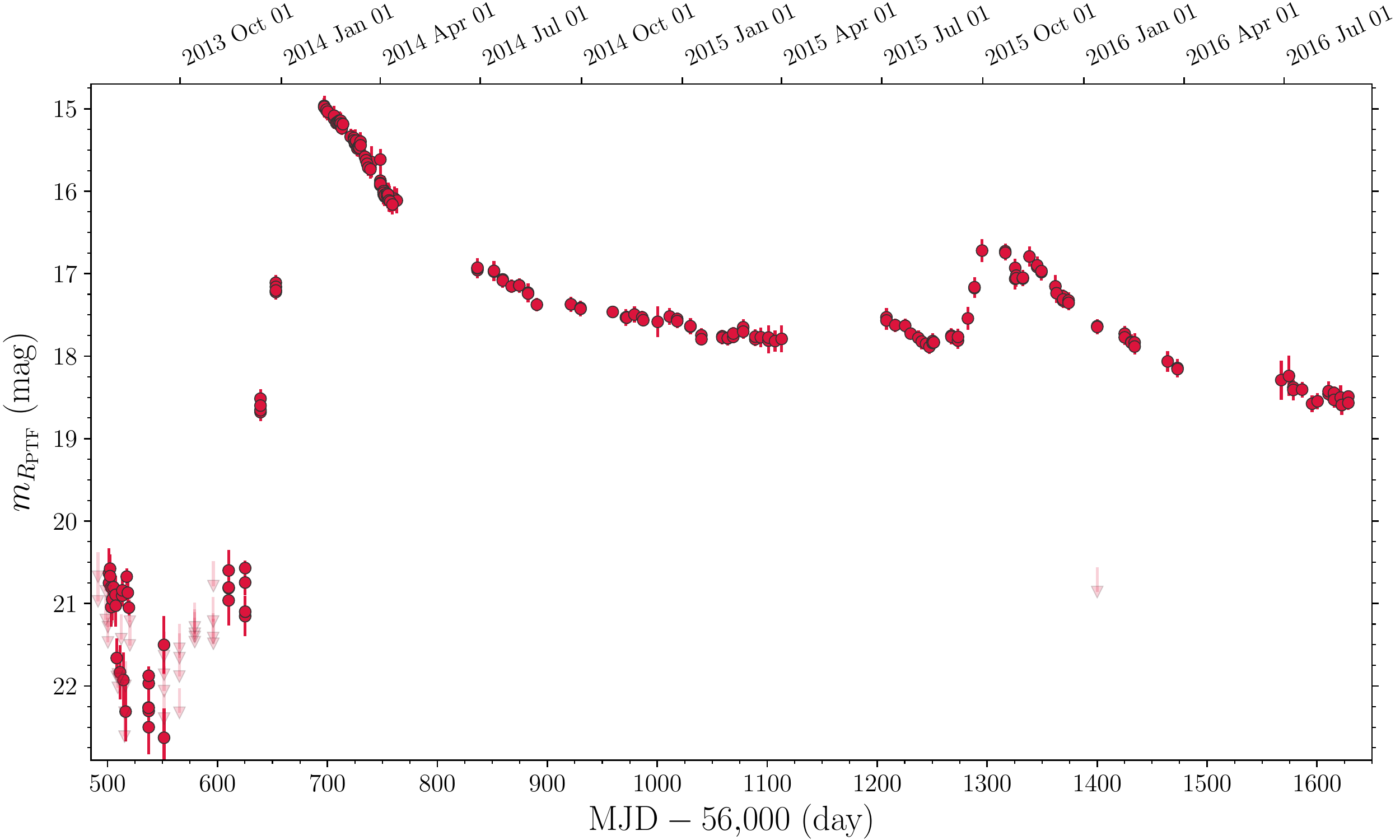}
\caption{
Left: PTF image over 5' x 5' in 2013 before the outburst (top panel), 
created from an image stack of available data, 
and on 2014 February 9, near the peak of the light curve (bottom panel). 
Right: 
$R_{PTF}$ lightcurve beginning in late 2013 and extending 
through late-2016.  There is evidence for a secondary maximum
occuring when the object was behind the sun in mid-2015,
and detection of a tertiary maximum during late 2015. 
}
\label{fig:lightcurve}
\end{figure}

\begin{figure}
\includegraphics[width=0.9\textwidth]{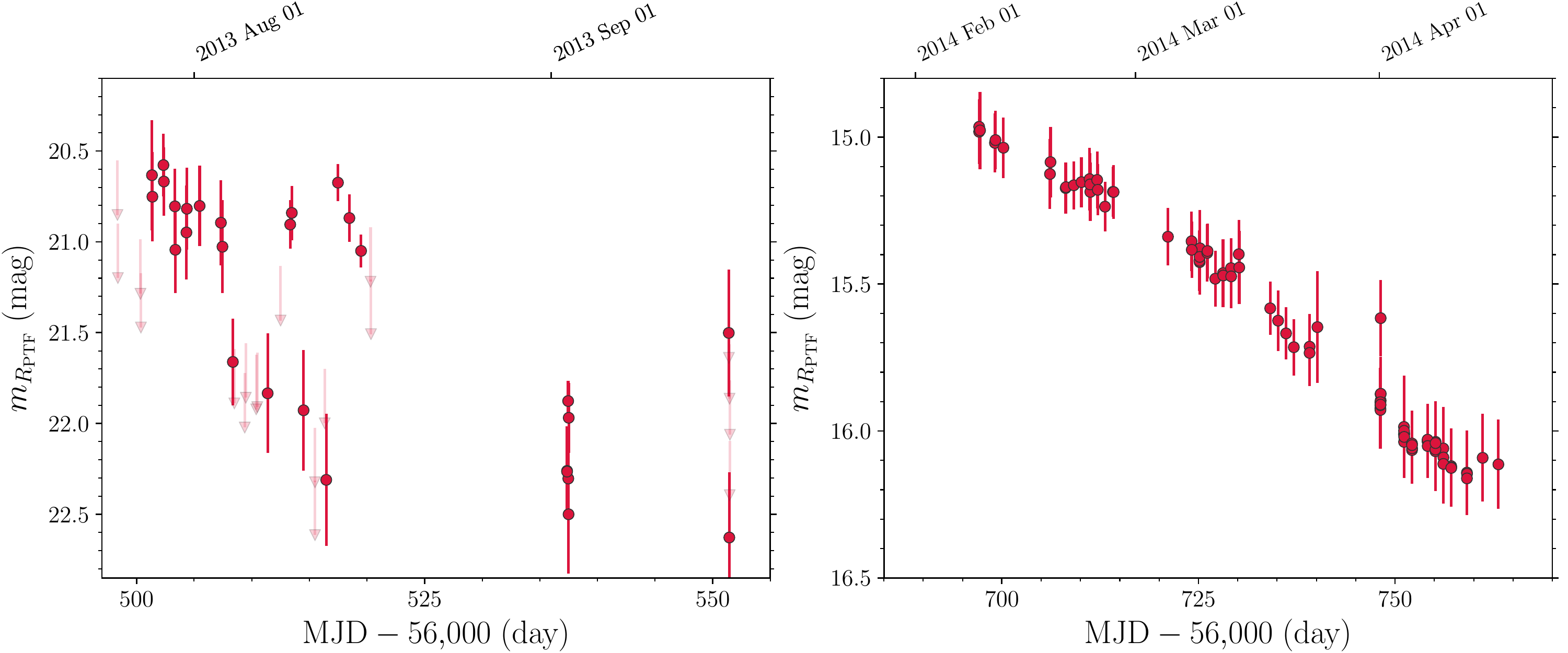}
\caption{
Expanded view of the Figure 1 lightcurve, highlighting 
the pre-outburst brightness variations in late 2013 (left panel) and the
initial decline from absolute lightcurve peak in early 2014 (right panel) 
which appears to change slope around day 730.
Note that the ordinate scaling is different between the two panels.
}
\label{fig:peak}
\end{figure}

\newpage

\section{Detection of the \jg\ Outburst Event}

The (intermediate) Palomar Transient Factory \citep{law2009,kulkarni2013}
monitored the W3 / W4 / W5 complex in the R-band
beginning in summer of 2013.  After initial experimentation with the
viability of this crowded field in Cassiopeia for automated processing and photometry,
low-cadence (once every two weeks) observations began in fall of 2013. 
A new source was identified 
as a candidate transient on 2014 January 24 and given the name \jg.  
We estimate that \jg\ achieved peak brightness around 2014 February 9 then
decreased in brightness (see Figures ~\ref{fig:lightcurve} and
~\ref{fig:peak}).

Faint photometry recovered \citep{laher2014}
from observations taken before the outburst show
that \jg\ exhibited significant variability during July through September of 2013 
at about the 1 mag level, between $R_{PTF} = 21-22^m$, including
several coherent dips in brightness on a time scale of about five days.
In December 2013, the source began gradually to increase its brightness
from $R_{PTF} < 21^m$ to $R_{PTF} \approx 15^m$ by mid-February 2014, 
i.e. a $>$6 mag rise in $\sim$70 days.
The light curve was subsequently sampled with nightly cadence, though 
interrupted due to the prescribed PTF switch from R-band observing 
to H$\alpha$ filter observing on the five days around full moon.  The realized cadence 
was further subject to some poor weather and continuing monthly gaps 
around new moon, as well as a longer gap when the object set for the 2013-2014 season. 
Nevertheless, the lightcurve has good coverage of its main features.  The cadence was
decreased for the 2014-2015 season and those beyond.

The photometry reported in Table~\ref{tbl:P48_phot} 
was obtained with the Palomar 48-in
telescope (P48) and the PTF Survey Camera, as part of routine iPTF operations.
The $R_{\rm PTF}$ band is approximately a Mould $R$ filter \citep{law2009}. 
Observations are calibrated relative to Sloan Digital Sky Survey (SDSS; \citealt{york2000}) 
stars, which are observed throughout the night (see \citealt{ofek2012} for 
further details). Photometry of \jg\ was measured using the custom point 
spread function (PSF) fitting routine PTFIDE
\citep{masci2017}. 

\begin{deluxetable}{rrr}
\tabletypesize{\small}
\tablecolumns{3}
\tablewidth{0pt}
\setlength{\tabcolsep}{6pt}
\tablecaption{{P48 $R$-band Observations} \label{tbl:P48_phot}}
\tablehead{\colhead{MJD} & \colhead{$m_{R_\mathrm{PTF}}$} & 
    \colhead{$\sigma_{\rm m}$} \\
    \colhead{} & \colhead{(mag)} & \colhead{(mag)} }
\startdata
.         & .     &  .   \\
\enddata
\tablenotetext{a}{4$\sigma$ statistical upper limit.}
\end{deluxetable}

\section{Archival Information}

A search of available image and catalog data from optical survey sources 
(e.g. DSS, iPHAS, UVEX) 
does not reveal a previous detection at the position of \jg. 
Pan-STARRS appears to have caught the burst, though individual-epoch measurements 
that potentially include pre-burst as well as in-burst 
magnitudes are not yet available in the DR1 catalog.
The source is not in the near-infrared 2MASS catalog \citep{cutri2003}. UKIDSS,
the most sensitive near-infrared catalog, missed covering this high declination field.
In the mid-infrared, \jg\ 
was not detected in the $ WISE$ all-sky survey \citep{cutri2012}, 
which had lower spatial resolution and sensitivity relative to Spitzer.

The source was observed as part of the GLIMPSE-360 survey with Spitzer \citep{hora2007, churchwell2009}. 
The data products available at NASA's IRSA\footnote{https://irsa.ipac.caltech.edu/data/SPITZER/GLIMPSE/overview.html}
show a detection within 0.44 arcsec in the GLIMPSE-360 Catalog, 
but only in the 3.6 $\mu$m band at $17.2 \pm 0.1$ mag.
However, the ``more complete,  less reliable" GLIMPSE-360 Archive
reports a [3.6]=17.21$\pm$0.12 and [4.5]=16.82$\pm$0.18 source.
The infrared color is therefore [3.6] - [4.5] = 0.39 $\pm$0.22. 
For the average pre-outburst optical brightness of $R_{PTF}=21.5$,
the optical-infrared color $R_{PTF}$ - [3.6] is therefore 4.3 mag.

Although red, the source would not have been identified as a 
clear candiate young star based on its infrared color alone.
At these faint magnitudes, a color of $[3.6]-[4.5] > 0.5$ 
would be required in order to be distinguishable from noise 
among the field star population for a source that is not also detected 
at longer infrared wavelengths.

Photosphere models for stars predict essentially zero 
$[3.6]-[4.5]$ color, and increasing to a maximum color of 
only 0.15 mag for low mass brown dwarfs.
For \jg, a color excess of 0.1-0.4 mag suggests 
the presence of circumstellar dust in the pre-outburst stage. 
A dust excess is also revealed in our more considered SED analysis below (\S5).

\section{Photometric and Spectroscopic Follow-up Observations}

Our follow-up to the dramatic optical brightening event in \jg\ includes 
photometric and spectroscopic observations, in addition to continued 
Palomar 48-in R-band monitoring.  Optical and near-infrared photometric data
were obtained at the Palomar 60-in telescope (ugriz) and Mt Abu 1.2m telescope (JHK$_s$), 
while optical and infrared spectra were obtained at: 
the Apache Point Observatory, Palomar Observatory, and Keck Observatory. 
We also acquired bluer wavelength ultraviolet and x-ray data from UVOT 
on board \textit{Swift}\footnote{\textit{Swift} was renamed in early 2018 as the \textit{Neil Gehrels Swift Observatory}.}
and redder
wavelength mid-infrared ($Spitzer$, $NEOWISE$) and millimeter ($CARMA$) observations.

Details of the follow-up observations appear in the Tables and are described
below.  All follow-up observations occurred after the lightcurve peak in early Feburary 2014.  

\subsection{X-ray and UV Photometry}

Space-based ultraviolet and optical observations were obtained with the UVOT 
instrument \citep{roming2005}. 
\jg\ was observed in all 6 UVOT filters on 28 March, 6 April, and 25 June 2014. 
Photometry was performed and calibrated using standard UVOT tools 
(see \citealt{breeveld2010}) and is 
reported in Table~\ref{tbl:uvot_phot}, on the AB mag system. 
In the x-ray channel, only an upper limit of 0.00235 counts/s
was derived from 6414 seconds of integration.

\begin{deluxetable}{crrrrrr}
\tabletypesize{\small}
\tablecolumns{7}
\tablewidth{0pt}
\setlength{\tabcolsep}{6pt}
\tablecaption{\textit{Swift} UVOT Observations\label{tbl:uvot_phot}}
\tablehead{\colhead{filter} & \colhead{$t_{\rm start}$} & 
    \colhead{$t_{\rm end}$} &
    \colhead{exp. $t$} & \colhead{mag} & \colhead{$\sigma_{\rm m}$ stat.} &
    \colhead{$\sigma_{\rm m}$ sys.}\\
    \colhead{} & \colhead{(MJD)} & \colhead{(MJD)} & 
    \colhead{(s)} & \colhead{(AB)} & \colhead{(AB)} & \colhead{(AB)} }
\startdata
UVW2 & 56744.04 & 56744.24 & 498.0 & $>$22.07\tablenotemark{a} & . . . & . . . \\
UVM2 & 56744.04 & 56744.24 & 935.8 & $>$22.28\tablenotemark{a} & . . . & . . . \\
UVW1 & 56744.04 & 56744.24 & 299.2 & 21.30 & 0.36 & 0.03 \\
U & 56744.04 & 56744.24 &  99.5 & 18.62 & 0.13 & 0.02 \\
B & 56744.04 & 56744.24 &  99.5 & 17.33 & 0.09 & 0.02 \\
V & 56744.04 & 56744.24 &  99.5 & 16.59 & 0.09 & 0.01 \\
UVW2 & 56753.63 & 56753.70 & 521.2 & $>$22.20\tablenotemark{a} & . . . & . . . \\
UVM2 & 56753.64 & 56753.71 & 928.5 & $>$22.38\tablenotemark{a} & . . . & . . . \\
UVW1 & 56753.63 & 56753.70 & 312.5 & 21.41 & 0.36 & 0.03 \\
U & 56753.63 & 56753.70 & 103.9 & 19.23 & 0.17 & 0.02 \\
B & 56753.63 & 56753.70 & 103.9 & 17.70 & 0.10 & 0.02 \\
V & 56753.64 & 56753.70 & 103.9 & 16.72 & 0.09 & 0.01 \\
UVW2 & 56833.55 & 56833.62 & 555.7 & $>$22.26\tablenotemark{a} & . . . & . . . \\
UVM2 & 56833.55 & 56833.62 & 834.8 & $>$22.25\tablenotemark{a} & . . . & . . . \\
UVW1 & 56833.54 & 56833.61 & 333.2 & $>$21.59\tablenotemark{a} & . . . & . . . \\
U & 56833.54 & 56833.61 & 110.7 & 19.90 & 0.22 & 0.02 \\
B & 56833.55 & 56833.61 & 110.7 & 18.52 & 0.13 & 0.02 \\
V & 56833.55 & 56833.62 & 110.8 & 17.71 & 0.15 & 0.01 \\
\enddata
\tablenotetext{a}{3$\sigma$ statistical upper limit.}
\end{deluxetable}

\subsection{Optical Photometry} 

\begin{figure}
\begin{center}
\includegraphics[scale=0.75]{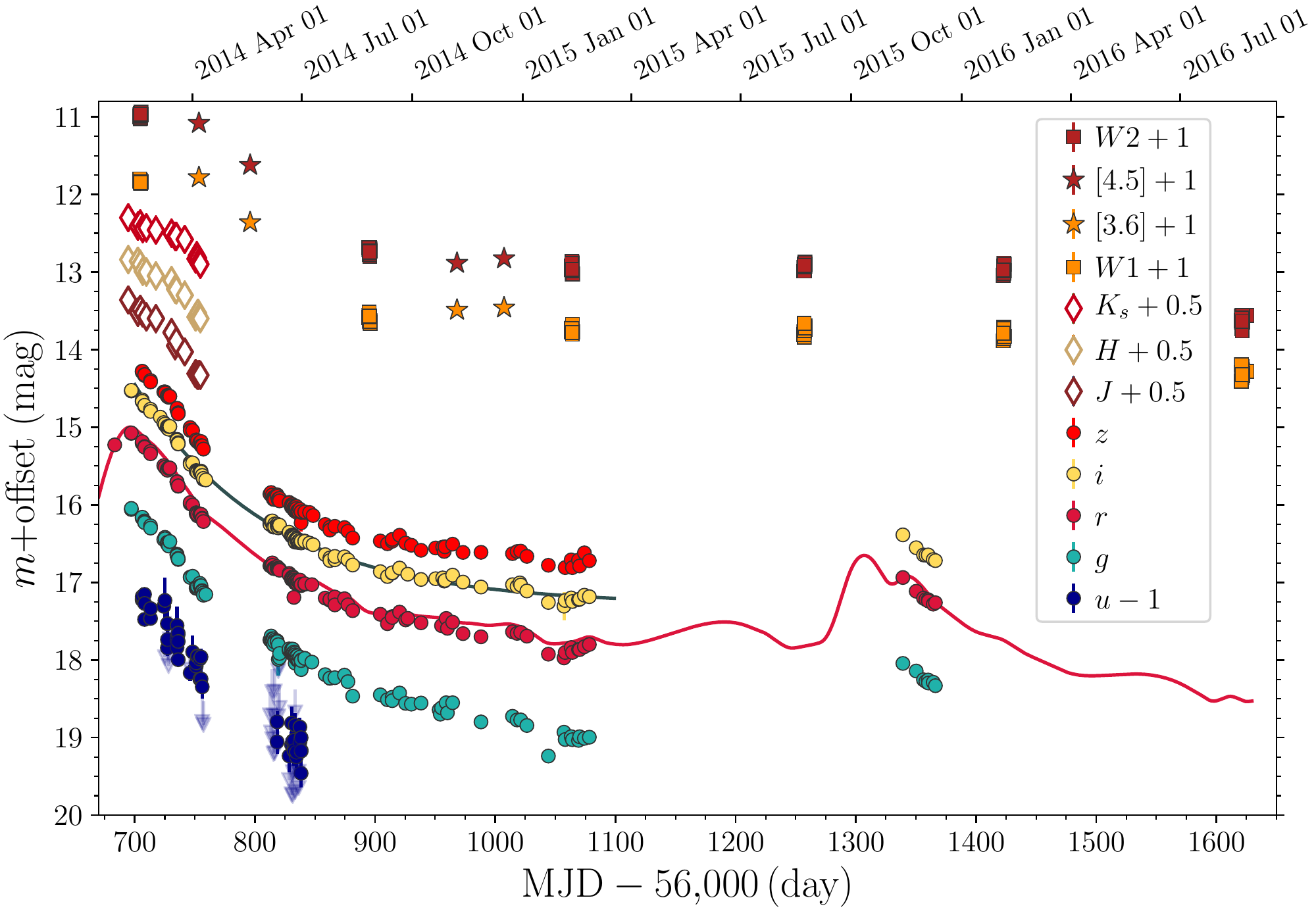}
\end{center}
\caption{
Post-outburst brightness evolution in the
optical $ugriz$ Palomar 60" telescope photometry, 
near-infrared $JHK$ Mt. Abu 1.2m photometry, 
and mid-infrared IRAC1, IRAC2 from $Spitzer$
and W1 and W2 from $NEOWISE$ photometry.  
Lightcurves begin in early 2014, near or just after the peak brightness 
of \jg.  The decline over 3 mag in brightness is relatively colorless 
for the duration of the time series, with color changes of no more than 0.2 mag
(see details in Figure \ref{fig:colorcurves});
the source is currently still 3-4 mag brighter than its pre-outburst state.
Solid red line overlaid on the $r$-band lightcurve is a gaussian-process model 
fit to the $R_{PTF}$ photometry of Figure~\ref{fig:lightcurve}.
}
\label{fig:lightcurves}
\end{figure}

\begin{figure}
\begin{center}
\includegraphics[scale=0.5]{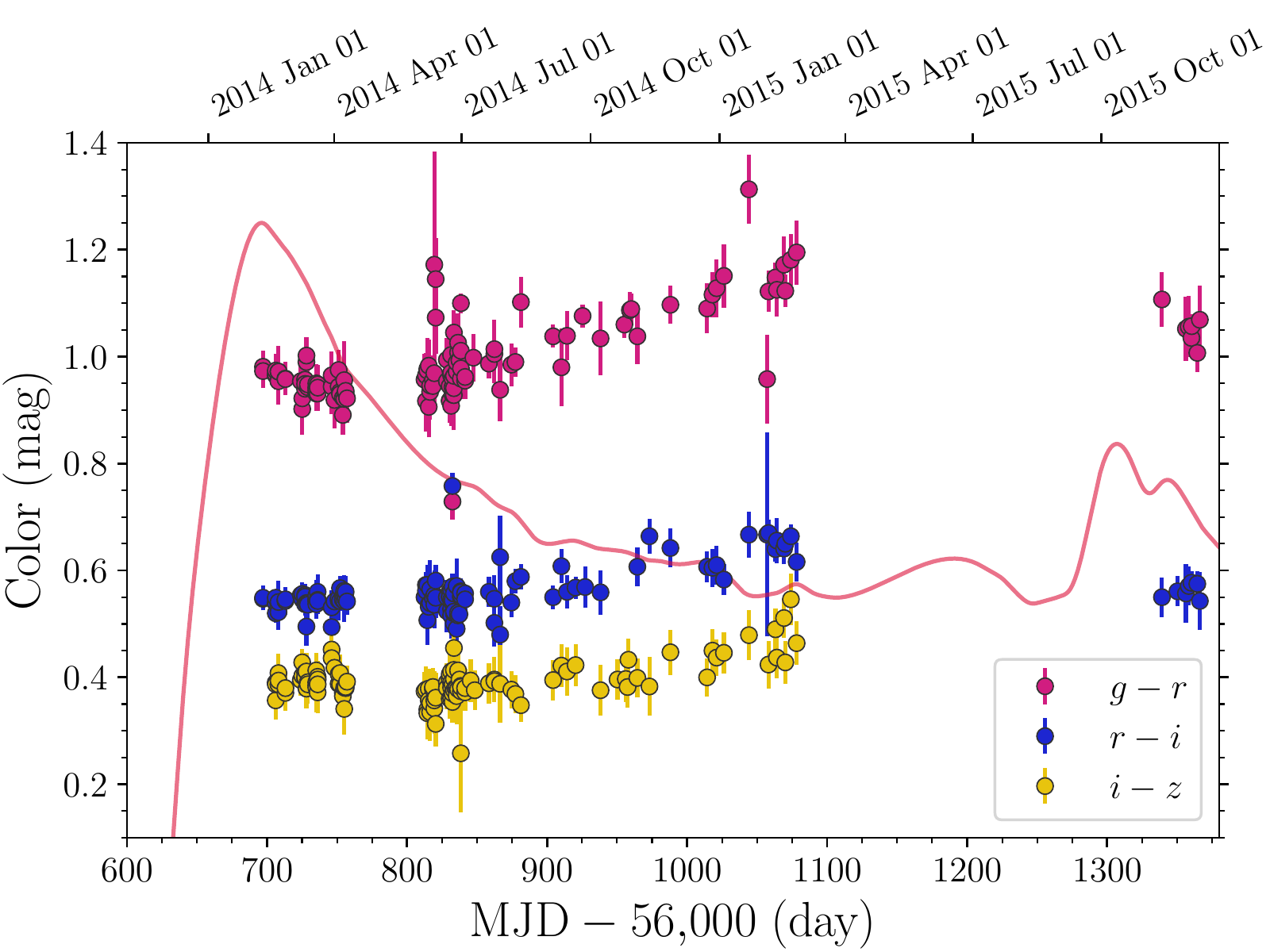}
\includegraphics[scale=0.5]{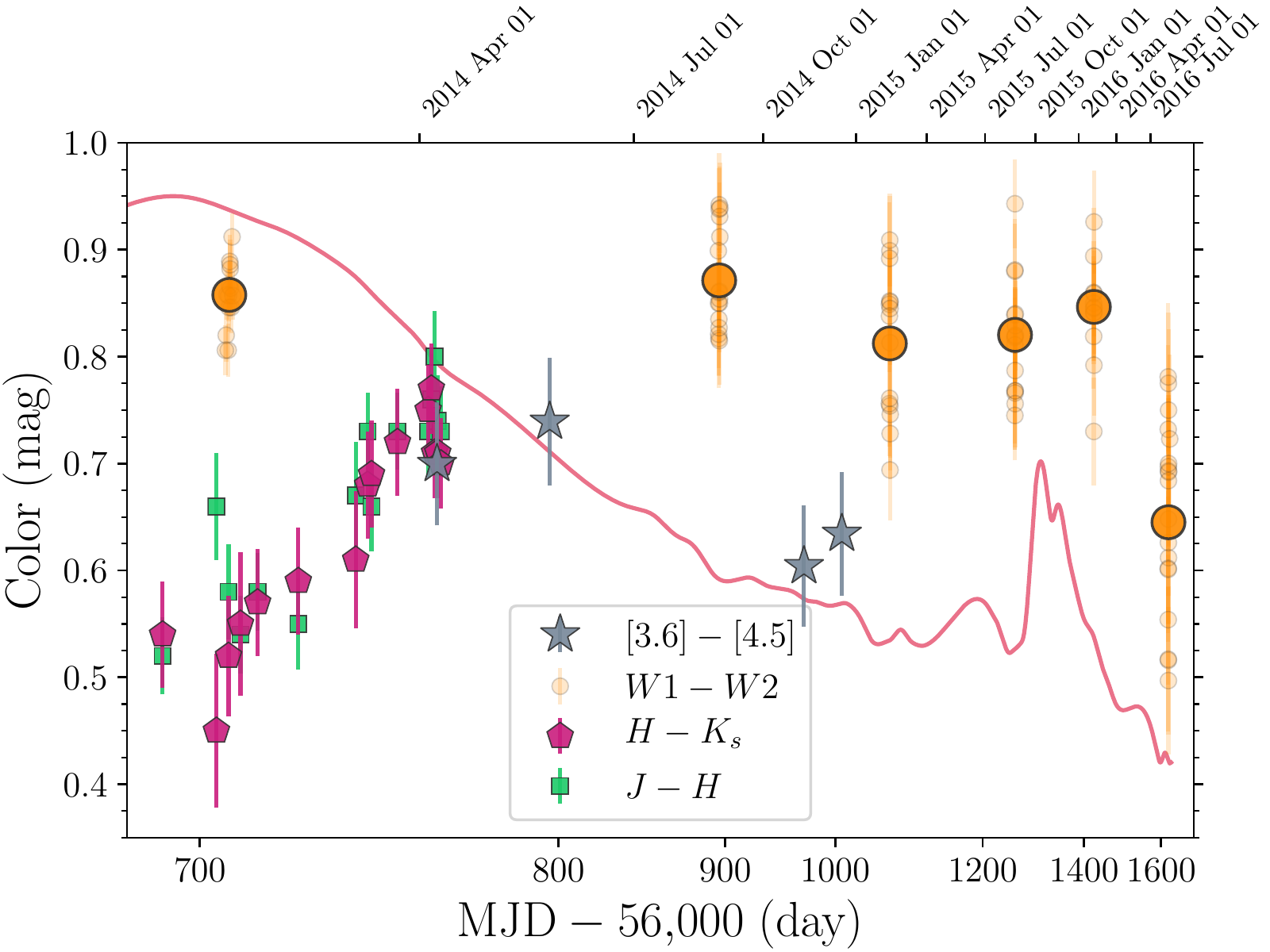}
\end{center}
\caption{
Color curves superposed on a scaled version of the gaussian-process model fit to the $R_{PTF}$ photometry of Figure~\ref{fig:lightcurve}. 
In the optical (left panel), colors were flat in the immediate post-peak epochs, but then  
the source became systematically redder as it faded beyond one e-folding time, 
with small overall color change.  Optical colors became relatively bluer around the tertiary maximum
(at MJD=57300 to 57400), and comparable to those around the first maximum.
In the near-infrared (right panel; note change of scale), by contrast, the colors became systematically redder immediately following the optical peak. 
The mid-infrared (right panel) color evolution is relatively flat in the post-peak period according
to the $NEOWISE$ data (where large point is the average of all measurements at a given epoch),
but the $Spitzer$ photometry appears to indicate blueing behavior.
}
\label{fig:colorcurves}
\end{figure}

Observations at the Palomar 60-in telescope 
were conducted using the ``GRB camera" \citep{cenko2006} 
in $ugriz$ bands, starting on 09 Feb 2014 UT. 
Aperture photometry was performed using SExtractor \citep{bertin1996} 
and is reported in Table~\ref{tbl:P60_phot}. 

\begin{deluxetable}{rrrrrrrrrrrrrrr}
\tabletypesize{\scriptsize}  
\tablecolumns{15}
\tablewidth{0pt}
\setlength{\tabcolsep}{3pt}
\tablecaption{P60 $ugriz$ Observations \label{tbl:P60_phot}}
\tablehead{\colhead{MJD$_u$} & \colhead{$u$} & \colhead{$\sigma_u$} &
    \colhead{MJD$_g$} & \colhead{$g$} & \colhead{$\sigma_g$} &
    \colhead{MJD$_r$} & \colhead{$r$} & \colhead{$\sigma_r$} &
    \colhead{MJD$_i$} & \colhead{$i$} & \colhead{$\sigma_i$} &
    \colhead{MJD$_z$} & \colhead{$z$} & \colhead{$\sigma_z$} \\
    \colhead{} & \colhead{(mag)} & \colhead{(mag)} & 
    \colhead{} & \colhead{(mag)} & \colhead{(mag)} & 
    \colhead{} & \colhead{(mag)} & \colhead{(mag)} & 
    \colhead{} & \colhead{(mag)} & \colhead{(mag)} & 
    \colhead{} & \colhead{(mag)} & \colhead{(mag)} }
\startdata
  .   &   .   &   .   &     .     &   .    &  .    &  .   &   .    &  .    &   .   &   .   &   .   &   .   &   .   &   .   \\
\enddata
\tablecomments{$griz$ photometry has been calibrated relative to SDSS stars (see text). 
    $u$-band photometry has been calibrated relative to \textit{Swift} UVOT observations of the same field.}
\tablenotetext{a}{5$\sigma$ upper limit.}
\end{deluxetable}

The $griz$ data were calibrated using 6 secondary standards, 
whose brightness was determined relative to SDSS stars observed by P60 on 
2014 Feb 11 UT, a photometric night. Based on the scatter in the photometric 
solution, there is a $\sim$0.04 mag systematic uncertainty for the reported 
$griz$ photometry. There was an insufficient number of P60 $u$-band observations 
of SDSS fields to determine secondary $u$-band standards, thus, P60 $u$ 
observations were calibrated using the \textit{Swift}/UVOT $U$-band observations 
of the field.  In the two epochs of UVOT observations there 
were 10 bright ($U < 18$ mag) stars that did not show significant variations, 
which we adopt as our $U$-band secondary standards.  The UVOT $u$ filter does 
not perfectly match the one employed by the P60, and thus we expect a 
$\sim$0.1 mag systematic uncertainty associated with these calibrations. 

Figure~\ref{fig:lightcurves} shows the multi-wavelength evolution in brightness,
and Figure~\ref{fig:colorcurves} the color evolution. 
No colors are available during the outburst, but the decline from peak brightness
shows little color change at optical wavelengths, until after about 1 e-folding time when
the source appears to become slightly redder by about 0.2 mag in $g-r$ and 0.1 mag in $r-i$.
Adopting the source extinction estimated below of $A_V=4.75$ mag and dereddening the colors
at the outburst peak results in intrinsic 
$u-g \sim 0.0$, $g - r \sim -0.7$, $r - i \sim -0.4$, $i - z \sim -0.5$.

\subsection{Near-Infrared Photometry} 

\begin{figure}
\includegraphics[scale=0.50]{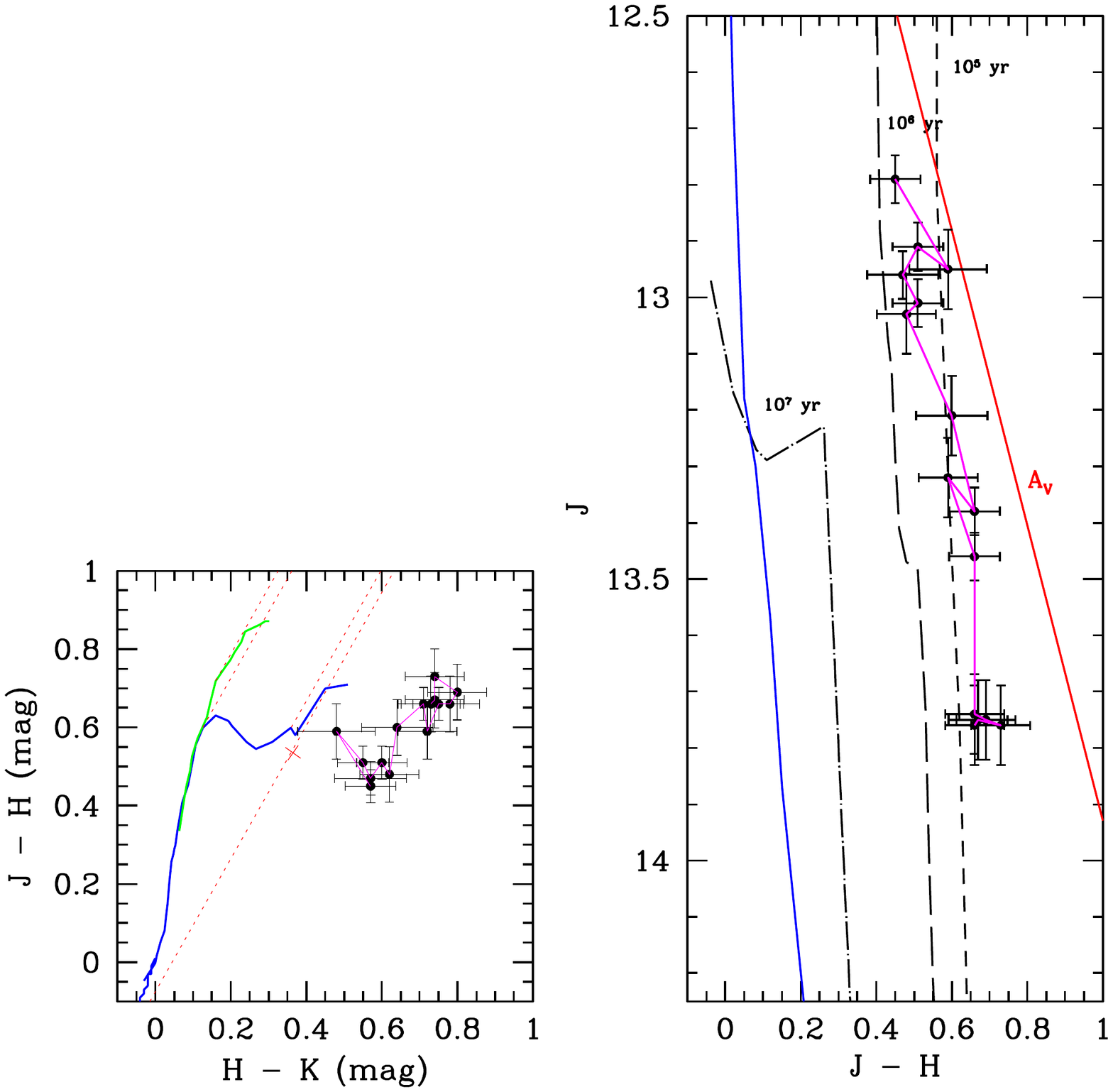}
\vskip -1truein
\caption{Near-infrared photometry in the post-peak outburst stage.  
The color-color diagram compares \jg\ to the sequence of normal dwarf and giant 
(solid lines in blue, green) stars, and illustrates the infrared excess.
The color-magnitude diagram compares 
to pre-main sequence isochrones assuming the 1.95 kpc distance (black lines), 
and illustrates the pre-main sequence nature of the source.
Magenta lines connect the time series and show \jg\ generally 
becoming redder over time as it faded, post-peak. The fade is somewhat steeper 
than the expectations from standard extinction (red lines), with 
less color change given the magnitude change.
} 
\label{fig:jhhk}
\end{figure}

Near-infrared photometric observations were carried out from the 1.2 m telescope
of Mount Abu Infrared Observatory 
using the Near-Infrared Camera/Spectrometer (NICS) equipped with a $1024 \times 1024$
HgCdTe Hawaii 1 array (Anandarao et al 2008; 
Banerjee \& Ashok, 2012). The camera has an unvignetted 8x8 arcminute square field and uses $JHK_s$ filters that conform to Mauna Kea Observatories (MKO) specifications. Frames in each filter were obtained in five dithered positions offset typically by 30 arcsec, with multiple frames (generally 5) being obtained in each dithered position.  Dark frames of same integration time as those used for the science frames were acquired.

Bad pixel masking was applied and corrections were made for cosmic ray hits. 
The dark-subtracted dithered frames were median combined to produce a sky flat frame. Individual frames were then divided with the sky flat normalized to unity.
The flat-corrected science frames were again median combined in each filter to produce a median sky frame, which was subtracted from the individual science frames in order to correct for bias, dark, and sky background. 
The flat-fielding and sky subtraction methods adopted are customary techniques in the near-infrared\footnote{e.g. {\url{www.cfht.hawaii.edu/Instruments/Detectors/IR/Redeye/Manual/chapter7.html}}}. 
Finally, the corrected science frames were co-added to produce an average frame on which photometry was done. 

Aperture photometry was derived using aperture photometry routines in $IRAF$ with the nearby stars 2MASS J02403063+6051272 and J02403914+6052201 
(avoiding the nearest bright star to \jg\ which appears to be a variable)
used for photometric calibration. Because NICS employs MKO filters, the 2MASS magnitudes of the calibrating stars were first converted to the MKO near-infrared photometric system using the transformation equations of \cite{leggett2006}. The resulting $JHK_s$ magnitudes of \jg, in the MKO photometric system, are listed in Table~\ref{tbl:NIR_phot} and the photometry is illustrated in
Figure~\ref{fig:jhhk}. 

\begin{deluxetable}{rrrrrrr}
\tabletypesize{\small}
\tablecolumns{7}
\tablewidth{0pt}
\setlength{\tabcolsep}{6pt}
\tablecaption{Mt.\ Abu NIR Observations\label{tbl:NIR_phot}}
\tablehead{\colhead{MJD} & \colhead{J} & \colhead{$\sigma_J$} &
    \colhead{H} & \colhead{$\sigma_H$} & \colhead{{$K_s$}} & \colhead{$\sigma_{K_s}$} \\
    \colhead{} & \colhead{(mag)} & \colhead{(mag)} & \colhead{(mag)} & \colhead{(mag)} & 
    \colhead{(mag)} & \colhead{(mag)}}
\startdata
56694.61&	12.79&		0.03&	12.34&		0.03&	11.77&		0.06 \\
56702.60&	12.95&		0.05&	12.36&		0.05&	11.88&		0.09 \\
56704.59&	12.91&		0.03&	12.40&		0.03&	11.85&		0.06 \\
56706.63&	12.96&		0.03&	12.49&		0.03&	11.92&		0.09 \\
56709.67&	13.01&		0.03&	12.50&		0.03&	11.90&		0.06 \\
56717.60&	13.03&		0.05&	12.55&		0.05&	11.93&		0.06 \\
56730.60&	13.21&		0.05&	12.61&		0.05&	11.97&		0.08 \\
56733.59&	13.38&		0.03&	12.72&		0.03&	12.01&		0.06 \\
56734.59&	13.32&		0.05&	12.73&		0.05&	12.01&		0.06 \\
56741.59&	13.46&		0.03&	12.80&		0.03&	12.05&		0.06 \\
56750.61&	13.74&		0.05&	13.08&		0.05&	12.30&		0.06 \\
56751.61&	13.75&		0.05&	13.06&		0.05&	12.26&		0.06 \\
56752.61&	13.76&		0.05&	13.03&		0.05&	12.29&		0.06 \\
56753.61&	13.75&		0.05&	13.08&		0.05&	12.34&		0.06 \\
56754.61&	13.76&		0.05&	13.10&		0.05&	12.37&		0.06 
\enddata
\tablecomments{Photometry has been calibrated relative to the MKO system (see text).}
\end{deluxetable}

The infrared burst colors are redder than can be explained
by reddened stellar photospheres, and consistent with the colors of 
young star-plus-circumstellar-dust systems.
Unlike in the optical, following its peak brightness,
\jg\ exhibited a color change in the near-infrared during its early decline 
(Figure~\ref{fig:colorcurves}), becoming redder. If interpreted as increasing 
extinction, the change corresponds to almost 0.4 mag in $A_K$  or 3-4 mag in $A_V$.
Considering the large near-infrared color changes relative to
the unchanging optical colors during this same time period could suggest 
scattering in addition to reddening. Alternately, that the near-infrared 
becomes redder while the optical has the same colors could be due to a dust echo.

\subsection{Mid-Infrared Photometry} 

The $WISE$ satellite, after not dectecting the \jg\ progenitor during
its main mission when observing the field in 2010 and 2011, 
did detect it in outburst during the warm $WISE$ mission 
re-dubbed as $NEOWISE$ \citep{mainzer2014}. 
A strong detection was recorded on 17 February 2014, around the optical peak, 
with W1 = 10.83 mag and W2 = 9.98 mag ($W1-W2= 0.85$ mag in color).
The available $NEOWISE$ time series (Figure~\ref{fig:colorcurves}, right panel) 
shows no evidence for color evolution until the last epoch, though the data are noisy.
This late-time blue-ing seems to occur 100-200 days later than the blue-ing
seen in the optical colors (Figure~\ref{fig:colorcurves}, left panel),
which is associated with the tertiary peak in the optical lightcurve. 

Mid-infrared photometry was also obtained with the $Spitzer$ Space Telescope,
based on a DDT allocation, over four epochs (2014 April, May, November, December). 
The new data are reported in Table~\ref{tbl:MIR_phot} and also illustrated in
Figures~\ref{fig:lightcurves} and Figure~\ref{fig:colorcurves}.
Compared to the pre-outburst $Spitzer$ color of [3.6]-[4.5] = 0.4$\pm$0.2 mag
reported above, the outburst color was redder at [3.6]-[4.5] = 0.7 mag 
but seems to have became slightly bluer as the source faded by nearly 2 mag 
in the mid-infrared, reaching [3.6]-[4.5] = 0.6 mag in the long plateau phase.  
If significant, this is the only color to exhibit a blueing trend
in the outburst period.

Referring to Figure~\ref{fig:lightcurves}, near the optical peak 
\jg\ was 0.1 mag fainter in W1, and 0.05 mag brighter in W2,
than recorded 2 months later in S[3.6] and S[4.5], respectively.  
Similarly, the next $NEOWISE$ epoch shows W1 fainter and W2 brighter than
the corresponding S[3.6] and S[4.5] data taken a few months later.
These small magnitude differences lead to the color differences between
$NEOWISE$ and $Spitzer$ that are apparent in Figure~\ref{fig:colorcurves}.
We believe that the $NEOWISE$-$Spitzer$ offsets are due to
color terms that arise for red sources like \jg.  
\cite{hillenbrand2018} report a similar finding and an empirically derived  
relation for transforming $Spitzer/IRAC$ photometry to the $NEOWISE$ filter system: 
$(W1-W2) = 1.62\times(I1-I2) - 0.04$ mag, with rms=0.24 mag.
This indeed brings the two colors for \jg\ into better agreement.

\begin{deluxetable}{rrrrr}
\tabletypesize{\small}
\tablecolumns{5}
\tablewidth{0pt}
\setlength{\tabcolsep}{6pt}
\tablecaption{$Spitzer$ Photometry\label{tbl:MIR_phot}}
\tablehead{\colhead{MJD} & \colhead{[3.6]} & \colhead{S/N}  & \colhead{[4.5]} & \colhead{S/N}}
\startdata
56753.394   &   10.72 &  2109     &10.03    &     2773 \\       56796.004   &   11.30 &  1812     &10.57    &     1918    \\    56968.136   &   12.42 &  1015     &11.83    &     1187      \\  57007.288   &   12.39 &  1212     &11.78    &     1213          \enddata
\end{deluxetable}

\subsection{Millimeter Photometry} 

$CARMA$ was used to observe \jg\ at 108 GHz (2.8 mm) 
on 15 February, 2014. There was no detection in the continuum
at the expected source position (or anywhere within $\pm$40").
The RMS noise level was 0.30 mJy and the beam size 5.7 $\times$ 3.6 arcseconds. 
Also, $^{12}$CO was not detected between VLSR velocities 
of -2.6 and -83.4 \kms, with RMS noise level of 0.35 K 
(in 1 \kms\ channels) and a beam size of 5.6 $\times$ 3.3 arcseconds. 

The 4-sigma upper limit to the flux density of 1.2 mJy implies 
an upper limit to the dust mass of 0.26 \msun, assuming: a distance of 2 kpc,
``standard" disk opacities of 2.3 cm$^2$/g at 230 GHz and $\beta=1$ \citep{beckwith1990},
a gas to dust ratio of 100, and a dust temperature of 25 K.
If the source is located at larger distance,
then our limits are higher by the distance ratio squared. 

\subsection{Low- and High-Dispersion Spectroscopy} 

As summarized in Table~\ref{tbl:spec}, a number of telescopes and instruments were 
used to characterize the spectroscopic evolution of \jg\,
as it reached its photometric peak and then faded in brightness.
The spectra were obtained at: 
the Apache Point Observatory (DIS optical spectra at R$\approx$2400, TripleSpec infrared spectrum at R$\approx$3500),
Palomar Observatory (DBSP optical spectra at R$\approx$2000, TripleSpec infrared spectrum at R$\approx$3000),
and Keck Observatory (HIRES optical spectra at R$\approx$34,000-48,000,
DEIMOS optical spectrum at R$\approx$2200, MOSFIRE infrared spectrum at R$\approx$3300).
Our first optical and infrared spectra at low dispersion were taken about a week before the peak, 
and the second optical spectrum was obtained about one week post-peak.
Spectroscopic monitoring continued during the initial decline and for the next several years  
during the lightcurve plateau phase, secondary peaks, and resumed declines.

In addition to the low-dispersion data, Table~\ref{tbl:spec}
also reports high-dispersion spectroscopic observations.
Keck/HIRES spectra were obtained on two occasions
by Howard Isaacson and processed through the standard 
California Planet Search image processing and spectral extraction pipeline.
Four additional Keck/HIRES spectra were acquired by LAH
and reduced using the MAKEE package (written by Tom Barlow). 
The S/N values for these spectra range from 15 to 5 around 5500 \AA, 
and from 35 to 15 around 7500 \AA. 

\begin{figure}
\begin{center}
\includegraphics[scale=0.75]{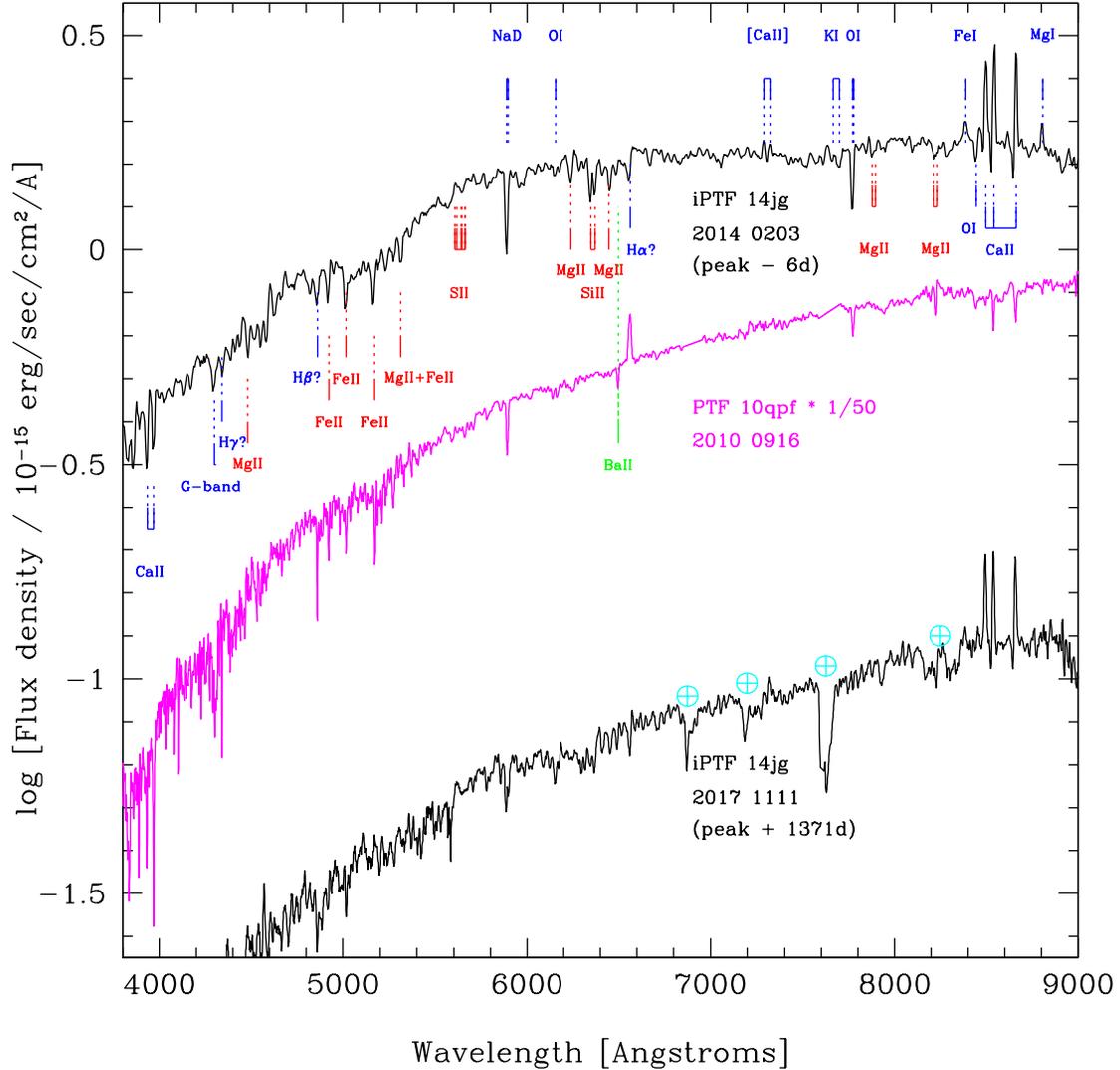}
\end{center}
\vskip-1.2truein
\caption{
Comparison of the \jg\ low-dispersion optical spectrum with that of PTF~10qpf
(HBC 722), a {\it bona fide} FU Ori star. 
Denoted spectral features are in blue when seen in young stars of
various categories, and in red if not typically seen in young stars.
The low-ionization \ion{Ba}{2}/\ion{Ca}{1}/\ion{Fe}{1} blend of lines 
at 6497 \AA\ that is commonly seen in FU Ori stars, but is weak in 
the hotter spectrum of \jg\ (and may in fact be an \ion{Fe}{2} line),
is also indicated.  This first \jg\ spectrum, exhibiting 
red continuum, weak hydrogen lines, and ionized metal absorption components,
initially appeared to be a reasonable albeit imperfect match 
to stars in the temperature range of F and G stars. 
However, there was additional absorption, notably within 6200-6500 \AA, 
that is not explained by such cool photospheres; the 4900-5200 \AA\
absorptions are likewise better attributed to hotter ionized lines than the
cooler neutral lines of FG stars.
The bottom spectrum, taken nearly four years after the initial rise, illustrates the
nearly colorless optical fade, yet still shows hot ionized line absorption;  
Telluric absorption regions are marked in cyan.
}
\label{fig:optspec}
\end{figure}

The initial \jg\ optical low-dispersion spectrum 
(see Figure~\ref{fig:optspec} where it is compared to 
PTF 10qpf = HBC 722 in outburst) showed a red continuum,  
with prominent absorption in \ion{Ca}{2} H \& K, the \ion{Na}{1} D doublet, \ion{O}{1} 7774 and 8446 \AA, and possibly weak \ion{Ba}{2} 6497 \AA. 
It initially compared well to late F and G 
giant/supergiant templates from Silva \& Cornell (1992), with the exception of 
H$\alpha$, which if present, was weak and blueshifted.
The \ion{Ca}{2} ``near-infrared" triplet exhibited an emission component 
as part of a P-Cygni type profile, indicating outflow.  
The [\ion{Ca}{2}] doublet at 7291, 7324 \AA\  was in emission, 
and there was also weak emission from \ion{Fe}{1} multiplet 60, 
most strong at 8387 \AA, as well as \ion{Mg}{1} 8807 \AA.

Moderate strength absorption from \ion{Si}{2} 6347 and 6371 \AA\ 
was also apparent in this earliest outburst spectrum. 
Aided by the high-dispersion spectrum obtained later,
we identified additional ``hot" lines visible at low dispersion, 
specifically from e.g.  \ion{Si}{2} 3858, 4128, 4815, 5958, 5979 \AA; 
from \ion{Mg}{2} 4481, 5228, 5311, 5954, 5982, 6239, probable 6343 and 6366 (contributing in the \ion{Si}{2} 6347 and 6371 regions), 
6447, 7877, 7896, 8214, 8235 \AA; 
and from \ion{Fe}{2} e.g. 4924, 5018, doublet 5169, 5316, 6433 \AA.
Many of these lines have blue asymmetries in their absorption profiles.

In the earliest high dispersion spectrum (discussed in detail below in \S7, 
a confounding mix of broad and deep absorption features, 
with no readily identifiable continuum,
plus the narrow emission-line component mentioned above.  
Examination at high dispersion of the deep lines seen in
the low resolution spectrum showed that many have strong outflow signatures.
This complicated the spectrum interpretation  -- especially as many of the 
lines could not even be uniquely identified given their breadth and blueshift. 

\begin{figure}
\begin{center}
\includegraphics[width=0.9\textwidth,trim={0.75cm 0 0.75cm 0},clip]{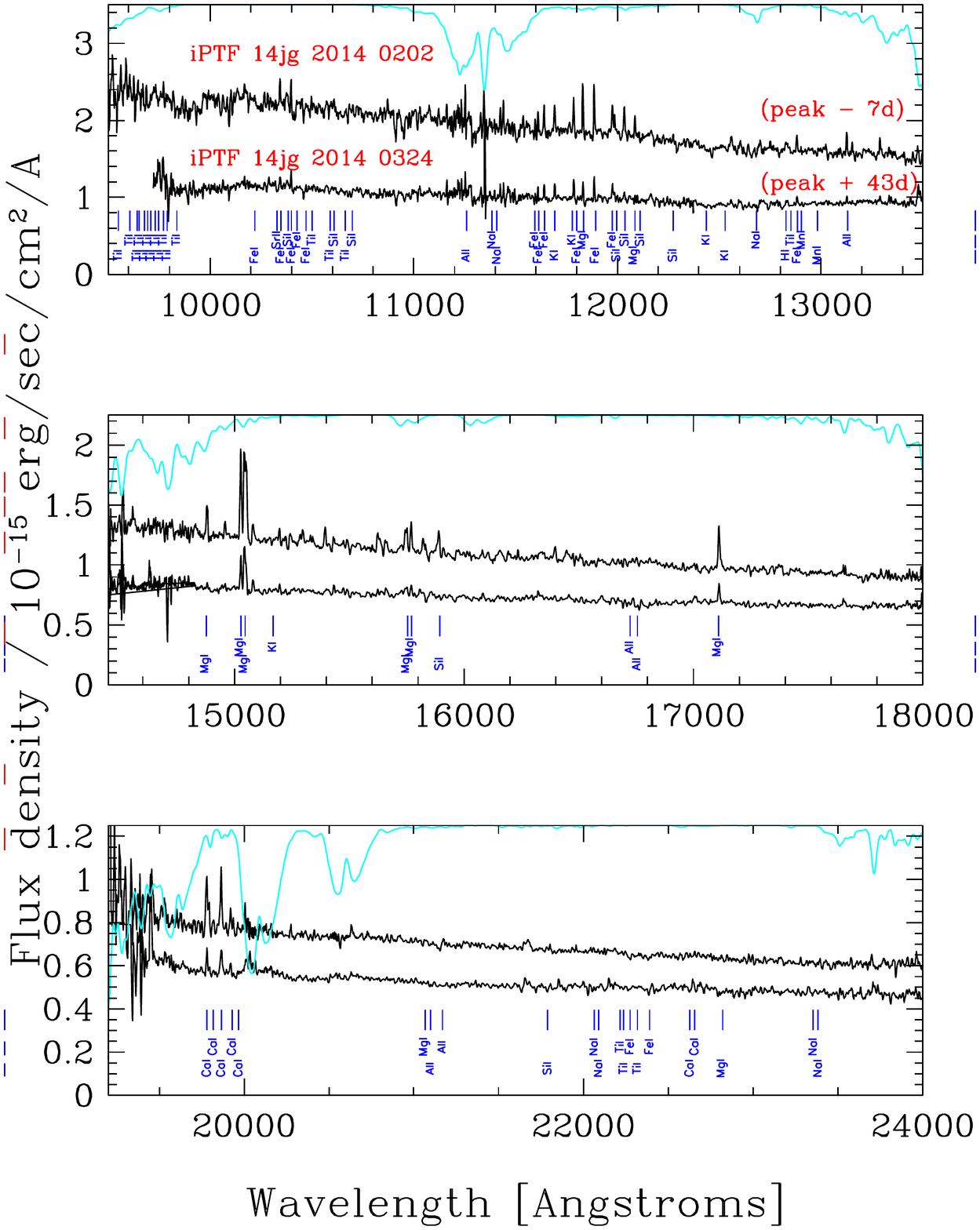}
\end{center}
\caption{
\jg\ infrared spectrum at two early epochs, illustrating a blue continuum with
superposed narrow emission lines (primarily \ion{Mg}{1}, \ion{Ca}{1}, and \ion{Fe}{1}) 
that faded within less than six weeks. 
For reference, atomic spectral lines that are seen in the cool stars 
presented by \cite{cushing2005} are marked in blue;
there are no identifiable molecular features.
The cyan line is a model atmospheric transmission spectrum 
plotted on a vertical scale from 0-100\%
and indicates regions where the telluric correction applied to the data 
is large and possibly uncertain.
}
\label{fig:irspec}
\end{figure}

The \jg\ near-infrared spectra (see Figure~\ref{fig:irspec}) 
exhibited a {\it blue} continuum, in contrast to the red optical continuum. 
There are no detectable absorption features anywhere, but  
several narrow metallic emission lines, 
consistent with the presence of such lines in the optical spectrum.
No forbidden line emission was seen, unlike in the optical where [\ion{Ca}{2}] 
was exhibited.  Nor was there any H$_2$ or CO bandhead emission (or absorption)
in the infrared spectrum; these molecular features are seen in some young stars
having only optical and near-infrared continua, with few absorption lines, 
such as \jg.

\begin{deluxetable}{clrlcc}    \tabletypesize{\small}
\tablecolumns{7}
\tablewidth{0pt}
\tablecaption{Summary of Spectroscopic Observations \label{tbl:spec}}
\tablehead{\colhead{UT Date} & \colhead{MJD} &\colhead{Post Peak} &\colhead{R$_{PTF}$}& \colhead{Telescope / Instrument} & \colhead{Observers / Reducers}  \\    \colhead{[YYYY MMDD]} & \colhead{[day]} &\colhead{[day]} &\colhead{[mag]}& \colhead{} & \colhead{}  }  \startdata
{\underline{Optical Spectra}}&&& \\
2014 0203 & 56691  &  -6   &15.05 & APO / DIS & M. Kasliwal / Y. Cao\\ 
2014 0218 & 56706  &  9   &15.11 & APO / DIS & M. Kasliwal/ Y. Cao \\
2014 0221 & 56709  &  12  &15.15 & Keck / HIRES & H. Isaacson et al.\\
2014 0225 & 56713  & 16   &15.20  & Palomar 200" / DBSP & A. Waszczak  \\
2014 0301 &  56717 &  20      & 15.25& Keck / DEIMOS & S. Tang \&  Y. Cao  \\ 2014 0324 & 56740  &  43   &  15.73 & Palomar 200" / DBSP & A. Waszczak \\ 2014 0404 & 56751  &  54   & 15.99 & Palomar 200" / DBSP & A. Waszczak \& A. Miller  \\
2014 0624 & 56832  &  135  & 16.92 & Palomar 200" / DBSP & I. Arcavi  \\
2014 0822 & 56891  &  194  &17.36 & Keck / HIRES & H. Isaacson et al.\\
2014 1123 & 56984  & 287   &17.54 & Palomar 200" / DBSP & L. Hillenbrand \& A.M. Cody  \\ 2014 1126 & 56987  & 290   &17.54 & APO / DIS & M. Kasliwal / Y. Cao \\ 2014 1209 & 57000  & 303   &17.55 & Keck / HIRES & L. Hillenbrand \\
2015 0723  & 57226  & 529   &17.67 & Palomar 200" / DBSP & Pavanman / Khazov  \\
2015 0724 & 57227  & 530   &17.68 & Keck / HIRES & L. Hillenbrand  \\
2015 1027 & 57322  & 625   &16.91 & Keck / HIRES & L. Hillenbrand \\
2015 1206 & 57362  & 665   &17.18 & Palomar 200" / DBSP & Lunnan, Bladgorodnova / Cao  \\
2016 0203 & 57421  & 724   &17.86 & Keck / HIRES & L. Hillenbrand \\
2016 0826 & 57626  & 929  &18.48 & Palomar 200" / DBSP & Cook / Knezevic \\ 2017 1111 & 58068  & 1371 & \nodata& Palomar 200" / DBSP & Ho, Kulkarni  \\
\hline
{\underline{Infrared Spectra}}&&& \\
2014 0202 &  56690 &  -7      &15.08 & APO / TSpec & M. Kasliwal / S. Tang  \\  2014 0323 & 56739  &  43    &15.71 & Palomar / TSpec & Y. Cao, D. O'Sullivan / J. Jencsen \\  \enddata
\end{deluxetable}

\section{Analysis of Outburst Colors and Spectral Energy Distribution}

In the optical, the (dereddened) burst colors of \jg\ 
are blue and imply ultraviolet excess relative to a normal 
stellar atmosphere. 
The colors are also much bluer than typical for young star accretion systems, 
and they would be on the extreme blue end 
of known cataclysmic variables (by 0.1-0.2 mag).

In the near-infrared, as shown in the left panel of Figure~\ref{fig:jhhk}, 
the burst colors of $J-H \approx 0.45-0.75$ mag and 
$H-K \approx 0.5-0.85$ mag are red, and denote an 
infrared excess relative to the colors expected from normal stars, 
in a manner that can not be explained by reddening.  
\jg\ resides in the color regime populated 
by standard low-mass T Tauri stars and higher mass Herbig Ae/Be stars
with disks. The colors are similar to, but slightly bluer than 
those of FU Ori stars, including FU Ori itself,  V1515 Cyg, and  V1057 Cyg,  
and are much bluer than those of the EX Lup stars.

The mid-infrared colors of the \jg\ outburst 
are also redder than can be explained by reddened stellar photospheres, 
with $K-L > 2$ mag, $[3.6]-[4.5]=0.7$ mag, and $W1-W2 \approx 0.85$ mag.  
In the pre-outburst phase, as mentioned above,
Spitzer measured an uncertain color of $[3.6]-[4.5] \approx 0.4$,
indicating moderate infrared excess (though formally only at the 2$\sigma$
level).  

\begin{figure}
\includegraphics[scale=0.45]{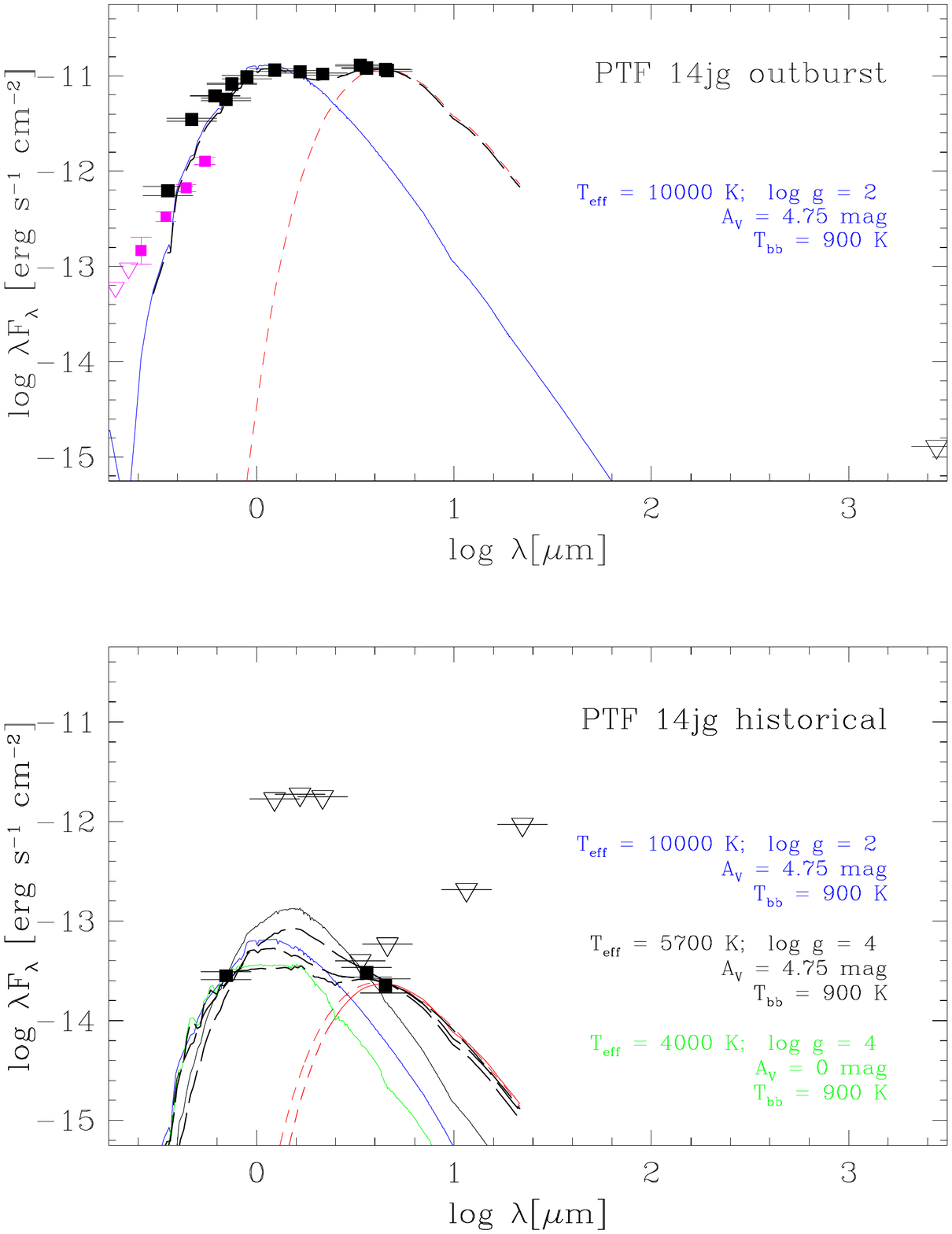}
\includegraphics[scale=0.45,trim={0 0 1.75cm 0},clip]{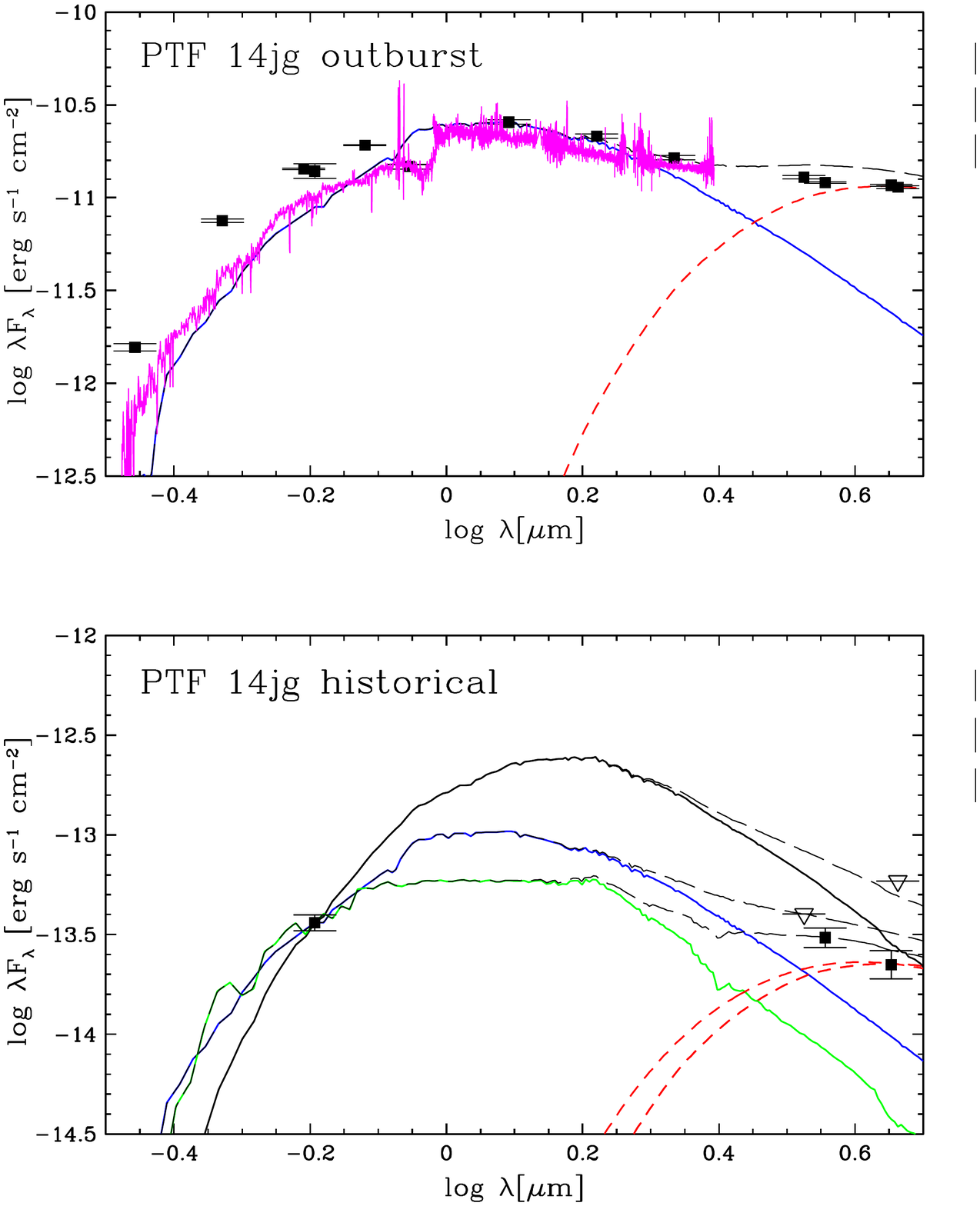}
\caption{Post-outburst (top panels) and pre-outburst (bottom panels)
spectral energy distribution compared to reddened model photospheres 
allowing for a blackbody infrared excess.  
Left panels have the same axis scaling. Right panels are an expanded view
around the SED peak, with magenta line showing the
outburst optical and near-infrared spectrophotometry.
Filled symbols are measurements 
(from UVOT, P60, P48/PTF, Mt. Abu, NEOWISE-R in outburst, and
from P48/PTF and Spitzer in the pre-outburst phase) 
while open triangles are upper limits
(from CARMA in outburst and from WISE and 2MASS 
in the pre-outburst phase).
Blue lines represent the hot supergiant extincted photosphere 
that is indicated by the outburst spectrum.
In the bottom panels, black line is the minimum temperature photosphere 
that can still match the pre-outburst SED at the same value of extinction.
Green line in the bottom panels represents a much cooler photosphere that
can fit the data only if the extinction is also much lower, with the 
minimum case of 0 mag shown.
Red lines indicate a 900 K blackbody matched to the longest wavelength
data point. 
}
\label{fig:sed}
\end{figure}

Figure ~\ref{fig:sed} shows the available pre-outburst and outburst 
spectral energy distribution compared to model photospheres.  
We have already hinted that that the outburst
source is spectroscopically hot, and in the next section 
we present our conclusions regarding an early type optical spectrum.
The observed spectral energy distribution is therefore modelled 
in outburst with an A0 supergiant stellar atmosphere.  
A hot (900 K) infrared excess is also required. and in order to model 
the full SED, extinction of $A_V=4.75$ mag is implied.   

We illustrate the same model fit to the pre-outburst photometry.
However, the limited number of pre-outburst data points
allows for a wider range of spectral type and extinction combinations 
to also provide a good match.  Specifically, 
photospheres warmer than $\sim$5700 K with the same extinction
and the same temperature blackbody dust excess as in outburst
can also fit the photometric measurements in R-band, 3.6 and 4.5 $\mu$m.  
Cooler photospheres demand less extinction, 
with e.g. a 4000 K photosphere and no extinction
plus the same temperature (but higher luminosity) blackbody excess 
also a good match.  
The nature of the pre-outburst object is not well-constrained.

In the outburst phase, the shape of the 1-2.5 $\mu$m spectrum has
a best-fit Planck function of temperature $T=$3417 K for the observed spectrum, and
$T=$ 9438 K for the spectrum de-reddened using the adopted $A_V$=4.75 mag value.
This rough color temperature estimate was already implied by the
broadband SED fitting above, but more directly demonstrates consistency 
between the hot spectral type
of the source derived from spectral absorption lines in the optical (see \S7), 
and the SED slope in the near-infrared.

The SED is totally unconstrained at long wavelengths.
The mass upper limit from the sub-millimeter observations is 
a factor of ten higher than the median mass of T Tauri disks.
It is thus not too surprising that we did not detect the source if it contains 
only a disk.  In the context of FU Ori stars, the upper limit is
at about the median mass according to the SED modelling by \citet{gramajo2014},
which includes both envelope and disk mass. 

Finally, as shown in the right panel of Figure~\ref{fig:jhhk}, \jg\ is 
consistent with a young pre-main sequence star at the assumed source distance,
regardless of the exact reddening value.  

\section{Analysis of Outburst Lightcurve}

The PTF lightcurve covers approximately 1130 days of the \jg\ pre-outburst and outburst evolution.
Prior to the major brightening, the source was variable about its mean magnitude at the 0.60 mag (rms) level.
After the outburst, however, variation about the fitted exponentials 
(see below) was only 0.09 mag (rms).

The immediate post-peak behavior of \jg\ (see Figure ~\ref{fig:peak})
was that of roughly linear decline at a rate of $\sim$0.5 mag/month.
After several weeks, the lightcurve 
increased its decline to $\sim$0.8 mag/month, which again lasted a few weeks,
as it then transitioned to a generally exponential shape. 
There are indications of several brief plateaus over the post-peak year. 
Then, between 1 and 3 years after the first peak,
several more substantial departures from the exponential decay occurred, 
with evidence for two local maxima (see Figure ~\ref{fig:lightcurve}).
These brief rises and the exponential decays from them resulted in 
the \jg\ lightcurve resuming 
its previous exponential decline from the main peak.

Using traditional nova nomenclature, the post-peak times $t_1$, $t_2$, and $t_3$ (corresponding to the
times at which the lightcurve has declined from its maximum by 1, 2, and 3 mag, respectively),
are 54, 139, and 747 days.  These are much longer timescales than typical of the standard novae categories.
Regarding the lightcurve shape, \jg\ would be similar to a P-Class or plateau nova, but it also
exhibits several C-Class type re-brightenings, albeit with very different shape compared to typical cusp novae \citep[see][]{strope2010}.

We fit the pre-outburst and outburst rise with a sigmoid function, coming to a peak 
of $R_{PTF}=14.96$ mag at a time $t=0$, and the post-peak decay with an exponential. 
The sigmoid has parameters $L/(1 + e^{(-k\times (t-t_o)/days)})$.  
For the first peak, our best-fit values are 
$t_o=-56.23\pm 0.52$ (meaning that the outburst starts 
$2\times t_o =  112.5$ days before its peak), 
$k=-0.066\pm 0.002$ (where the $1/k$ value represents a timescale for the rise, 
a 15.2 day e-folding), and $L=6.73\pm 0.04$ 
(representating the amplitude of the rise in magnitudes).
The exponential decay has parameters $a \times  e^{-b \times  t/days} + c$
with our best-fit values for the first peak $a=-2.92\pm 0.01$, $b=0.0090\pm 0.0001$, and $c=2.72\pm 0.01$.
The $a$ value indicates the post-peak decline in magnitudes.
The $1/b$ value represents a timescale of 111 days for an e-folding.

The additional brightening and exponential decay that is well-sampled in the PTF data, called the tertiary peak above, was also fit.  The resulting values 
are similar to those above, with $a=-2.34\pm 0.03$, $b=0.0058\pm 0.0002$, and $c=2.15\pm 0.04$.  The amplitude is smaller than the main peak by 0.5 mag, 
and the $1/b$ timescale is longer at 173 days.

\section{Analysis of Outburst Spectroscopy}

The low resolution optical and infrared spectra were shown in Figures~\ref{fig:optspec} and ~\ref{fig:irspec}.
Portions of the first optical high dispersion spectrum appear in
Figure~\ref{fig:hires} and Figure~\ref{fig:profiles}, the latter 
highlighting the velocity profiles of particularly illustrative lines.
Subsequent figures show the spectral evolution as \jg\ initially cooled
from its lightcurve peak.

\subsection{Radial Velocity}

We took advantage of the presence of narrow emission lines in the first
high-dispersion spectrum to estimate a radial velocity for \jg,
which was not discernable otherwise from the broad absorption lines.
To find an approximate velocity, we inverted the \jg\ spectrum 
and compared it to the G supergiant HR 8414 
with HRV = 6.63 \kms\ \cite{soubiran2008}. 
The measured heliocentric velocity was -26.6 \kms, assuming that 
the weak narrow-line emission spectrum is located at the systemic velocity
of the star.  This value guided our spectral analysis efforts for some time.

However, a re-asssessment of the radial velocity became possible when 
a less-wind-dominated absorption line spectrum emerged 
by the time of the 2015 1027 high dispersion spectrum.
At that point, with a narrower absorption-line spectrum presented, 
we were able to compare to an accepted set of true 
radial velocity standards 
that we have used in our studies of eclipsing binaries.  
We derived $-38.1 \pm 1.2$ \kms\  
using standards in the FGK spectral type range. 
This more robust radial velocity value is entirely consistent with 
location of \jg\ in the Perseus spiral arm.

\subsection{Spectral Type: A Composite Spectrum}

As described above,
the early low resolution optical spectrum (Figure~\ref{fig:optspec})
appeared roughly consistent with a reddened late F or early G supergiant spectrum.
This assessment was based on the classical signature of roughly equal strength
\ion{Ca}{2} H and K lines, which are also stronger than the very weak Balmer \ion{H}{1} lines,  
and the presence of weaker metallic features plus CH near H$\gamma$.
The lack of strong hydrogen appeared to exclude types earlier than late F. 
All K and later low-gravity spectral types 
were discounted based on the lack of expected strong absorption features from e.g. \ion{Ca}{1} 
for the K or TiO for the M star ranges.  
A luminosity classification of $I$ was justified by details 
such as the lack of the 4700 \AA\ absorption expected from luminosity class III
FG stars, and the presence of features such as 6238 \AA\ and 6486 \AA\ that are seen in 
luminosity class I but not III FG stars.  
However, early A- and B-type features were also present in the spectrum,
including the numerous \ion{Fe}{2}, \ion{Mg}{2}, and \ion{Si}{2} features listed earlier. 
Given the evidence for a hot temperature spectral component, 
it was somewhat surprising that there is no 
Balmer jump visible in the low-dispersion data, and 
that the HI lines are generally absent or very weak. 
Overall, the spectrum of \jg\ eluded strict classification.

\begin{figure}
\includegraphics[scale=0.4,trim={0.75cm 0 0.75cm 0.75cm},clip]{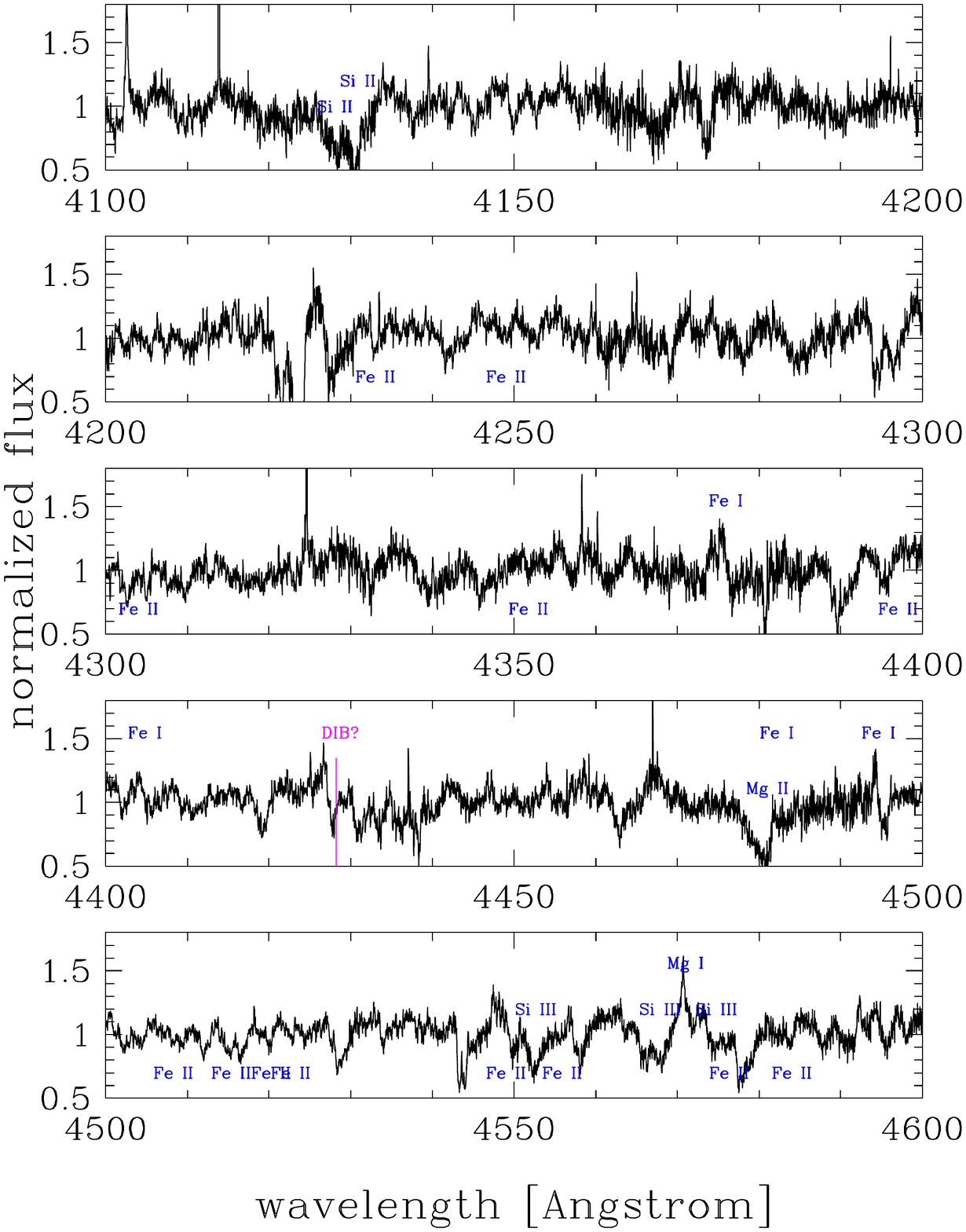}
\includegraphics[scale=0.4,trim={0.75cm 0 0.75cm 0.75cm},clip]{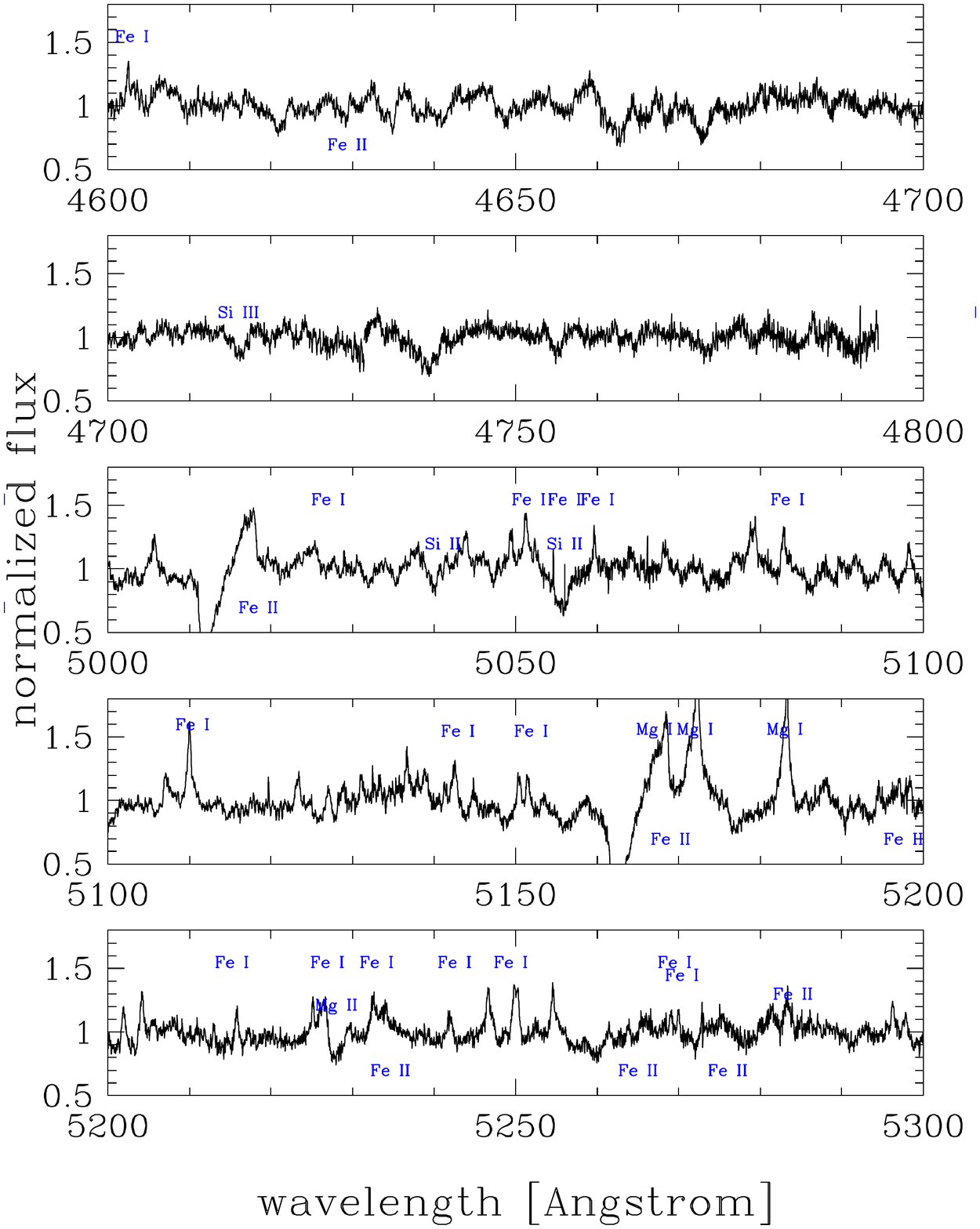}
\\
\includegraphics[scale=0.4,trim={0.75cm 0 0.75cm 0.75cm},clip]{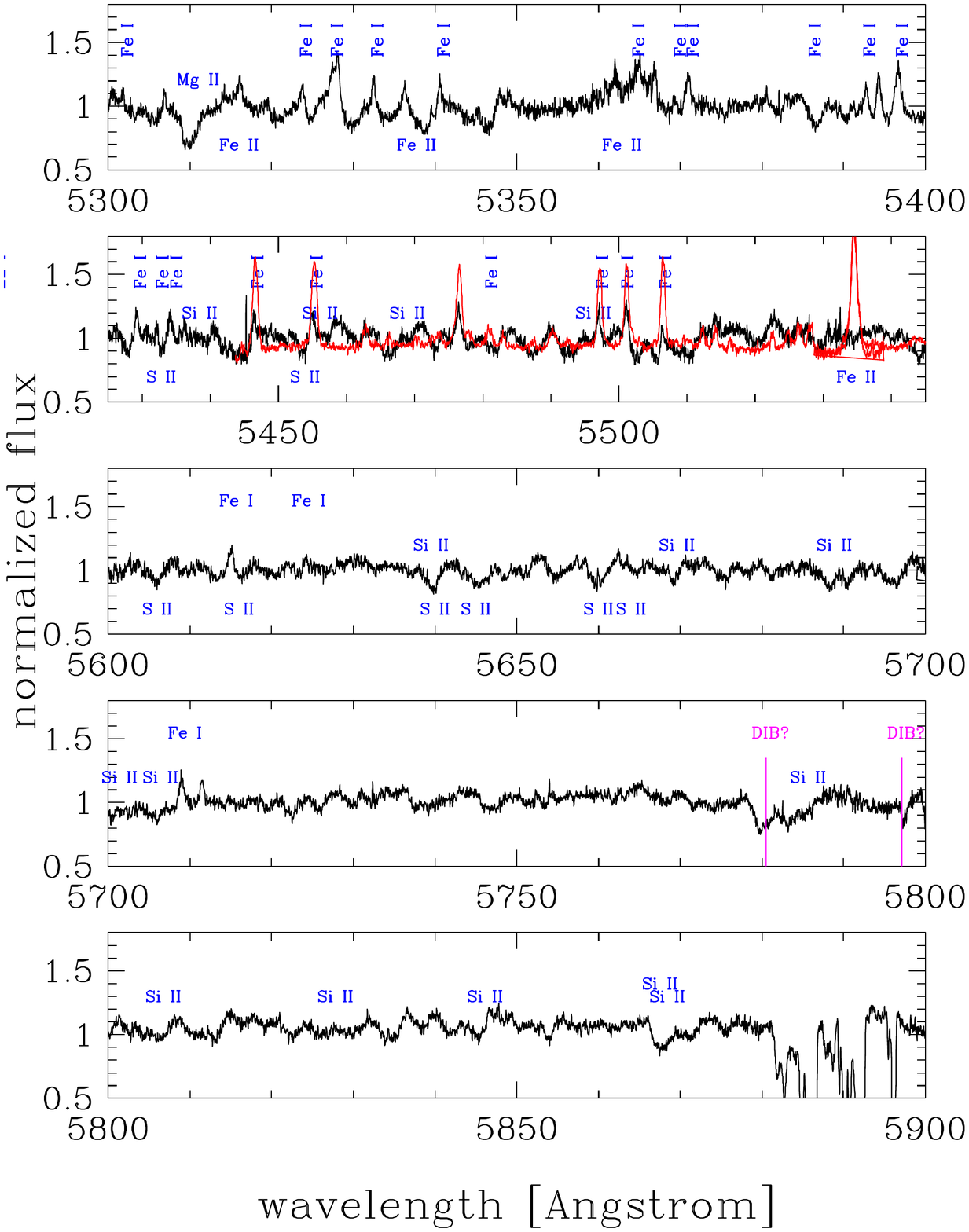}
\includegraphics[scale=0.4,trim={0.75cm 0 0.75cm 0.75cm},clip]{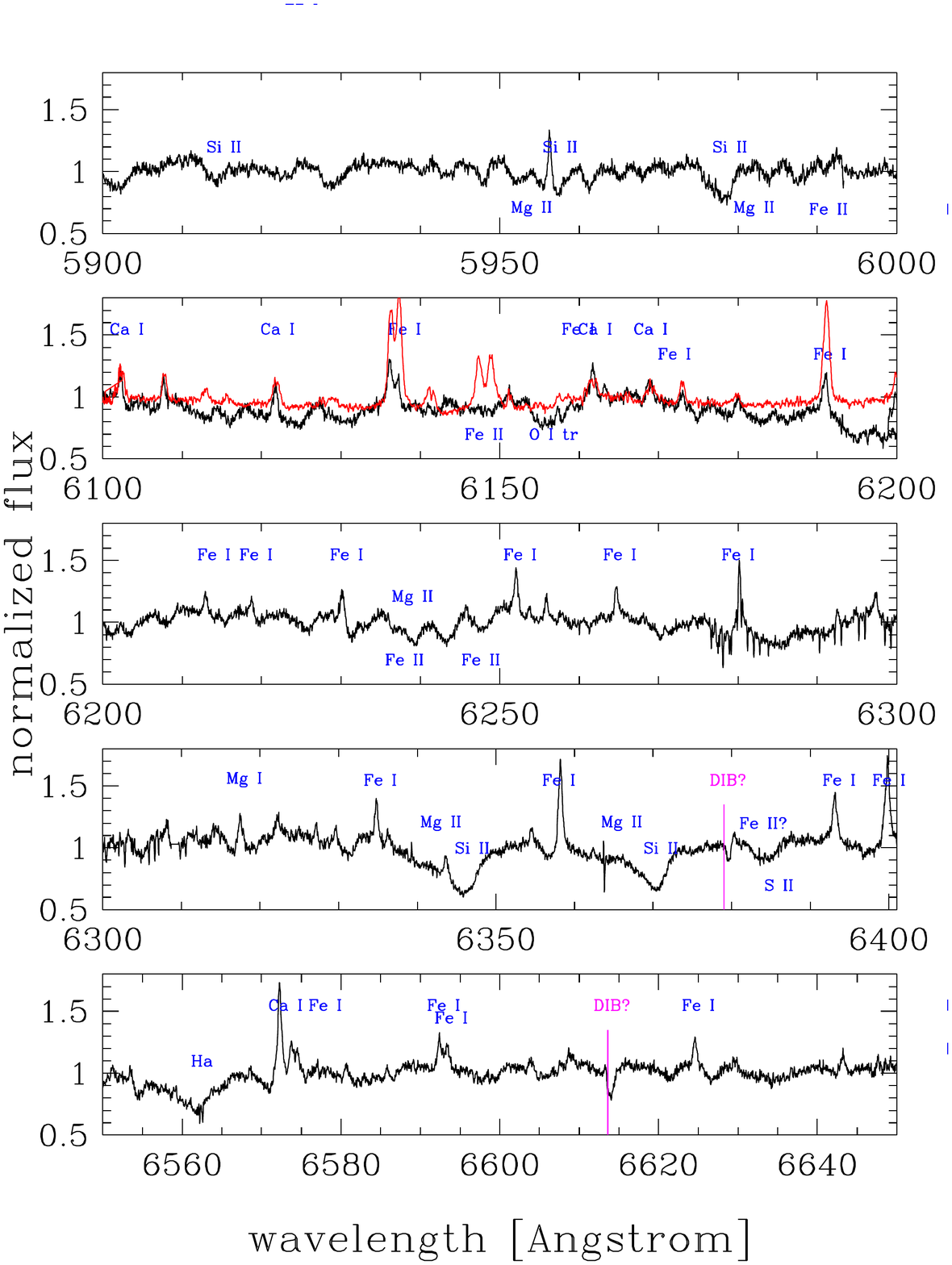}
\caption{Portions of the first high-dispersion spectrum of \jg, 
taken on 2014 0221 or 12 days after the estimated lightcurve peak.
We note the broad and deep absorption, especially towards the 
blue end of the spectrum, and the superposed narrow emission lines.
The well-known continuum$+$emission-line young stellar object V1331 Cyg 
is also shown in several orders (in red), for comparison. 
\jg\ shares the neutral species emission, but not the 
(higher excitation) ionized emission that is exhibited by V1331 Cyg.
Many of the absorption lines in \jg\ are blueshifted from the
indicated line center position.  Some DIBS are present. 
We have been unable to identify all of the broad absorption 
and narrow emission contributors to this spectrum.
}
\label{fig:hires}
\end{figure}

\begin{figure}
\includegraphics[scale=0.6]{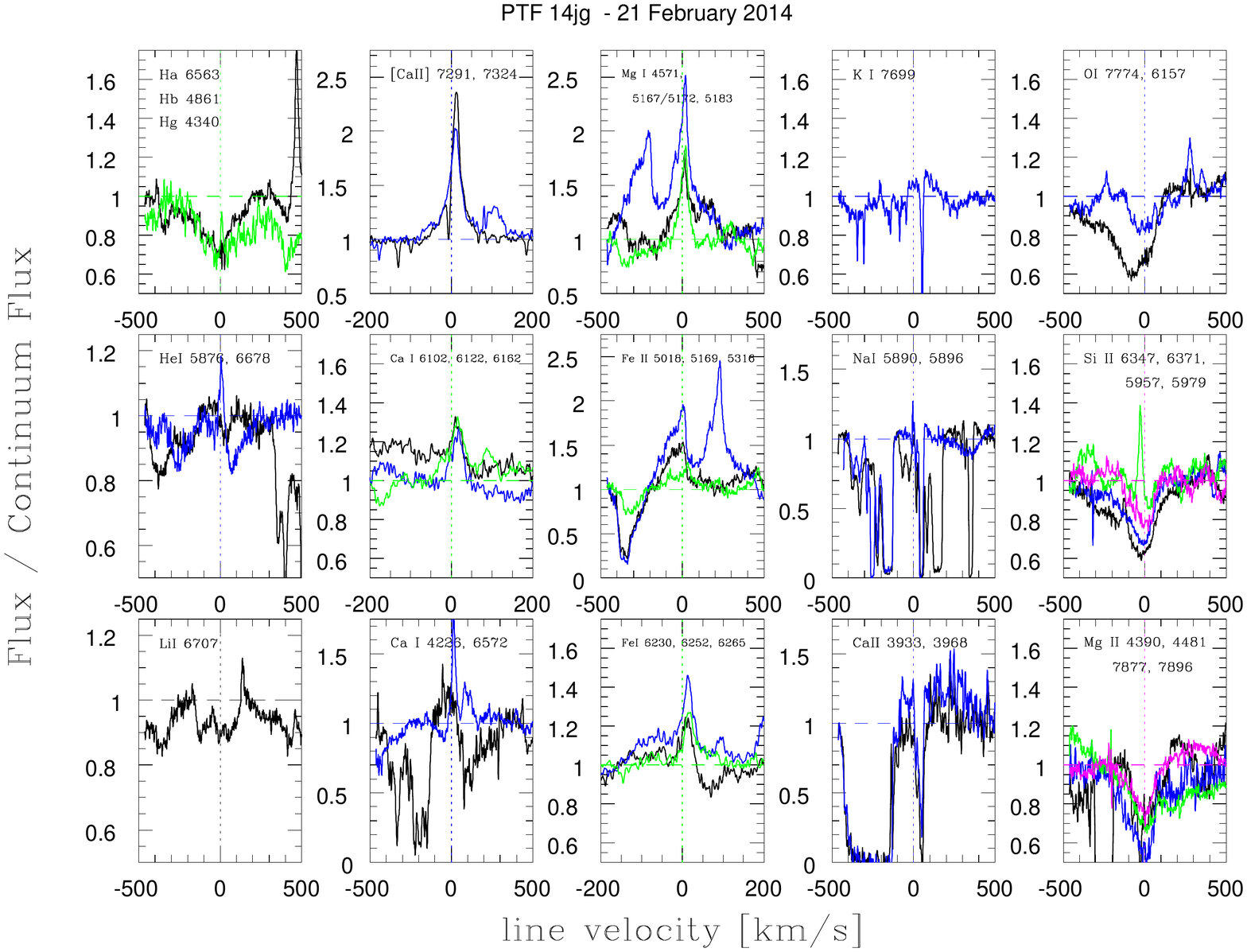}
\caption{Line profiles from the first epoch of high-dispersion data
taken 12 days after the lightcurve peak.  
Where multiple profiles are overlayed, the color corresponds to the 
denoted line, in the order: black, blue, green, magenta.
The normalization is based on values at the edges of the velocity range
and is non-optimal in some cases.
A heliocentric velocity shift of 38.1 \kms\  
has been applied, so as to bring the lines to zero velocity
relative to the absorption-line spectrum that developed later.
Note that the narrow-peak emission line spectrum 
(represented here in e.g. \ion{Ca}{1}, \ion{Fe}{1}, \ion{Mg}{1}, as well as [\ion{Ca}{2}]) 
would require a shift of only 26.6 \kms, i.e. the narrow emission lines 
have a redshift of about 11.5 \kms\ with respect to the systemic velocity
of the photosphere.
}
\label{fig:profiles}
\end{figure}

Further consideration of the \jg\ spectral type based on high-dispersion spectroscopy
(Figure~\ref{fig:hires}) only exacerbated our confusion about the nature of the source.   
The first Keck/HIRES spectra showed some features that are seen 
only at very low gravity in early-type supergiant and giant standards, 
specifically those characteristic of late F superigants,  but also features
more similar to those of late B supergiants (B6-A0).  However, 
these lines are much stronger in \jg\ than in standard stars, which we now attribute to 
enhanced line contributions from the outflow  (see e.g. right panels of Figure~\ref{fig:profiles}).
In the early epoch data, particularly towards the blue, 
the broad and deep lines were not resolved from one another, even at high dispersion,
and therefore were not readily identifiable.
Many of these ``hot" lines exhibiting prominent outflow signatures 
persisted as narrower absorption features later, 
as the source faded photometrically and the outflow velocities decreased.

\begin{figure}
\includegraphics[width=0.5\textwidth]{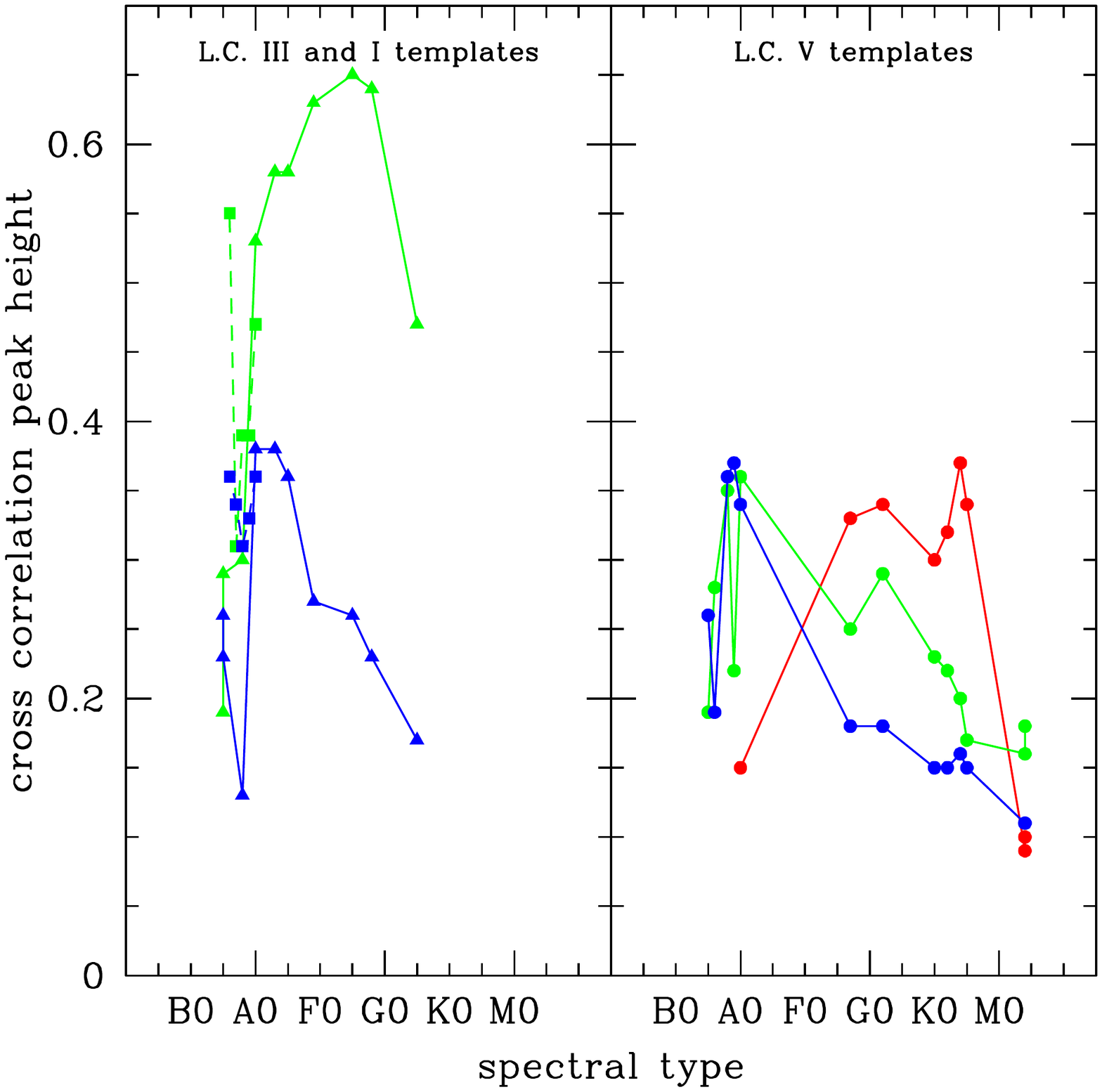}
\vskip-0.75truein
\caption{
Cross correlation peak heights as a function of spectral type for low-gravity (left panel)
and dwarf (right panel) templates.  The analyzed \jg\ spectrum is that from 
2015 1027 (shown in Figure~\ref{fig:comparev1515}),  
taken 625 days after the estimated lightcurve peak.  Blue points are from the
5200-5285 \AA\ region, green 6230-6300 \AA, and red 7370-7480 \AA.
The correlation peak heights change systematically with wavelength, i.e.
the bluer wavelengths correlate best with early A spectral types, while
the redder wavelengths correlate best with late F through K spectral types.
The overall strongest correlations are with supergiants of 
late-A to late-F spectral type in the ``green" part of the spectrum.
}
\label{fig:crosscorr}
\end{figure}

In an attempt to decipher the spectrum, we cross-correlated 
with a suite of spectral template stars
taken with the same Keck/HIRES settings as our \jg\ spectra.  
There was only weak correlation with KM type supergiants or dwarfs, 
but decent correlation power with FG supergiant standards. 
We also correlated \jg\ with a grid of BA supergiants, giants, and dwarfs 
taken from the Elodie grid, since we do not possess our own standards for these earlier types.
For supergiants and giants, the peaks were highest overall for types 
in the early A to late F spectral range, with a clear trend with wavelength. 
Blue wavelengths correlated best
with early A stars, while green wavelengths correlated best with late F stars,
and red wavelengths best matching GK type stars.  
Figure~\ref{fig:crosscorr} shows the correlation strengths at a much later 
epoch, when the strong outflow signatures had somewhat subsided, and 
the spectrum may have cooled as the lightcurve declined; however,
the basic patterns described above for the early epoch spectrum are even more apparent.

No line absorption at all was seen in the early infrared spectra (Figure~\ref{fig:irspec}).
Thus we can not derive an infrared spectral type near the burst peak.
Despite the blue continuum, there is no hydrogen Paschen nor Brackett series lines, 
consistent with the lack of strong Balmer lines in the optical spectrum.  
Likewise, there is an absence of strong spectral features from
cool temperature gas, such as CO absorption.  

There is no overall best-fitting spectral type for the early outburst
spectrum of \jg.  Instead, the optical spectrum exhibits evidence of a 
composite absorption spectrum, with a temperature that changes with wavelength.
The other dominant aspect is a multi-temperature wind, with both cool
species e.g. Ca H\&K and Na D, as well as hot species such as 
\ion{Fe}{2}, \ion{Mg}{2}, and \ion{Si}{2} seen in blue-shifted absorption.
There is very little \ion{H}{1} or \ion{He}{1} signature apparent in \jg\ --
the only strong absorption is from wind-produced metal lines.

\subsection{Broad Hot Absorption Spectrum}

As noted above in our attempts at spectral typing, many of the 
spectral features in \jg\ seem atypical for a single temperature normal stellar photosphere.
The initial high-dispersion optical spectrum (Figure~\ref{fig:hires})
exhibited very strong lines, 
having equivalent widths in most lines exceeding 1 \AA,  
and highly broadened lines, having typical FWHM $\sim$100-150 \kms.

Some features associated with late F supergiants could be identified, 
such as higher excitation (3-4 eV) lines of \ion{Fe}{1} and \ion{Ni}{1}, 
and the 11 eV lines of \ion{Fe}{2}.  However, there is a preponderance 
of even hotter lines, such as are seen in late B stars.  
The \ion{O}{1} 8446 \AA\ singlet at 9.5 eV is present.
And we were able to associate some lines with \ion{Si}{1} 
(intermediate excitation potential of 6 eV, e.g. 7405, 7409, 7415, 7742 \AA\ lines), 
or \ion{Si}{2}, \ion{Mg}{2}, or \ion{Fe}{2} 
(high excitation of 10.5-12 eV), and even \ion{Si}{3} 
(19-20 and 26 eV, e.g. 4552, 4567, 4574, and 4716 \AA\ lines)!  
Placing an upper bound on the temperature,
there is no \ion{Si}{4} at e.g. 4088, 4116, 4212 \AA\ (24 and 35-40 eV lines).

The hot spectrum lines seem to indicate an 11,000-15,000 K, log g = 2-3 
photospheric component of the spectrum, as assessed using the MILES 
stellar population synthesis tool (Vazdekis et al. 2003). 
From \ion{Si}{3} / \ion{Si}{2} ratios and
guided by \cite{lefever2010}, a temperature of 15,500 K is inferred
for the absorption spectrum. These temperature values are consistent with 
the cross-correlation analysis in the bluer parts of the spectrum, discussed above. 

We conclude that the blue-wavelength absorption spectrum of \jg\ 
has a moderately warm temperature, and a low surface gravity.  However, 
it is possible that much of the temperature and surface gravity information 
we have inferred from the absorption spectrum actually characterizes 
the wind launching region, rather than a stellar or a disk photosphere.

Supporting this is that the observed hot absorption lines are very deep, 
reaching between 90-70\% of the continuum   
instead of the typical 99-95\% of continuum for
normal yellow supergiants/hypergiants (specifically those in the Elodie archive).
In addition, many of the hot lines are affected by outflow kinematics,  
and are even broader than the 100-150 \kms\ FWHM that we believe characterizes 
lines defining the at-rest photosphere.  
For example, \ion{Si}{2} 6347 \AA\ and 6371 \AA\ equivalent widths 
are off the scale of the temperature calibration by Miroshnichenko et al. 
-- due to the influence of the strong outflow (even accounting for 
probable contamination on the blue side by the nearby 
\ion{Mg}{2} 6343 \AA\ $+$ 6366 \AA\ lines). 

\subsection{Blueshifted Absorption Features Indicative of a Wind}

Strong signatures of outflowing material are seen in many spectral lines,
The descriptions below are based on the first high-dispersion spectrum, 
taken 12 days past maximum light, corrected for the inferred -38.1 \kms\ 
systemic velocity.  Figures~\ref{fig:hires} and ~\ref{fig:profiles}
are relevant to the discussion.

-- \underline{\ion{Ca}{2} H\&K} doublet lines
are saturated and highly blueshifted. 
The saturated component extends from about -125 out to
-400 \kms, departing the continuum at all velocities between 
about -100 and -450 \kms.  A separate narrow component located at 
zero velocity also reaches zero intensity.  
The 8498, 8542, 8662 \AA\ triplet lines, 
which share upper levels with the doublet lines, 
are in emission. Although not covered in the first two HIRES observations,
in later data these lines peak blueward of the rest velocity, 
around -40 \kms, and extend to -200 \kms. 
While the early-epoch low-dispersion data exhibit P-Cygni line structure
in the triplet lines, none of our (later) high-dispersion data capture this structure.

-- \underline{\ion{Na}{1} D} doublet lines are saturated between -125 and -200 \kms\ 
and also have multiple unsaturated absorption components extending out to 
about -400 \kms. Like in the \ion{Ca}{2} H\&K lines, there is a separate 
narrow component at zero velocity, that also reaches zero intensity.

-- \underline{ \ion{K}{1} 7699 \AA}  (the companion 7665 \AA\ line is between 
spectral orders) 
exhibits two weak and broad blueshifted absorption components 
that seem to correspond to components at similar velocities in NaD, 
extending to -450 \kms.  There is also a strong narrow zero-velocity component,
but unlike in \ion{Ca}{2} H\&K and \ion{Na}{1} D, it is unsaturated. 

-- \underline{\ion{Li}{1} 6707 \AA} has a broad absorption that appears
asymmetric towards the blue, but the interpretation is complicated by the
narrow and weak \ion{Fe}{1} emission that is nearby. 
Several other lines appear to have a similar shallow and highly bluefhifted 
component to their profiles between -300 and -450 \kms 
(in addition to the \ion{K}{1} already discussed, see also the description below
of H$\alpha$, \ion{He}{1} 5876, \ion{O}{1} 7774, and \ion{Fe}{2} 5018, 5169 lines). 
We are thus inclined to believe that we are seeing outflow signatures
in \ion{Li}{1} as well.
The velocity-centered part of the line is very broad, FWHM = 200 \kms. 
By December 2014, a narrower component to the \ion{Li}{1} profile had developed,
consistent with the developing narrow $H\alpha$.  In February 2016, a   
double-peaked \ion{Li}{1} absorption line was seen.

-- \underline{\ion{Ca}{1} 4226 \AA} has multiple absorption components from -150 all the way 
out to -400\kms. The profile appears quite similar to the 
\ion{Ca}{2} H\&K and \ion{Na}{1} D profiles, though is not saturated.  
It is also similar to the multiple \ion{K}{1} 7699 and possibly 
\ion{Li}{1} 6707 profile components.  
The other ground-state transition of \ion{Ca}{1} is at 6572 \AA;  this profile
lacks the blueshifted absorption of the 4226 line, exhibiting only 
narrowly peaked emission that is also seen in the weaker precursor 
6102, 6122, and 6162 \AA\ lines.

-- \underline{\ion{Mg}{1} 5172, 5183 \AA} and \underline{\ion{Mg}{1} 4571 \AA}
all show blueshifted emission between about zero velocity and -200 \kms.
The 5167 \AA\ component of the \ion{Mg}{1} 5167, 5172, 5183 triplet 
is confused with the 5169 \AA\ \ion{Fe}{2} line discussed below.  
The 4571 \AA\ line is a singlet that immediately follows the triplet 
in cascade to the ground state. 
These lines as well as the 8806 \AA\ line that is seen in the low-dispersion
spectrum (but not covered by HIRES) are intermediate excitation, 
in the 3-6 eV range.

-- \underline{ \ion{O}{1} 7774 \AA\ triplet} 
is very broad, extending beyond -500 \kms\ on the blue end, and redward 
to about +75 \kms.  However, as the velocity scale is set at 
the central wavelength of the middle triplet component, 
the extreme red and blue velocities compared to other features 
are due to the presence of the multiplet.
The \ion{O}{1} 6157 \AA\ triplet line seems narrower; its three lines span only 2.2 \AA\ rather than 3.5 \AA\ for the 7774 \AA\ triplet.

-- \underline{\ion{Fe}{2} 5018 \AA\ and 5169 \AA} 
are similar to the \ion{Mg}{1} in
showing blueshifted emission between about zero velocity and -200 \kms.
In addition, there is deep absorption between -200 and -450 \kms, 
reaching maximum absorption strength at -350 to -400 \kms.
These are low-excitation (3 eV) iron lines.   A higher excitation (10 eV) 
line at 5316 \AA\ has similar high velocity structure but is much weaker.

-- \underline{\ion{Si}{2} 6347, 6371 \AA, 
3853, 3856, 3862 \AA, 4128, 4130 \AA, 4815, 5041, 5056, 5957, 5978 \AA} 
are predominantly blueshifted with a wing out to -450 \kms,
but have absorption extending redward of line center.  
These are all high excitation (8-10 eV) lines that are not commonly seen 
in young star winds. They are present in supernovae spectra,
however, and suggest that a hot shell-like feature 
may be associated with the photometric brightening of \jg.
No lines such as \ion{C}{1} with similar excitation potential 
are prominent in our spectra.

-- \underline{\ion{Mg}{2} 4481, 7877, 7896 \AA}
is similar in morphology to the \ion{Si}{2} lines, 
which is not surprising given the similar excitation potentials,
though the lines do not extend to as high a blueshift.

-- \underline{\ion{S}{2} 5606, 5616, 5640, 5645, 5660, 5664, 6386 \AA}
is present, though bluer lines of this species are not.  At 13.6 eV this is
another typical supernova line (seen in type Ia events
but not in core collapse events).

-- \underline{ H$\alpha$} is in absorption, seemingly blueshifted, 
and appears similar to \ion{O}{1} in profile, though less deep.  
H$\beta$ is not covered in the first two high-dispersion spectra 
(but is in our later data); 
H$\gamma$ mimics the H$\alpha$ profile in both shape and strength.

-- \underline{ \ion{He}{1} 5876 \AA\ and 6678 \AA} appear to have blueshifted absorption between 
-200 and -500 \kms, though this identification is not strong given the confused
nature of the spectrum.  The emission component in the 6678 \AA\ profile 
is likely the \ion{Fe}{1} line at the same wavelength.  

Summarizing, at low velocities, between 0 and -150 \kms, there is absorption by
\ion{H}{1}, \ion{O}{1}, \ion{Si}{2}, and \ion{Mg}{2} with full line widths extending
further to the blue than the photospheric line width, up to -500 \kms.
Over these same 0 to -150 \kms\ velocities, 
there is broad \ion{Ca}{2} triplet and \ion{Fe}{2} 5018 \AA\ emission,
but P Cygni absorption in these lines at the higher velocities out to -400 - -500 \kms. 
Beginning at -150 \kms and going to -450 or so, there is a strong multi-component 
absorption seen in \ion{Ca}{2} H\& K, \ion{Ca}{1} 4226 \AA, \ion{Na}{1} D, \ion{Li}{1} 6707 \AA, and likely \ion{K}{1} 7665,7699 \AA,  
as well as the continued absorption in \ion{Ca}{2} triplet and \ion{Fe}{2} 5018 \AA.
Given the strength and breadth of the \ion{Ca}{2} H\&K and the \ion{Na}{1} D lines, 
the absence of stronger \ion{K}{1} 7665,7699 \AA\ and any \ion{Sr}{2} 4077 \AA\ 
blueshifted absorption seems notable.

\subsection{Prominent Emission Features}

\begin{figure}
\includegraphics[scale=0.44,trim={1.75cm 0 0 0},clip]{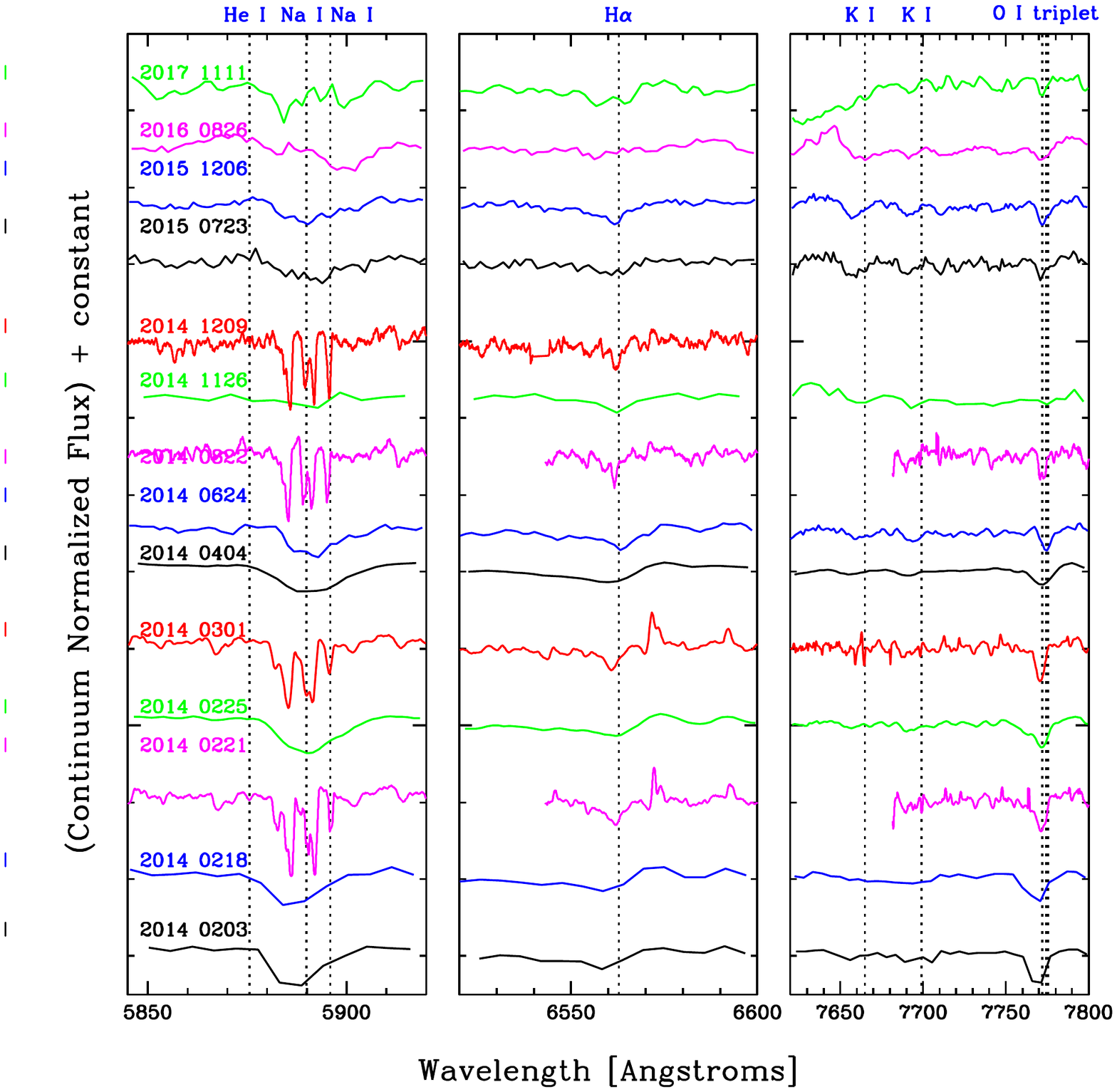}
\includegraphics[scale=0.44,trim={1.75cm 0 0 0},clip]{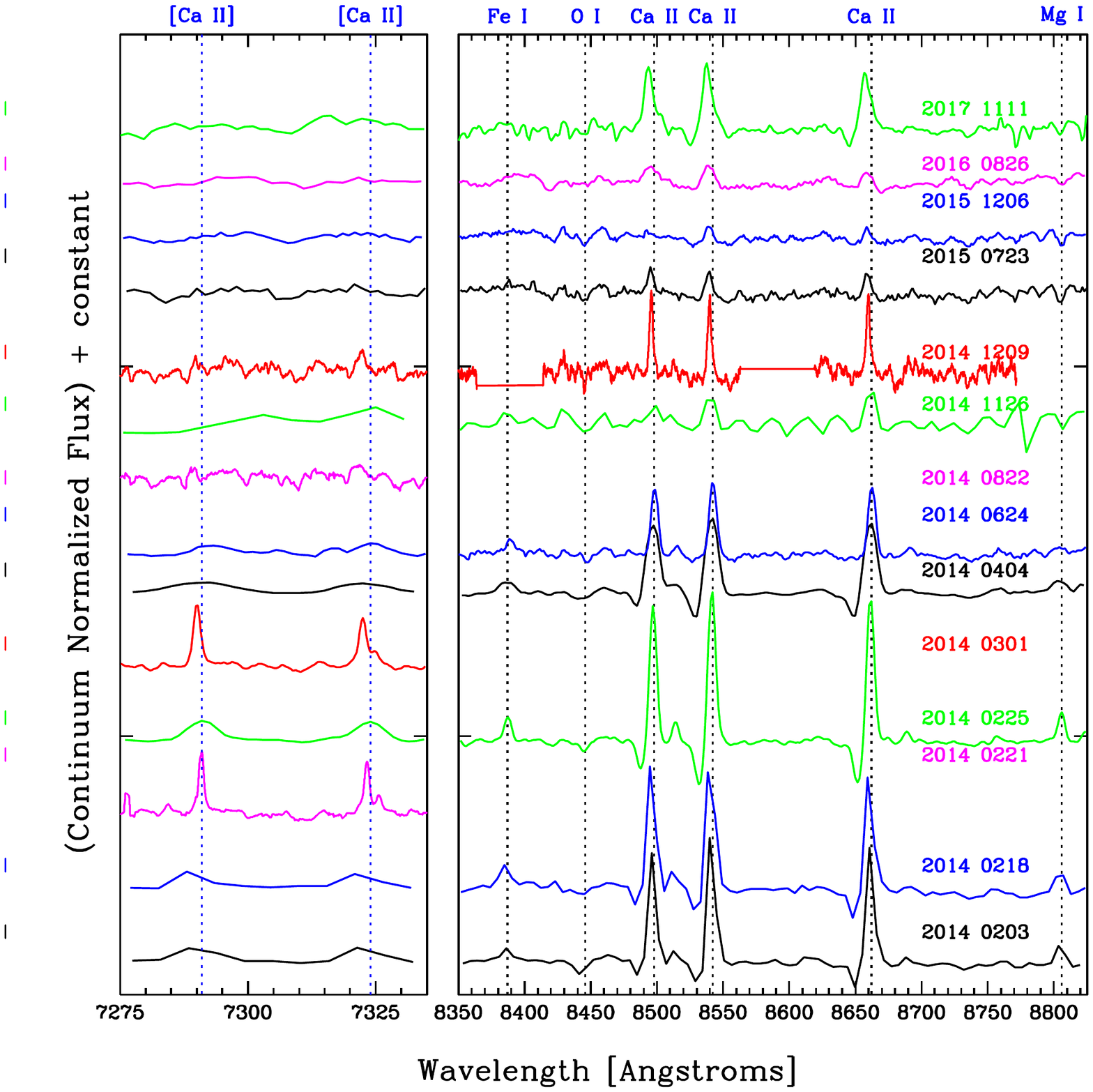}
\vskip -0.75truein
\caption{Evolution of spectral features in the \jg\ optical spectrum.
Left panels show
absorption in H$\alpha$ (weak), \ion{Na}{1}, \ion{K}{1}, \ion{O}{1}. 
Right panels show prominent emission lines in \ion{Ca}{2} and [\ion{Ca}{2}].
The resolution varies among the spectra, with those
from 2014 0221, 2014 0301, 2014 0822, and 2014 1209 
significantly higher than the others (and smoothed here, by a factor of 31).
Even accounting for the differences in spectral resolution, 
the variability over time in the \ion{Ca}{2} and [\ion{Ca}{2}] 
line strengths is real, as is the disappearance of the 
early-stage \ion{Fe}{1} and \ion{Mg}{1} emission.
}
\label{fig:lines}
\end{figure}

As illustrated in Figure~\ref{fig:optspec},
the only strong emission lines in the optical spectrum of \jg\ come from calcium. 
Both the \ion{Ca}{2} 8498, 8542, 8662 \AA\ triplet lines 
and the [\ion{Ca}{2}] 7291 \AA\ and 7324 \AA\ doublet are present. 
Their spectral evolution is highlighted in the right panels of Figure~\ref{fig:lines}.
These calcium lines are related, with the 7291 \AA\ forbidden line
the ground state follow-on transition to the permitted 8542 \AA\ 
de-excitation, while the 7324 \AA\ forbidden line is the ground-state transition 
that follows the 8498 and 8662 \AA\ de-excitations \citep{merrill1943}.
The permitted triplet emission is quite common in accreting young stars;
the forbidden doublet emission is rare, however.

Near the outburst peak, and for at least several months after it,
the triplet lines were moderately strong 
($W_{8542}=$ -12 to -15 \AA), and they showed clear classical P Cygni 
profiles that were visible even in the low resolution data.  However,
by the time of the first high-dispersion spectral epoch covering this wavelength range (Figure~\ref{fig:profile_evol})
at 303 days post-peak, the measured $W_{8542}$ was only -5.5 \AA,  and there was no evidence 
for P Cygni structure.  The profiles at this later time had $\sim$120 \kms\ widths 
and the peaks were distinctly blueward of line center, by about $\sim$40 \kms.  The
profiles extended to -200 \kms\ on the blue side but were missing their shoulder
on the red side.  This blue flux asymmetry is 
reminiscent of magnetospheric accretion line profile models 
\citep[][and references therein]{azevedo2006,kurosawa2012} in a system having
moderate inclination. \cite{konigl2011} argues that such geometries can also
be present in high accretion FU Ori systems, with the magnetosphere located simply
closer to the star, but not crushed entirely.

As just discussed, the \ion{Ca}{2} triplet lines are broad and blueshifted.
However the forbidden [\ion{Ca}{2}] 7291 and 7324 \AA\ doublet that follows
the triplet lines exhibits narrow lines, with $\sim$25 \kms\ FWHM widths
(Figure~\ref{fig:profiles}).  
This doublet shares the same radial velocity as the series of
weak and narrow {\it permitted} emission lines that also appeared in the early spectra of \jg\
(discussed in the next section).  
The moderately strong [\ion{Ca}{2}] lines have a clearly lorentzian line shape, and equivalent width -1.2 \AA.  
We note that [\ion{Ca}{2}] is the only forbidden line species 
detected in \jg.  No forbidden emission from the so-called 
``nebular" lines of [\ion{O}{1}], [\ion{N}{2}], or [\ion{S}{2}] 
that are quite common in young accretion/outflow systems, was obvious at any epoch.

The lack of H$\alpha$ (Figure~\ref{fig:lines})
is puzzling, as this line is always strongly in emission when strong 
\ion{Ca}{2} triplet emission is detected in young stars.  However, the H$\alpha$ line
in \jg\ is apparent only in shallow weak absorption ($W_{H\alpha} \approx 1.8$ \AA), 
with probable P Cygni structure, or at least a blue-side asymmetry.
The line structure could also be explained by contamination from a nearby
but unidentified high-excitation species that appears in absorption.  
Similar blueshifted and very weak absorption profiles {\it may} characterize 
the lines of \ion{He}{1} at 5876, 6678 and 7065 \AA\ as well, in the first few spectral epochs. 

Figure~\ref{fig:lines} also illustrates 
in the early low-dispersion spectra of \jg\
emission from \ion{Mg}{1} 8806 \AA, \ion{Fe}{1} 8787 \AA\ (multiplet 60), 
and a few other even weaker \ion{Fe}{1} lines. 
As the source fades from peak, these lines become weaker, 
with the \ion{Mg}{1} 8806 \AA\ line even going into absorption 
by about 9 months post-peak.

In the infrared (Figure~\ref{fig:irspec}), the most prominent emission lines 
in the early low-dispersion spectra are from \ion{Mg}{1}, with features at 
1.1831 (strong), 1.2087, 1.4880, 1.5029 (strong), 1.5044 (strong), 1.5052 (strong), 1.5745, 1.5753, 1.5767, 1.5892, and 1.7111 (strong) $\mu$m.  
Weak emission was also detected in various \ion{Ca}{1}, \ion{Fe}{1} 
and possibly \ion{Si}{1} lines.
Despite the lack of \ion{H}{1} Balmer emission associated with the outburst, 
there is a hint of emission in the \ion{H}{1} Brackett gamma line with
$W_{21661} =$ -2.0 \AA\ and FWHM=500 \kms; however,
problems with telluric correction that might introduce this effect can not be
ruled out.
We have only two early-stage spectra of \jg, and have not been able to 
follow the infrared spectral evolution as we did in the optical.

\subsection{Weak and Narrow Cool Emission Spectrum Near the Outburst Epoch}

The early high-dispersion spectrum, taken only 12 days post-peak, 
revealed a weak and narrow $\approx$ 25-30 \kms\ FWHM 
emission line spectrum, superposed on the broad and confused absorption 
spectrum.  
Figure~\ref{fig:hires} illustrates the coincidence of many of these narrow lines 
with those that are seen more clearly against a flatter continuum in the 
somewhat extreme young stellar object V1331 Cyg. 
Figure~\ref{fig:profiles} includes a few of these narrow line profiles.

The early-stage narrow emission in \jg\ is mostly in neutral species, 
e.g. \ion{Fe}{1}, \ion{Mg}{1}, and \ion{Ca}{1}, though no \ion{Ti}{1} is identified, whereas that of V1331 Cyg has both neutral and ionized species. 
When a narrow metallic emission spectrum is seen in young stars, 
it is more typically comprised of ionized species such as \ion{Fe}{2} and \ion{Ti}{2},
not the neutral species.

Correlating the inverted spectrum with a grid of spectral standard stars 
resulted in consistently good correlation heights 
and correlation coefficients 
over a range of temperatures from early M through early G types.
We take this as an indication that the narrow emission spectrum 
corresponds to temperatures between $\sim$3700-6000 K.  

Weak emission was also detected in the  early-stage infrared spectrum,
with the same \ion{Mg}{1}, \ion{Ca}{1}, and \ion{Fe}{1} 
species as seen in the optical present in the near-infrared.  

The cool gas likely sits above the underlying broad absorption 
in the photosphere, which comes from a hotter spectral component.
As noted above, there is an offset of (-26 - -38) = +11.5 \kms\ 
of the narrow-line emission spectrum with respect to the systemic velocity
of the photosphere.  One possible scenario is a nearby interstellar cloud,
or a cirumstellar cloud located at high-latitude towards the
pole of the star/disk system, that is radiatively excited by
the new influx of photons from the outburst.  We mention again that the
weak emission spectrum lasted no more than six months (based on the second
Keck/HIRES spectrum), but at least six weeks (based on the second
Palomar/TripleSpec spectrum), after the outburst peak.

\subsection{Spectrum Changes During the Fade}

\begin{figure}
\includegraphics[scale=0.6, angle=-90]{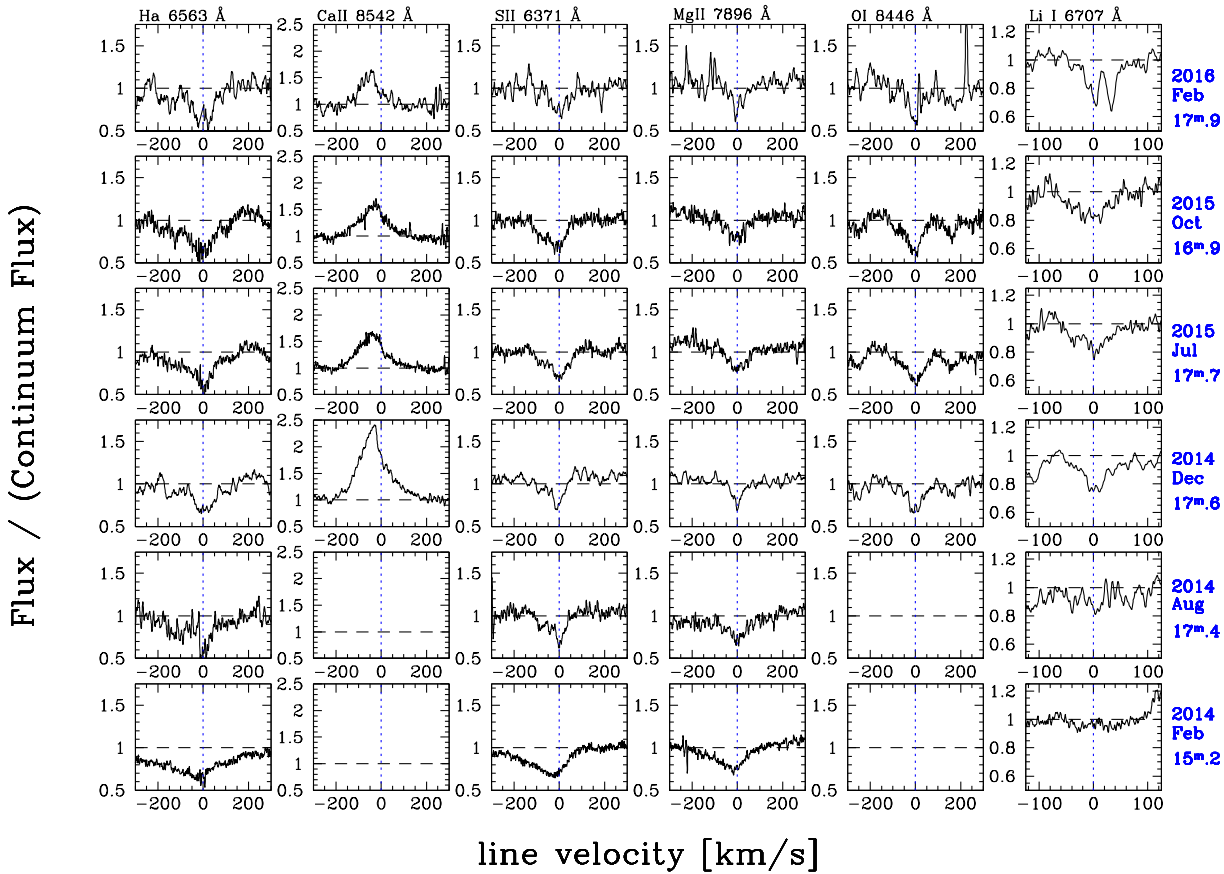}
\caption{Evolution of permitted line profiles. 
Spectra are sequenced from bottom to top with source magnitudes also given, 
at right.  The velocity scale is $\pm$300 \kms\ for all panels except the
rightmost, which is $\pm$125 \kms.  A velocity shift of 38.1 \kms, 
has been applied to correct for the systemic radial velocity
inferred from the late-epoch absorption line spectrum. 
Narrow core $H\alpha$ emission may have developed in the most recent spectrum,
though the \ion{Ca}{2} triplet lines are still broad and have an asymmetric
profile that is peaked at negative velocity.
Blueshifted velocities in \ion{Si}{2} and \ion{Mg}{2} have decreased as
the source has faded.  
\ion{Fe}{2} and \ion{O}{1} profiles are essentially unchanged. 
}
\label{fig:profile_evol}
\end{figure}

\begin{figure}
\includegraphics[scale=0.6,angle=-90]{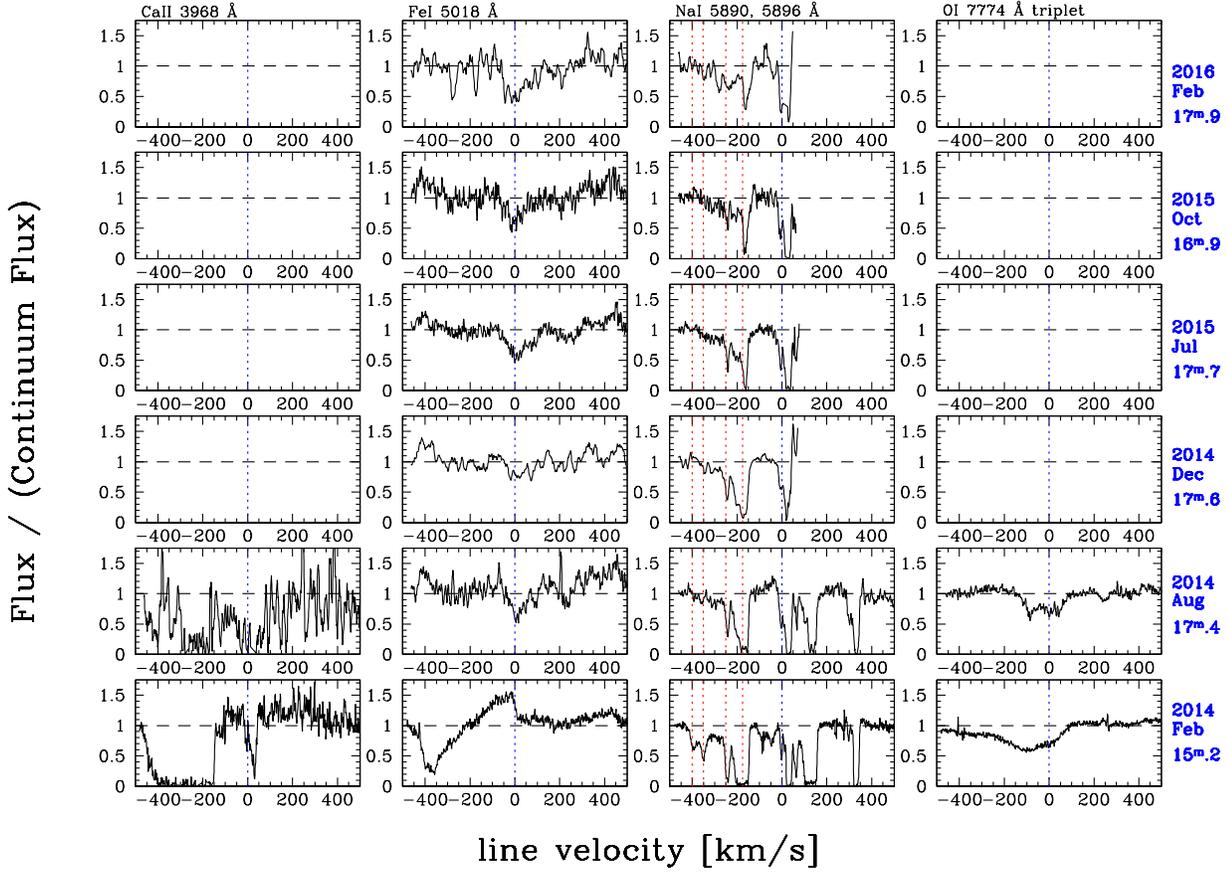}
\caption{Evolution of wind-dominated line profiles. 
Spectra are ordered as in Figure~\ref{fig:profile_evol}, with the same
adopted systemic velocity.  
Here, the velocity scale is $\pm$500 \kms\ for all panels.
Red vertical lines on the \ion{Na}{1} D2 panels indicate semi-stable absorption components
at -175, -250, -350, and -400 \kms.  The two additional features  
at $\sim$ -50 and -100 \kms are actually the -400 and -350 \kms components
of the \ion{Na}{1} D1 line.  The \ion{Ca}{2} K (and H) lines exhibit continuous saturated
absoption over this entire velocity range, extending at the terminus to
$\sim$ -530 \kms.
}
\label{fig:wind_evol}
\end{figure}

\begin{figure}
\includegraphics[angle=-90, scale=0.4]{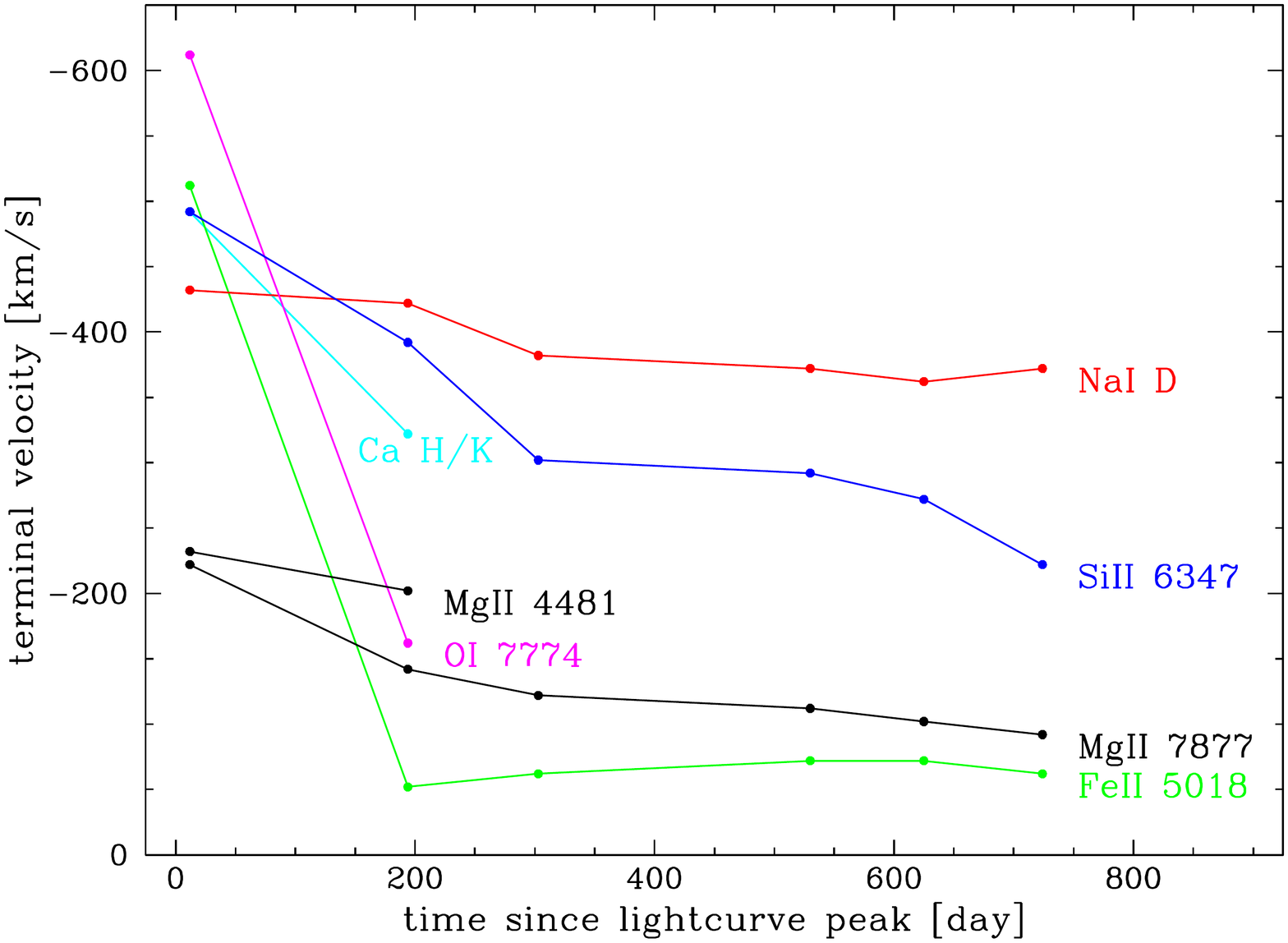}
\caption{Evolution of blue-side terminal velocities for 
absorption lines with wind signatures in 
Figures~\ref{fig:profile_evol} and ~\ref{fig:wind_evol}. 
Velocities are in the rest frame of the source and have
estimated measurement errors of 10-20 \kms.  The wind speeds
were highest at outburst, then declined at different rates for different lines,
mostly within the first 1-2 e-folding times of the lightcurve.
The \ion{Na}{1} D lines exhibited the least change.
The \ion{O}{1} measurement is over-estimated due to the triplet nature of the line.
}
\label{fig:terminal}
\end{figure}

Considering the entire set of spectroscopic follow-up observations, there was some, but not dramatic
spectral evolution over the nearly 4 post-peak years.  As the lightcurve faded
by several magnitudes,  the optical continuum shape remained essentially 
the same, consistent with the only modest 
broadband color changes reported in \S4.2. 
There was also little change inferred in the temperature of the absorption 
spectrum during the early fade, with the high resolution spectrum essentially 
unchanged in terms of the absorption species -- until nearly two years post-peak. 
At that time, having emerged from the previous wind-dominated hotter absorption spectrum,
a cooler absorption component in the spectrum was revealed.

The modest evolution in the profiles of various strong lines 
is illustrated in Figure \ref{fig:profile_evol}. 
Overall, the broad and hot absorption spectrum exhibited only minor 
morphology changes over time, retaining its general features for the most part. 
Specifically, as \jg\ faded photometrically, 
some lines became narrower, losing power from their highest velocity absorption.
Line depth closer to line center remained constant, however. 

Evolution of the strong outflow lines is shown in Figure \ref{fig:wind_evol}. 
Figure \ref{fig:terminal} demonstrates that the outflow velocities declined
sharply as the lightcurve faded through one e-folding time (111 days). 
The \ion{Na}{1} D lines have the shallowest slope, meaning that the maximum
velocity did not change appreciable as the source brightness declined,
though there is a general decrease in the line depth in nearly all
velocity ranges (Figure \ref{fig:wind_evol}).
Several of the distinct components in the blueshifted absorption profile remain
steady. The highest velocity deep components disappear, however. 
This behavior suggests that much of the \ion{Na}{1} D 
absorption comes from stationary locations through which moving gas 
(ejected in the outburst and decelerating) is flowing.

The P Cygni structure in the \ion{Ca}{2} triplet 
apparent in the early low-dispersion data also disappeared over time 
(Figure~\ref{fig:lines}), consistent with the weakening wind signature
in other lines.  
In spectra taken more than $\sim$180 days after the lightcurve peak,
the emission part of the line was observed to vary in strength with
$W_{8542}=$ -7 to -4 \AA.  
Unfortunately this red spectral range was not covered at high dispersion 
until our third HIRES spectrum, taken $\sim$300 days after maximum light, 
by which time both the P Cygni feature of the triplet
and the overall weak narrow emission component of the spectrum had disappeared.  
Figure~\ref{fig:profile_evol} illustrates strength and morphology changes 
in the (blueshifted) emission part of the line during the epochs of 
high-dispersion coverage.  
However, a later low-dispersion spectrum - taken almost four years 
after the initial rise -- once again had P Cygni structure in the \ion{Ca}{2} triplet
(see Figure~\ref{fig:lines}), with emission strength comparable to 
the early burst stage. We have no contemporaneous photometric monitoring 
of the source at this late time. 

The narrow emission spectrum in neutral species such as \ion{Fe}{1}, \ion{Mg}{1}, and \ion{Ca}{1} 
was present only in the first high-dispersion spectrum,
taken within two weeks of maximum light,
and had disappeared by the next high-dispersion spectrum six months later.
The [\ion{Ca}{2}] doublet emission weakened over the same time period.

The stronger \ion{Fe}{1} 8787 \AA\ and \ion{Mg}{1} 8806 \AA\ 
that were visible in the early low-dispersion data also gradually weakened, 
with the \ion{Mg}{1} line actually evolving to apear in absorption 
some time after 4.5 months post-peak (see rightmost panel of Figure~\ref{fig:lines}).

Returning to the absorption spectrum, 
there was evidence by around $\sim$300 days post-peak
for development of a narrower neutral absorption component,
specifically in \ion{Fe}{1} and \ion{Mg}{1} (e.g. 4481 and 6318 \AA). 
By 625 days post-peak, signatures of a K-type absorption spectrum 
began to emerge in the high-dispersion data
(see Figure~\ref{fig:comparev1515} below for illustration).
In addition, we call attention to the clear presence of \ion{Li}{1} 6707 \AA\ in \jg\ at this late time.

By nearly four years after the outburst, in late 2017, the low-resoluton 
absorption spectrum of \jg\ did evolve somewhat, with significant weakening 
of \ion{Si}{2} 6347, 6371 \AA\ and the nearby \ion{Mg}{2} absorption 
(see Figure~\ref{fig:optspec}).  This may indicate further, continued
cooling of the wind launching region.
The composite spectrum nature of the \jg\ absorption spectrum was maintained 
over this time period, as illustrated in the cross correlation analysis 
of Figure~\ref{fig:crosscorr}.

\section{Synthesis of Evidence to Date: A Possible FU Ori Star}

In this section we discuss \jg\ as a likely young star outburst. 
Given its location in the Galactic Plane and projection near 
a known star forming region, 
and the evidence for ultraviolet/infrared excess as well as \ion{Li}{1}, 
this is a reasonable hypothesis to explore.
Furthermore, the observed properties of the \jg\ outburst are not a good match 
to other categories of plausible outburst objects, 
as discussed in the subsequent \S9.
We believe the source to be a likely FU Ori star, though larger amplitude and hotter
than those members of the class identified heretofore.

\subsection{Location and Environment }

\jg\ is projected on the sky about a degree away from the young massive star
cluster IC 1805, and outside the \ion{H}{2} region.
In a similar fashion, FU Ori itself is well north and west of the active
star forming clouds in Orion. And both V1515 Cyg and V1057 Cyg 
are likewise many degrees away from the molecular clouds assocated with 
the North America Nebula. The recently discovered {\it bona fide} 
FU Ori star 2MASS J06593158-0405277 (V960 Mon) is also
several degrees north of 
the well-studied star forming regions in Canis Majoris, with which
it has been associated.
It is perhaps more the rule than the exception that FU Ori eruptions
are found outside the main active regions of star formation. 
However, 
given that most FU Ori events were identified at optical wavelengths,
these locations could reflect a bias towards events associated with 
less obscured, and perhaps thus older, young stellar objects.
A contrasting case is that of PTF 10qpf / HBC 722, which is in the heart of 
the main embedded cluster associated with the North America Nebula region.

In this general direction on the sky, only the mainly local clouds 
at $v_{LSR} =0$ \kms, and the 2 kpc Perseus arm at $v_{LSR} = -40$ \kms\ are seen in CO emission, with little beyond.  
The -38.1 $\pm$ 1.2 \kms\ heliocentric velocity of \jg\ corresponds to 
a $v_{LSR}$ of -34.7 \kms, which would place the object
on the near side of the 2 kpc arm.  
Indeed, Georgelin \& Georgelin (1970) measure H$\alpha$ velocities 
around the nearby \ion{H}{2} regions, reporting $v_{LSR} = -36.6\pm$ 2.6 \kms.

There is an outer spiral arm 
beyond the Perseus Arm, located at 6 kpc and $v_{LSR} = -100$ \kms. 
Although this arm is not apparent at the exact position of \jg,
it is seen about 10 degrees closer to the inner Galaxy \citep{heyer1998}.
At the exact position of \jg\ there is no clear CO emission 
at any velocity. 
The cloud catalog of \cite{heyer2001} shows the closest cloud in projection to be HC 3927 
which is located between $l$=135.92 to 136.03 and $b$=0.70 to 0.84 deg,  
at a velocity of -75.92 km/s.  Its centroid is about 9.3 arcmin away from \jg, 
corresponding at the kinematic distance of 9.71 kpc to 26 pc. 
The radial velocity of \jg\ is not consistent with these larger 
distances, or anything much further than about 2 kpc.

Examination of 2MASS ($JHK_s$) and WISE (W1, W2, W3, W4) 
infrared colors of stars in a 15\arcmin\ $\times$ 15\arcmin\ box 
shows general similarity in various color-color diagrams
between a field centered on \jg, and a field centered
within the nearby \ion{H}{2} region (but south of the IC 1805 cluster).  
There is not a dominant population of infrared excess sources
at either location.

\subsection{Lightcurve}

\begin{figure}
\includegraphics[scale=1.0]{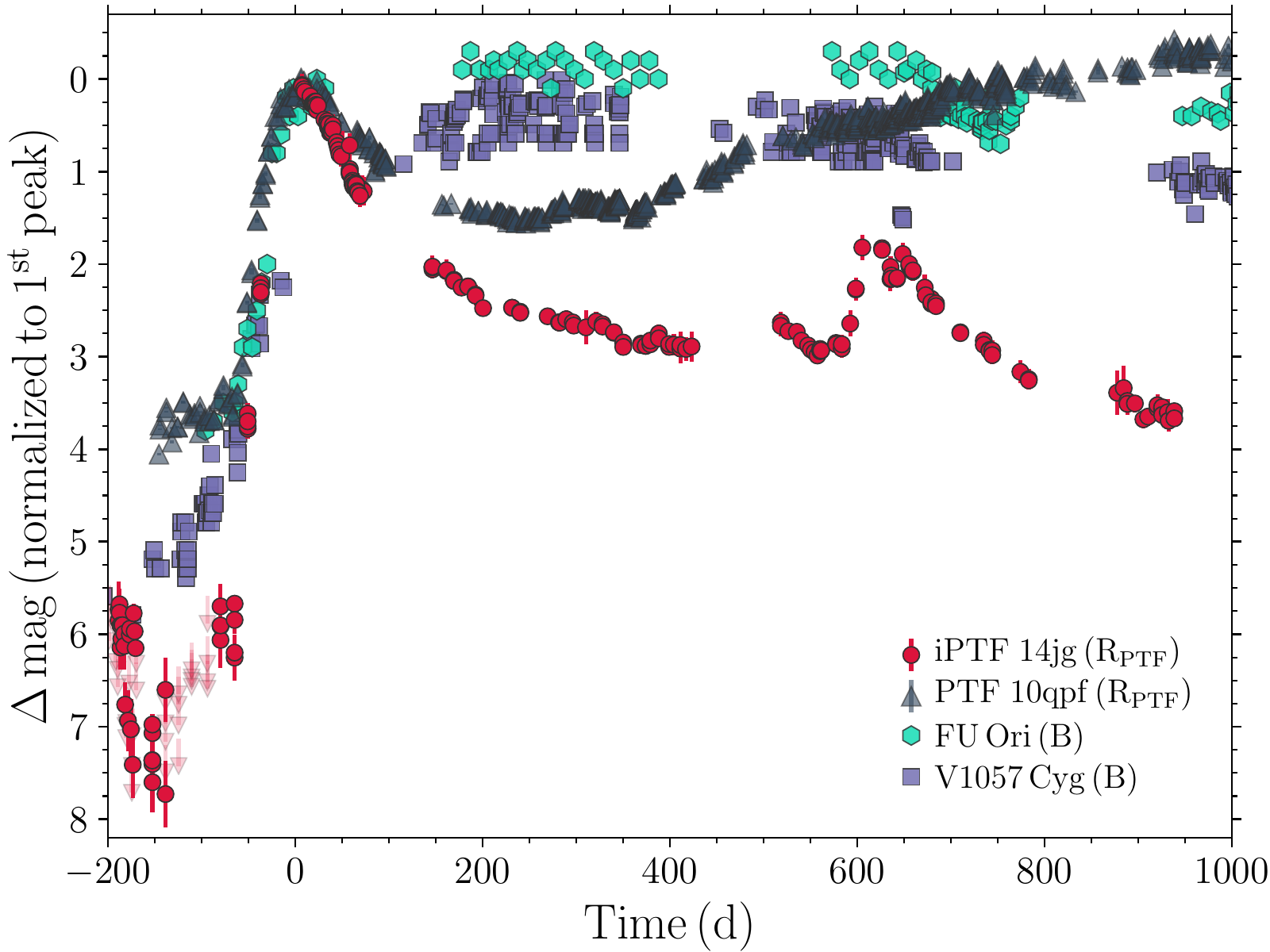}
\caption{Comparison of outburst lightcurves of \jg, PTF 10qpf = HBC 722, 
V1057 Cyg, and FU Ori.  The data are shifted along both axes to match
at their peaks. Note that FU Ori and V1057 Cyg are B-band data,
while the two more recent outburst sources are R-band data.
The rise phases have similar slope among the objects, 
albeit different amplitudes.
The post-peak lightcurve shapes are very diverse.
}
\label{fig:comparelc}
\end{figure}

The well-characterized brightening of \jg\ is large in amplitude,  
$>$6 mag in the red optical and $>4$ mag in the near-infrared.
For comparison, the recent PTF 10qpf event that was also captured by PTF 
(Miller et al. 2011; see also Semkov et al. 2010) in the star 
LkH$\alpha$ 188 G4 (also known as HBC 722, and now as V2493 Cyg) 
was smaller, at about 4 mag in the red optical and 3 mag in the near-infrared.
The classical events FU Ori and V1057 Cyg were observed only in the blue
optical, and both had amplitudes around 5.5 mag \citep{herbig1977}.

The outburst of \jg\ is much larger (6.7 mag according to our sigmoid fit)
and has lasted much longer (nearing five years at the time of this writing) 
than the temporary brightness increases associated with 
EX Lup type burst events; these are also accretion-driven brightenings
of young stars, 
typically only 1-2 mag in amplitude and only months- to year-long in duration. 

During its few month rise, the form of the lightcurve of \jg\ was concave 
during approach to peak brightness. This is consistent in terms of both 
time scale and shape with other FU Ori stars.
In Figure~\ref{fig:comparelc} we compare
the early lightcurve behavior among the recent and well-documented outbursts 
of \jg\ and PTF 10qpf (HBC 722), and also show the available historical data 
for the outburst of the prototype FU Ori, and the later V1057 Cyg.
The rise time of \jg\ is quite similar to that of PTF 10qpf (HBC 722),
as well as that of V960 Mon (not shown), 
and comparable to that of FU Ori, though shorter than that for V1057 Cyg.
The other canonical FU Ori type star, V1515 Cyg, had a much slower rise and
a less well-defined ``peak", as well as a much lower amplitude.

The time scale of the \jg\ rise, approximately 112 days, 
corresponds to a size scale in a classical accretion disk 
of about 32 $R_\odot$, 
assuming thermal diffusion of ($\tau_{thermal}/day)=(r/R_\odot)^{11/8}$ 
and estimating $R_\ast \approx R_\odot$ for the \jg\ progenitor, which should be
accurate to within a factor of two assuming a young pre-main sequence
progenitor. If a young star, the rise time thus
suggests a $\sim 0.15$ AU origin for the outburst,
perhaps driven by an inner disk instability mechanism.

The detailed behavior of 
FU Ori star brightness declines from maximum
is diverse (Figure~\ref{fig:comparelc}).
The morphology of the immediate post-peak lightcurve of \jg\ 
does share some similarities with other FU Ori stars.  
Like \jg, PTF 10qpf (HBC 722) exhibited an immediate fade, 
but then a smooth and slow rebrightening.  
In contrast, the implied re-brightening in \jg\ after $\approx$400 days 
and the detected re-brightening and tertiary local maximum 
after $\approx$600 days, were both more abrupt.  
See figures in \cite{clarke2005} for further comparison with early-stage behavior 
in the lightcurves of V1057 Cyg and FU Ori.

Over time, however, the \jg\ lightcurve became more consistent with an exponential shape. 
As described in \S6, the measured e-folding time is 111 days (0.3 years), 
with the photometric variation about the fitted exponential only 0.09 mag (rms).
Several other recent FU Ori stars with high quality outburst data, such as PTF 10qpf 
(HBC 722) and V960 Mon, also exhibit little scatter in the outburst photometry.  In contrast,
lower-amplitude, shorter-lived outbursts in objects like EX Lup and V1647 Ori, 
show variability at the $\approx$ 0.5-1 mag level in their outburst phases. 

Notably, an exponential decline also fits the post-peak lightcurve of V1057 Cyg,
the most rapidly declining classical FU Ori type star,
but with a much longer e-folding time of 2307 days (6.3 years) 
derived utilizing the data published in \cite{clarke2005}.  The post-peak 
times $t_1$, $t_2$, and $t_3$ for V1057 Cyg are $\sim$650, 1950, and 4400  
days -- factors of $\sim$10, 10, and 5 longer than those of \jg\ 
(54, 139, and 747 days, as described in \S6).  
Notably, \cite{kraus2016} reported 
that the long-accepted FU Ori object V346 Nor had decayed on just a few decades 
time scale, rather than the nominal century 
that is typically assumed for FU Ori event cooling.
Although there are substantial timescale differences 
among members of the FU Ori class, the lightcurve of 
\jg\ is most similar to that of V1057 Cyg in terms of 
the basic exponential profile of the decline.  

In addition to its similarity to the FU Ori star V1057 Cyg, 
the exponential decline in the \jg\ lightcurve resembles that of classical novae. 
However, the time scale of the \jg\ fade is much longer than that of novae.
Nova outbursts are caused by thermonuclear runaway and detonation, 
rather than by instability in an accretion disk. 
Type Ia (luminous) supernovae also exhibit exponential fades.  In these objects,
the decline is mapped onto the radioactive decay of particular isotopes 
following the explosion.  Destructive explosion scenarios 
fail overall for \jg\ in several ways (see \S9).

In terms of color comparisons, only the recent PTF 10qpf (HBC 722) 
has been observed well enough for definitive statements.  \cite{semkov2017}
show that the burst itself was a blue-ing event, with $V-I$ color
decreasing by $\sim$1 mag, then increasing in the immediate post-peak
period by $\sim$0.5 mag before again becoming slightly bluer after about
1 year, and finally settling around 0.5 mag bluer than the pre-burst
colors.  For \jg, there is no pre-burst optical color information, 
but the source clearly has very blue ultraviolet and optical colors
in the immediate post-peak period. Unlike PTF 10qpf,
the intial fade of \jg\ is relatively colorless in the optical, but does redden
slightly after a few months (Figure~\ref{fig:colorcurves}).

\subsection{Outburst Luminosity}

The luminosity of \jg\ can be estimated from the peak magnitude 
of $R_{PTF}=$ 14.93, an assumed distance, and an extinction estimate.  

Our radial velocity measurement appears to confirm the association
with the nearby W3/W4/W5 complex at 1.95 kpc.  However, as noted above, 
while this is the only known region of recent star formation 
anywhere close to the direct line of sight, there is also a further spiral arm 
beyond the Peseus Arm
in this general direction, that is not detected at the exact location of \jg, 
but can be also considered as a plausible distance.  

Using the extinction derived from the SED fitting
of $A_V$ = 4.75 mag or $A_R$ = 3.56 mag, and adopting the bolometric
correction appropriate to an A0 star (about -0.25 mag), 
a source located at 2 or 6 kpc would have luminosity 114 or 1014 \lsun. 
Changing the dominant spectral type would alter the bolometric correction
by only a few tenths.  Simple scaling to solar values 
also yields 100 \lsun\ for the preferred 2 kpc distance.
The luminosity can also be estimated by integrating the limited range 
that we have sampled photometrically near the peak of the spectral 
energy distribution.  
Fitting a model that accounts
for extinction (see Figure~\ref{fig:sed}) yields $\sim$130 \lsun,
modulo the exact temperature and extinction used in the fit.
Typical FU Ori luminosities are several hundred \lsun, 
though V900 Mon is also only $\sim$100 \lsun\ and PTF 10qpf (HBC 722) 
is a notable outlier at only $\sim$12 \lsun.
The recently announced Gaia 17bpi \citep{hillenbrand2018} is 
even lower at only $\sim$7.5 \lsun.

We believe that the large luminosity of \jg\ is produced in an accretion disk.
However, if instead of a disk, the luminosity is 
(hypothetically) generated by a normal stellar
photosphere, then assuming 114 (or 1014) \lsun\ and the 9700 K temperature 
corresponding to the A0 spectral type, a radius of 3.8 (11.3) \rsun\ 
is implied for the \jg\ photosphere. The location in the HR diagram of such a source
would be on (near) the main sequence, implying a mass of 3 (5) \msun.
This radius would also be consistent with both the large rotation rate 
implied by the 100-150 \kms\ line broadening that we observe, 
and with the high terminal velocity seen in the outflow
presuming this is also the escape velocity from the star.  
For the supergiant classification of the spectrum, 
instead of the 10$^{2-3}$ \lsun\ and 3-5 \msun\ calculated above,
such an object should \citep{verdugo1999,abt1995}
have 10$^{4-6}$ \lsun\ and 10-40 \msun, with much larger radius (30-100 \rsun) 
as well as slower rotation (20-40 \kms) and smaller escape velocity ($<$275 \kms). 
Another factor of several in distance -- beyond the 6 kpc arm -- 
is required to get above 
$10^4$ \lsun\ and into the lower luminosity regime of normal A0 supergiants, 
or roughly a factor of ten in distance to come close to the $10^5$ \lsun\ 
realm more typical of early A supergiants.   

The lack of consistency of the above logic 
lends credence to an accretion disk origin for the high source luminosity.
The spectrum of \jg\ is very clearly dominated by a hot component, 
but the photosphere also has low surface gravity.  
Yet no single-temperature
``normal" stellar source can fit all of the observational constraints.

\subsection{Absorption Spectrum}

The \jg\ infrared spectrum is pure continuum, lacking 
HI lines as in the optical spectrum
but also the characteristic CO absorption that is a hallmark of FU Ori stars.
This is consistent with a hot photosphere\footnote{The FU Ori + Herbig Ae/Be binary ZCMa 
also has no CO absoption in composite spectra, but when spatially resolved
as in \cite{hinkley2013} the absorption in the B component FU Ori source is apparent.}.

The optical low-dispersion spectrum of \jg\ (Figure~\ref{fig:optspec})
initially appeared a reasonable FU Ori match,
displaying many lines that are seen in FU Ori stars 
and having similarity to an FG giant or supergiant spectrum, 
but with the notable addition of strong Si II, Mg II, and Fe II absorption 
that is characteristic of late B and early A stars, but not FG stars.  
At high dispersion, besides atypically hot species,
the absorption lines exhibited by \jg\ are also atypically strong, 
with EW typically exceeding 1 \AA.
In the redder parts of the optical spectrum, the lines reach 
87-90\% depth instead of the 92-98\% depth seen in other FU Ori stars.
In addition, at FWHM $\sim$ 100--150 \kms, the early absorption spectrum 
was broader than that of all other known FU Ori stars.  
The hot (high excitation) absorption lines are likely
coming from above the stellar surface, perhaps the inner wind.

We note that most other FU Ori stars have been observed at high dispersion
only at much later stages in their outbursts.   
Even the recent PTF 10qpf (HBC 722) and V960 Mon were sampled at high 
dispersion for the first time many weeks to months after their outburst peaks.
Indeed \jg\ exhibited significant spectral evolution from 
Figure~\ref{fig:hires}, taken just 12 days post-peak.

\begin{figure}
\includegraphics[scale=0.3,trim={0 0 0.75cm 0},clip]{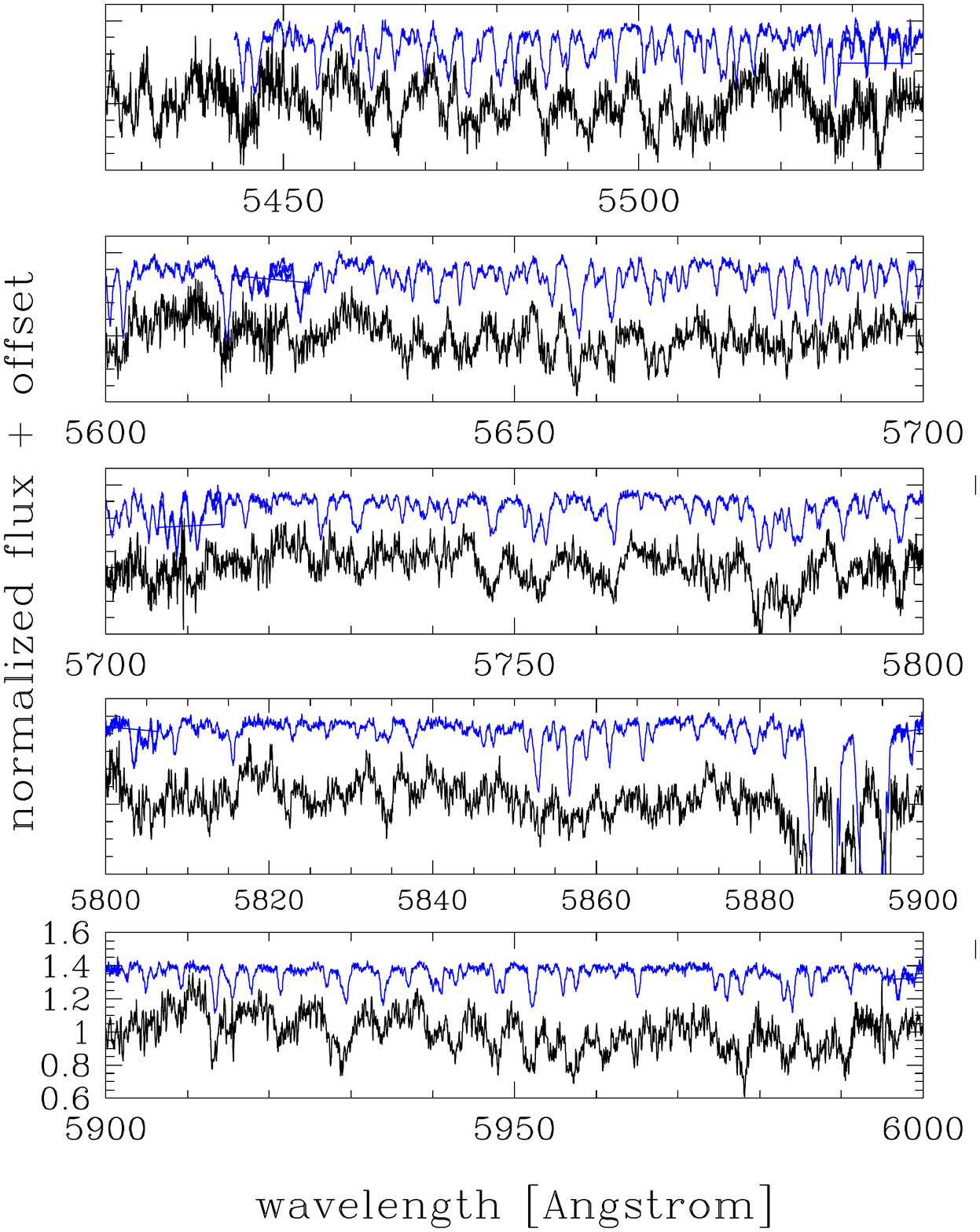}
\includegraphics[scale=0.3,trim={0 0 0.75cm 0},clip]{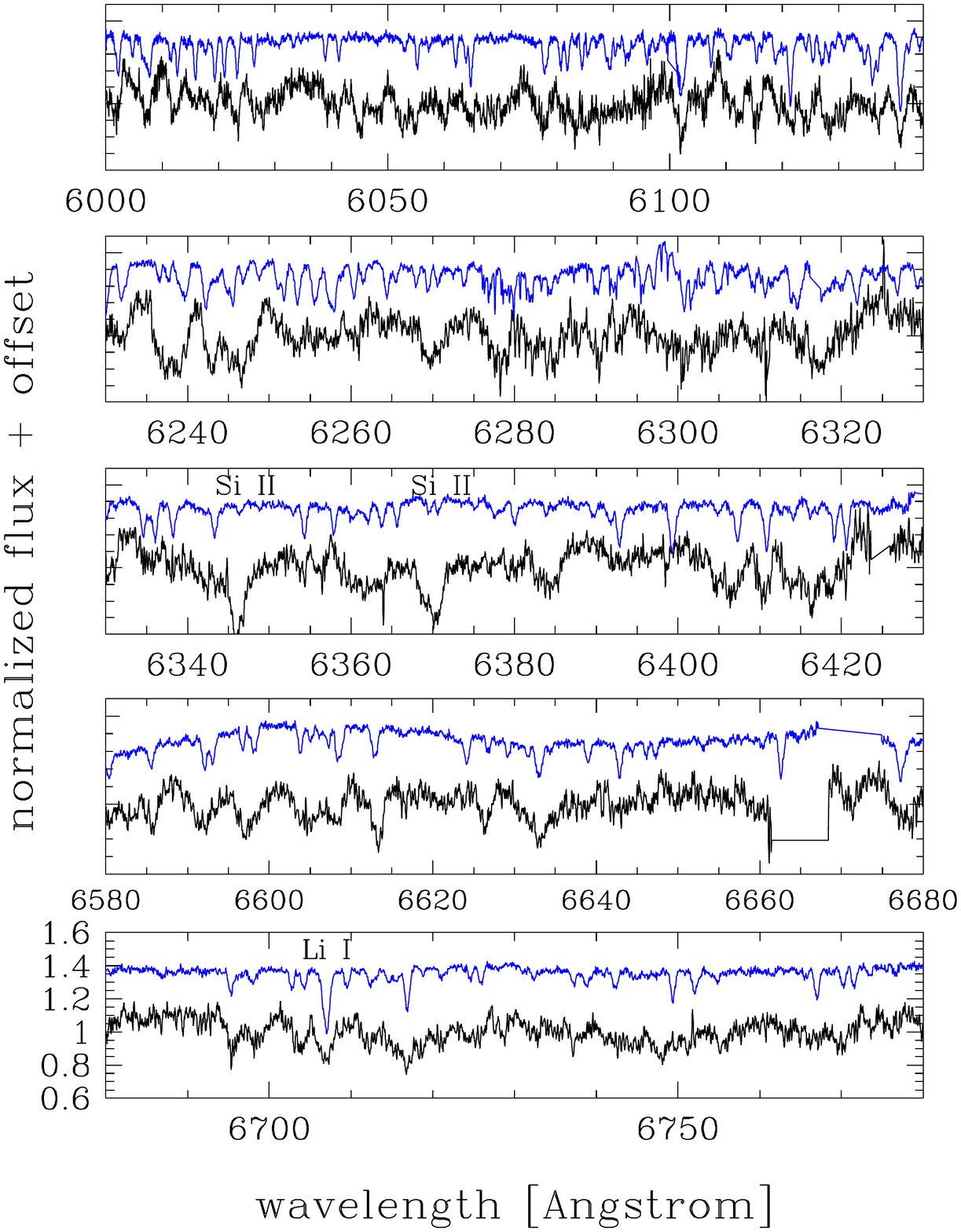}
\includegraphics[scale=0.3,trim={0.75cm 0 0.75cm 0},clip]{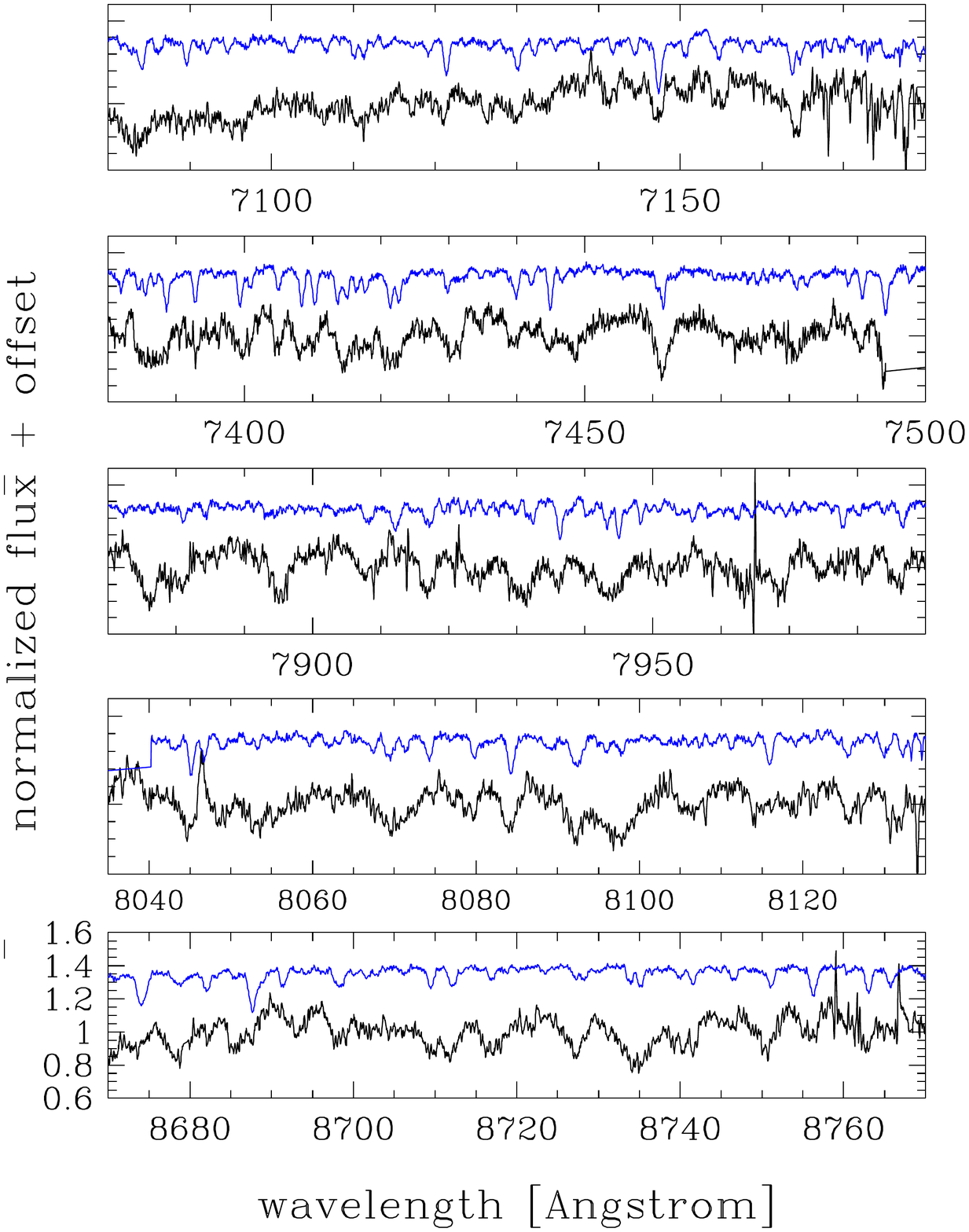}
\caption{Comparison of the \jg\ Keck/HIRES spectrum (in black) 
from 2015 1027 when the source was 625 days post-peak, 
to that of V1515~Cyg (in blue) a {\it bona fide} FU~Ori star. 
The normalization is non-optimal in some orders, 
and \jg\ is clearly broader, with many mainly high-excitation lines 
still present (most notably \ion{Si}{2} at e.g. 6347, 6371 \AA\, as well as
\ion{Mg}{2} and \ion{Fe}{2}). 
Nevertheless, the spectral match is reasonable, including 
the now-clear \ion{Li}{1} 6707 \AA\ signature in \jg.
This post-burst evolution of the optical absorption spectrum indicates 
development of a cooler photosphere relative to that displayed in
Figure~\ref{fig:hires}, from just 12 days post-peak.
}
\label{fig:comparev1515}
\end{figure}

Figure~\ref{fig:comparev1515}
shows a comparison of the late-time high dispersion spectrum of \jg, 
taken 1.7 years post-peak, with V1515~Cyg. 
The match is quite good, but not exact.
Most of the absorption that is present in \jg\ but not V1515~Cyg is due
to the high-excitation \ion{Si}{2}, \ion{Mg}{2}, \ion{Fe}{2} mentioned above.
\jg\ also has more highly blueshifted \ion{Na}{1} D than V1515~Cyg. 
V1515~Cyg is notably the least line-broadened member of the FU Ori class,
and is interpreted as having lower inclination than V1057 Cyg and FU Ori
which have much broader absorption lines.
The approximate match of widths in the non-wind absorption lines
may suggest that \jg\ is seen at a low-inclination viewing angle. 
However, the \ion{Ca}{2} triplet line morphology 
(Figure~\ref{fig:profile_evol}) seemed to suggest a higher viewing angle.

Also apparent in Figures~\ref{fig:profile_evol} 
and ~\ref{fig:comparev1515} is that, 
although there was a lack of clear and strong Li I in the earliest data on \jg, 
as the lightcurve faded, a narrow \ion{Li}{1} component did emerge.  
The early-stage \ion{Li}{1} profiles that prevented us from making a much earlier
conclusion regarding the young star nature of this object 
were likely dominated by the outflow.  We note that
several other accepted FU Ori stars also have ambiguous Li I profiles
(e.g. V1057 Cyg) that may be similarly affected by outflow. 
At early times, the \ion{Li}{1} profile seems similar 
to that of Par 21 or V883 Ori, while at late times
the double-peaked \ion{Li}{1} line is reminiscent of profiles
exhibited by FU Ori, V582 Aur, and PTF 10qpf / HBC 722.

Finally, we recall that absorption at 6614 \AA\ is present in \jg.  
This line is prominent in all FU Ori spectra and is also seen in FGK supergiant spectra, 
but not in the higher gravity giant or dwarf objects. While it seems
promising as a gravity constraint, this wavelength is also identified 
with a DIB (diffuse interstellar band) and is therefore more likely 
to be seen in more luminous sources.  Several other DIBs are also seen 
in the \jg\ spectrum (see Figure~\ref{fig:hires}) with 
narrower profiles at $\approx 40$\kms, compared to the generally broad 
absorption features.
Strong narrow atomic interstellar absorption is also identifiable in
the \ion{Ca}{1}, \ion{Na}{1}, and \ion{K}{1} profiles, whose outflow and zero-velocity
components were discussed above.

\subsection{Wind/Outflow Signatures}

As discussed in detail in \S7.4,
\jg\ exhibited a wind-dominated optical spectrum over the first 
six months of the outburst, the signatures of which gradually decreased in strength
over the subsequent several years, as the source faded photometrically.
The wind was shown to exhibit different absorbing components over a wide
range of velocities, 
presumably dictated by the density and temperature in 
the surrounding medium that allows different lines to form at different distances 
from the star.

For example, the optical light is coming through a region that is 
clearly optically thick in \ion{Na}{1} D and \ion{Ca}{2} H\&K. 
Both features show an inner edge to the absorption at -100 \kms\ (Figure~\ref{fig:wind_evol}), 
yet this region between the rest velocity and -100 \kms\ is emitting 
in the \ion{Ca}{2} triplet and -- early on -- in the \ion{Fe}{2} 5018 \AA\ line.
Hotter lines such as \ion{Si}{2} show continuous absorption from the rest
velocity out to a terminal velocity that is similar to the
\ion{Na}{1} D and \ion{Ca}{2} H\&K terminal velocity (Figure~\ref{fig:terminal}).
\ion{Mg}{2} likewise exhibits continuous absorption from the rest
velocity, but reaches a much lower terminal velocity.

The standard scenario for radiatively-driven outflow is governed by either the Eddington factor 
($2\times 10^{-5} ~L/M$, in solar units) under the most basic considerations or, 
for line-driven radiative winds, 
a relation between the wind momentum ($\dot{M}v_{\infty}$)
and the stellar parameters ($\sqrt{L^3/R}$ in simplified form).  As noted
above, the luminosity of \jg\ seems too low for radiative processes
to be important in producing the observed outflow.  Furthermore, the wind
terminal velocity of around 500 \kms\ for \jg\ is higher by a factor of two
than those typical of A supergiants, albeit in the range of values 
observed for mid-B supergiants \citep[e.g.][]{kudritzki1999}.

In the FU Ori scenario, the outflow is accretion driven, and emanates from
the inner disk region. It then propagates through the dense circumstellar and
nearby interstellar medium.  

Comparing to low dispersion spectra of FU Ori stars, 
the obvious P Cygni structure in the \ion{Ca}{2} triplet emission of \jg, 
(Figure~\ref{fig:lines}) is consistent with those objects,
indicating strong outflow. Notable is the rapid evolution from early epochs 
to disappearance of the blueshifted absorption by about 4.5 months post-peak, 
in favor of a purely blueshifted-emission spectrum in \ion{Ca}{2} 8542 \AA\ 
(Figure~\ref{fig:profile_evol}).
Rapid evolution of the line structure was also observed
in the early stages of the PTF 10qpf outburst, with the red emission side 
of the 8542 \AA\ profile decreasing significantly in strength 
over the first few years, while the blueshifted absorption remained present.
The late-time morphology of the \ion{Ca}{2} 8542 \AA\ 
feature in \jg\ is perhaps most similar to V1735 Cyg.

Despite the strong wind signature in \ion{Ca}{2}, there was never
strong emission or clear P Cygni absorption in H$\alpha$ from \jg. 
Among FU Ori stars, the H$\alpha$ profile of \jg\ seems most similar 
to that of V883 Ori and possibly Par 21.
It also resembles the higher Balmer lines in PTF 10qpf that 
had more broadly blueshifted absorption than exhibited in
the H$\alpha$ profile \citep{miller2011}.  

The overall influence of the outflow on line absorption profiles in \jg\ 
is much more complicated than in the established FU Ori stars. 
\jg\ has the strongest and the hottest outflow signatures.  For example,  
we are not aware of any other FU Ori stars that exhibit any hint of a 
\ion{Si}{2} feature.  
This line has been documented in a few other young stars, 
e.g. \cite{grinin2001} illustrates its presence in UX Ori 
but does not discuss the line. It is also generally present in early F III stars. 

The most comparable object may be V1057 Cyg. 
Like \jg, V1057 Cyg has a broad, nearly saturated, 
multi-component outflow in \ion{Na}{1} D 
with the wind also visible in the \ion{K}{1} 7699 \AA\ line.
In both \jg\ and V1057, the \ion{Li}{1} profile 
has a zero velocity narrow {\it emission} component, 
along with blueshifted absorption out to about -100 \kms; the phenomenon
is discussed in detail in \cite{herbig2009} for V1057 Cyg.
Notably, the high-excitation \ion{Si}{2} lines that are broad and blueshifted 
in \jg\ are also weakly seen at zero velocity in V1057 Cyg, 
as well as in early spectra of V960 Mon (though weakening significantly 
a year post-peak in that object).

\subsection{Early-Time Emission Line Spectrum}

In the FU Ori interpretation, narrow permitted emission lines at an early stage of
the outburst could originate in a region of high density
along the poles, that is newly illuminated by the hot outburst.
Another location for the narrow emission lines could be 
in the magnetosphere -- if the large increase in accretion rate is slow enough 
that normal magnetospheric emission continues for several weeks or months.  
As noted above, there is a slight redshift of the narrow emission spectrum
by about 11.5 \kms\ relative to the inferred systemic velocity from
the late-time absorption line spectrum.
If the outburst accretion rate 
is low enough, the magnetosphere could be maintained instead of crushed. 
Only weak lines would be seen as narrow emission, while the strong lines 
would have their narrow magnetospheric emission swamped by the wind signatures,
consistent with our observations.

Unusual for FU Ori stars, is the presence of forbidden emission at [\ion{Ca}{2}]
7291, 7324 \AA\ in \jg.  This is a high density line relative to the usual suite
of ``nebular" forbidden lines.  Although not too common in young stars\footnote{
It can also be seen in some high luminosity blue or yellow supergiants, 
that may be post-RSG or post-AGB stars.}  
this line is weakly present in extreme emission-line objects like 
V1331 Cyg, V2492 Cyg (PTF 10nvg), and RW Aur, all of which have ubiquitous 
permitted and forbidden line emission.  None of the {\it bona fide} FU Ori stars show any 
[ ]-line emission\footnote{Exceptionally, V1057 Cyg does exhibit very weak [\ion{O}{1}].} 
(or the \ion{He}{1} that characterizes many of the more regularly accreting 
T Tauri stars).
This means that the FU Ori winds must be more dense than the typical TTS winds
thought to originate near the magnetospheric region. 

\cite{hartigan2004} discuss the [\ion{Ca}{2}] doublet in young stars
in detail, reporting that the critical density is $5\times 10^{7}$ cm$^{-3}$ and that the expected equilibrium ratio of 7291 \AA\ : 7324 \AA\ $=1.5$, 
relatively independent of density.  This is roughly consistent with the 
measured ratio $\sim$1.2 in \jg.  As also noted earlier,
all of the \ion{Ca}{2} lines that we observe in \jg\ are related.  
The H\&K doublet
(which we see in strong and broadly blueshifted absorption) has 
an upper level that is also the ``infrared" triplet line 
(having classical P Cygni structure) upper level, with its lower level 
then the forbidden doublet's (narrow emission) upper level.
The forbidden doublet's lower level (the ground state) is in common with 
the H\&K lower level.  The deep absorption that we see in the H\&K lines
could power the radiative de-excitation causing the triplet line emission.
That the forbidden doublet is also in emission, implies continued radiative
de-excitation before the normal collisional de-excitation 
(or, less likely, collisional re-excitation) 
occurs from this level of the permitted line.  
However, these phenomena are occurring at different velocities in 
the different \ion{Ca}{2} lines in \jg, and thus the
relation between the lines is unclear.

The observed ratio of $\sim$9 in the emission peaks of \ion{Ca}{2} 8542 \AA\ 
to [\ion{Ca}{2}] 7291 \AA\ (Figure~\ref{fig:lines}, showing various epochs) 
corresponds to an electron density of several times $10^{8}$ cm$^{-3}$,
independent of temperature, according to Figure 8 from \cite{nisini2005}. 
Figure 2 from Ferland and Persson (1989) suggests a few times $10^{9}$ cm$^{-3}$.
The implied high density suggests that collisions are
more important than radiation in forming the emission lines, but again, 
the very different morphologies of the permitted vs. forbidden lines 
may invalidate the calculation.  The low velocity gas -- where the forbidden
(and other neutral atomic) emission arises -- may be at low density,
whereas the higher velocity gas may have higher density.
This would be consistent with a wind that is accelerated 
within a dense disk (forming the broad, permitted triplet emission)
and launched into a less dense circumstellar medium where it decelerates
(forming the narrow forbidden doublet emission).

\subsection{Progenitor}

The pre-outburst nature of \jg\ is poorly constrained. 
Considering the immediate pre-outburst brightness of 
$R_{PTF}\approx$ 21.5, and the plausible range of spectral types 
and extinctions that are consistent with the limited SED information, 
the luminosity estimate is $\sim$0.1-0.3 \lsun\ for the 2 kpc distance.
These values would be typical for low mass ($<$0.5 \msun) 
young pre-main sequence stars.

Using the 
\cite{baraffe1998} evolutionary models implemented within TADA
\citep{dario2012}, the faint pre-outburst PTF and Spitzer/IRAC photometry 
would correspond to just $\sim$ 0.05 to 0.2 to 0.5 $M_\odot$ objects 
for ages of 1, 10, and 100 Myr of age. 
Adopting the source extinction estimate of $A_V=4.75$ mag has little
effect on these mass numbers.  Even if an older field object in the Perseus Arm,
and not a young source associated with W3 / W4 / W5,
the \jg\ progenitor is implied to have sub-solar mass given its brightness.

Using another line of argument, if the $\approx$500 \kms\ 
that is observed as the maximum terminal velocity in the various 
(accretion-driven) wind-dominated lines can be interpreted as the
escape speed, this would correspond to 3.0-0.8-0.5 \msun\  
stars for the radii that are implied at ages of $\approx$1-3-10 Myr.

\section{\jg\ Characteristics in the Context of Other Possible Interpretations}

Besides the FU Ori event interpretation discussed above, 
what other type of large amplitude photometric rise would have characteristics 
like what we observe for \jg?  The salient features of the outburst are:
\begin{itemize}
\item
a progenitor of $\sim$0.1-0.3 \lsun\ and, if a pre-main sequence star,
approximately $0.2-0.5 M_\odot$. 
\item
source brightening by 6-7 mag over a few month time period,
\item
exponential decline from lightcurve peak with an e-folding time 
of $\sim$111 days, reaching a plateau $\sim$3 mag below peak, 
and $\sim$3.5-4 mag above quiescence, 
\item
initially colorless fade in the optical, though exhibiting a later 
reddening trend in optical and near-infrared colors, 
and possible blueing in [3.5]-[4.5] color,
\item
peak luminosity of 100-130 $L_\odot$,
\item
ultraviolet, near-infrared, and mid-infrared excess, 
\item
a low gravity (supergiant), composite spectrum with a systematically changing
temperature going from bluer optical wavelengths where it best matches an
early A spectral type (though lacking strong hydrogen lines), 
to redder optical wavelengths where it is best matched to a G or even K0 
spectral type,
\item
line broadening of $\sim$150 \kms\ in the optical absorption spectrum,
\item
only very weak and blueshifted absorption in HI and HeI lines, 
\item
strong and highly blueshifted (500 \kms\ terminal velocity)
absorption in the CaII H\&K resonance lines, NaD, and OI, 
which are often seen in winds, as well as less typical species
such as low-excitation Fe II and higher excitation SiII and MgII,  
\item
moderate CaII triplet and forbidden [CaII] emission, 
with early-time kinematic structure,
\item
narrow ($\sim$25-30 \kms) and symmetric neutral species atomic emission from 
e.g. Fe I, Mg I, Ca I that was observed initially in the 
optical and near-infrared, for at least six weeks post-peak, 
but disappearing by about six months after the light curve peak,
\item
an outburst lifetime lasting at least five years.
\end{itemize}

We have considered a wide range of explanations that are summarized in 
Table~\ref{tbl:explain}.  
Most known categories of large amplitude instabilities or explosive events 
can be ruled out by the long rise time and relatively slow decay time of \jg,
along with its only moderate outburst luminosity.   
Other explanations are ruled out by the lack of a dust formation phase
and/or a late-time emission spectrum phase. 

We also note that over the several years of high-dispersion follow-up data, 
there is no evidence for radial velocity variation 
in the emerging absorption spectrum. 
This eliminates a certain parameter space of binarity, 
and thus close binary interaction as an explanation for the photometric outburst.




\movetabledown=1.75in
\begin{rotatetable}

\begin{deluxetable}{c|c|cccccccccccccc}
\tabletypesize{\footnotesize}
\tablecolumns{16}
\tablewidth{0pt}
\tablecaption{PTF 14jg Hypothesis Matrix\label{tbl:explain}}
\tablehead{
\multicolumn{1}{c|}{Phenomenon} & 
\colhead{Examples} &  
\multicolumn{14}{|c}{Is the Observed Property in PTF\,14jg Consistent with the Category?} 
}
%
%
\startdata
{} & 
{} &
{pre-} &
{rise} &
{$\Delta$mag} &
{blue} &
{colorless} &
{decay} &
{2nd } &
{low } &
{infrared} &
{wind} &
{weak } &
{low eV} &
{high eV} &
\multicolumn{1}{c}{spec} 
\\
{} & 
{} &
{burst} & 
{time} &
{} &
{$+$hot } &
{fade} &
{time} &
{peak} &
{lum.} &
{excess} &
{speed} &
{H$\alpha$} &
{narrow} &
{broad } &
\multicolumn{1}{c}{evol.} 
\\
{} & 
{} &
{variab.} & 
{} &
{} &
{burst} &
{} &
{} &
{} &
{} &
{} &
{} &
{abs.} &
{emis.} &
{abs.} &
\multicolumn{1}{c}{} 
\\
\hline
\multicolumn{16}{c}{\textbf{Accretion-Related Events}} \\
\hline 
Young Star outburst & 
V1057\,Cyg  & 
\multirow{2}{*}{\checkmark} &     
\multirow{2}{*}{\checkmark} &     
\multirow{2}{*}{$\times$} &           
\multirow{2}{*}{$\times$} &       
\multirow{2}{*}{\checkmark} &     
\multirow{2}{*}{$\times$} &           
\multirow{2}{*}{\checkmark} &             
\multirow{2}{*}{\checkmark} &     
\multirow{2}{*}{\checkmark} &     
\multirow{2}{*}{\checkmark} &     
\multirow{2}{*}{$\times$} &       
\multirow{2}{*}{$\times$} &       
\multirow{2}{*}{\checkmark} &     
\multirow{2}{*}{$\checkmark$}     
\\
(FU Ori event) & 
PTF\,10qpf     & 
\multicolumn{13}{c}{} &
{} 
\\
\hline
\multirow{2}{*}{Symbiotic Binary nova} & 
V694 Mon  & 
\multirow{2}{*}{} &     
\multirow{2}{*}{} &     
\multirow{2}{*}{\checkmark} &           
\multirow{2}{*}{\checkmark} &       
\multirow{2}{*}{} &     
\multirow{2}{*}{\checkmark} &           
\multirow{2}{*}{} &             
\multirow{2}{*}{$\times$} &     
\multirow{2}{*}{\checkmark} &     
\multirow{2}{*}{\checkmark} &     
\multirow{2}{*}{$\times$} &       
\multirow{2}{*}{} &       
\multirow{2}{*}{$\times$} &     
\multirow{2}{*}{$\times$}     
\\
{} & 
FG Ser     & 
\multicolumn{13}{c}{} &
{} 
\\
\hline
\multirow{1}{*}{Classical nova} & 
RR\,Pic& 
\multirow{1}{*}{$\times$} &     
\multirow{1}{*}{$\times$} &     
\multirow{1}{*}{\checkmark} &           
\multirow{1}{*}{\checkmark} &       
\multirow{1}{*}{} &     
\multirow{1}{*}{$\times$} &           
\multirow{1}{*}{\checkmark} &             
\multirow{1}{*}{$\times$} &     
\multirow{1}{*}{\checkmark} &     
\multirow{1}{*}{} &     
\multirow{1}{*}{$\times$} &       
\multirow{1}{*}{} &       
\multirow{1}{*}{} &     
\multirow{1}{*}{$\times$}     
\\
{} & 
V4739\,Sgr& 
\multicolumn{13}{c}{} &
{} 
\\
{} & 
V3890\,Sgr& 
\multicolumn{13}{c}{} &
{} 
\\
{} & 
V4643\,Sgr& 
\multicolumn{13}{c}{} &
{} 
\\
\hline
\multirow{5}{*}{Slow nova} & 
T\,Pyx  & 
\multirow{5}{*}{} &     
\multirow{5}{*}{$\times$} &     
\multirow{5}{*}{\checkmark} &           
\multirow{5}{*}{\checkmark} &       
\multirow{5}{*}{$\times$} &     
\multirow{5}{*}{} &           
\multirow{5}{*}{\checkmark} &             
\multirow{5}{*}{$\times$} &     
\multirow{5}{*}{} &     
\multirow{5}{*}{} &     
\multirow{5}{*}{$\times$} &       
\multirow{5}{*}{} &       
\multirow{5}{*}{} &     
\multirow{5}{*}{$\times$}     
\\
{} & 
V2540\,Oph     & 
\multicolumn{13}{c}{} &
{} 
\\
{} & 
V5558\,Sgr     & 
\multicolumn{13}{c}{} &
{} 
\\
{} & 
HR\,Del     & 
\multicolumn{13}{c}{} &
{} 
\\
{} & 
V723\,Cas     & 
\multicolumn{13}{c}{} &
{} 
\\
\hline
\multirow{2}{*}{Pre-cataclysmic nova} & 
U\,Gem  & 
\multirow{2}{*}{} &     
\multirow{2}{*}{$\times$} &     
\multirow{2}{*}{\checkmark} &           
\multirow{2}{*}{$\times$} &       
\multirow{2}{*}{$\times$} &     
\multirow{2}{*}{$\times$} &           
\multirow{2}{*}{} &             
\multirow{2}{*}{$\times$} &     
\multirow{2}{*}{} &     
\multirow{2}{*}{} &     
\multirow{2}{*}{$\times$} &       
\multirow{2}{*}{} &       
\multirow{2}{*}{} &     
\multirow{2}{*}{$\times$}     
\\
{} & 
SS\,Cyg     & 
\multicolumn{13}{c}{} &
{} 
\\
{} & 
Z\,Cam     & 
\multicolumn{13}{c}{} &
{} 
\\
\hline
\multicolumn{16}{c}{\textbf{Nuclear Burning Instabilities}} \\
\hline
\multirow{1}{*}{Helium-Flash instability} & 
V445\,Pup & 
\multirow{1}{*}{} &     
\multirow{1}{*}{} &     
\multirow{1}{*}{} &           
\multirow{1}{*}{\checkmark} &       
\multirow{1}{*}{} &     
\multirow{1}{*}{} &           
\multirow{1}{*}{} &             
\multirow{1}{*}{$\times$} &     
\multirow{1}{*}{\checkmark} &     
\multirow{1}{*}{} &     
\multirow{1}{*}{\checkmark} &       
\multirow{1}{*}{\checkmark} &       
\multirow{1}{*}{$\times$} &     
\multirow{1}{*}{$\times$}     
\\
\hline
\multirow{2}{*}{Final Helium flash} & 
FG\,Sgr  & 
\multirow{3}{*}{$\times$} &     
\multirow{3}{*}{\checkmark} &     
\multirow{3}{*}{\checkmark} &           
\multirow{3}{*}{} &       
\multirow{3}{*}{$\times$} &     
\multirow{3}{*}{$\times$} &           
\multirow{3}{*}{} &             
\multirow{3}{*}{} &     
\multirow{3}{*}{} &     
\multirow{3}{*}{} &     
\multirow{3}{*}{\checkmark} &       
\multirow{3}{*}{} &       
\multirow{3}{*}{} &     
\multirow{3}{*}{$\times$}     
\\
{} & 
V4334\,Sgr     & 
\multicolumn{13}{c}{} &
{} 
\\
{(``born again" star)} & 
V605\,Aql    & 
\multicolumn{13}{c}{} &
{} 
\\
\hline
\multicolumn{16}{c}{\textbf{Massive Star Behavior}} 
\\
\hline
\multirow{2}{*}{Red Supergiant} & 
Mira  & 
\multirow{2}{*}{$\times$} &     
\multirow{2}{*}{\checkmark} &     
\multirow{2}{*}{\checkmark} &           
\multirow{2}{*}{$\times$} &       
\multirow{2}{*}{} &     
\multirow{2}{*}{\checkmark} &           
\multirow{2}{*}{$\times$} &             
\multirow{2}{*}{$\times$} &     
\multirow{2}{*}{\checkmark} &     
\multirow{2}{*}{} &     
\multirow{2}{*}{$\times$} &       
\multirow{2}{*}{} &       
\multirow{2}{*}{} &     
\multirow{2}{*}{$\times$}     
\\
{LPV pulsator} & 
V566\,Cas     & 
\multicolumn{13}{c}{} &
{} 
\\
\hline
\multirow{1}{*}{Yellow Hypergiant burst} & 
$\rho$ Cas  & 
\multirow{1}{*}{} &     
\multirow{1}{*}{} &     
\multirow{1}{*}{} &           
\multirow{1}{*}{\checkmark} &       
\multirow{1}{*}{} &     
\multirow{1}{*}{} &           
\multirow{1}{*}{} &             
\multirow{1}{*}{$\times$} &     
\multirow{1}{*}{} &     
\multirow{1}{*}{$\times$} &     
\multirow{1}{*}{} &       
\multirow{1}{*}{} &       
\multirow{1}{*}{} &     
\multirow{1}{*}{$\times$}     
\\
\hline
\multirow{1}{*}{Blue Supergiant} & 
$\eta$ Car  & 
\multirow{3}{*}{\checkmark} &     
\multirow{3}{*}{$\times$} &     
\multirow{3}{*}{\checkmark} &           
\multirow{3}{*}{$\times$} &       
\multirow{3}{*}{} &     
\multirow{3}{*}{\checkmark} &           
\multirow{3}{*}{} &             
\multirow{3}{*}{$\times$} &     
\multirow{3}{*}{} &     
\multirow{3}{*}{$\times$} &     
\multirow{3}{*}{$\times$} &       
\multirow{3}{*}{$\times$} &       
\multirow{3}{*}{$\times$} &     
\multirow{3}{*}{$\times$}     
\\
{LBV outburst} & 
P\,Cyg     & 
\multicolumn{13}{c}{} &
{} 
\\
{} & 
AG\,Car     & 
\multicolumn{13}{c}{} &
{} 
\\
\hline
\multirow{1}{*}{Supernova Imposter} & 
UGC\,2773-OT  & 
\multirow{1}{*}{$\times$} &     
\multirow{1}{*}{\checkmark} &     
\multirow{1}{*}{$\times$} &           
\multirow{1}{*}{\checkmark} &       
\multirow{1}{*}{} &     
\multirow{1}{*}{} &           
\multirow{1}{*}{} &             
\multirow{1}{*}{$\times$} &     
\multirow{1}{*}{} &     
\multirow{1}{*}{$\times$} &     
\multirow{1}{*}{$\times$} &       
\multirow{1}{*}{$\times$} &       
\multirow{1}{*}{$\times$} &     
\multirow{1}{*}{$\times$}     
\\
\hline
\multicolumn{16}{c}{\textbf{Merger Events}} 
\\
\hline
\multirow{2}{*}{Star-Star merger} & 
V1309\,Sco & 
\multirow{2}{*}{\checkmark} &     
\multirow{2}{*}{$\times$} &     
\multirow{2}{*}{$\times$} &           
\multirow{2}{*}{$\times$} &       
\multirow{2}{*}{} &     
\multirow{2}{*}{$\times$} &           
\multirow{2}{*}{$\times$} &             
\multirow{2}{*}{$\times$} &     
\multirow{2}{*}{\checkmark} &     
\multirow{2}{*}{$\times$} &     
\multirow{2}{*}{$\times$} &       
\multirow{2}{*}{$\times$} &       
\multirow{2}{*}{$\times$} &     
\multirow{2}{*}{$\times$}     
\\
{} & 
V838\,Mon     & 
\multicolumn{13}{c}{} &
{} 
\\
{} & 
V4332\,Sgr     & 
\multicolumn{13}{c}{} &
{} 
\\
\hline
\multirow{1}{*}{Star-Planet merger} & 
theoretical & 
\multirow{1}{*}{} &     
\multirow{1}{*}{\checkmark} &     
\multirow{1}{*}{} &           
\multirow{1}{*}{$\times$} &       
\multirow{1}{*}{} &     
\multirow{1}{*}{$\times$} &           
\multirow{1}{*}{} &             
\multirow{1}{*}{$\times$} &     
\multirow{1}{*}{} &     
\multirow{1}{*}{} &     
\multirow{1}{*}{} &       
\multirow{1}{*}{} &       
\multirow{1}{*}{} &     
\multirow{1}{*}{$\times$}     
\\
\enddata
\tablecomments{
The table can be interpreted as follows:
\\ $\checkmark$ = the observed properties in PTF\,14jg and the phenomenon being considered are consistent. 
\\ $\times$ = the phenomenon under consideration does not match the observed behavior of PTF\,14jg.
\\
Young star outbursts have the most properties in common with PTF\,14jg. Almost all of the other hypotheses fail to replicate the low luminosity or weak absorption spectrum observed in PTF\,14jg. 
}

\end{deluxetable}
\end{rotatetable}


\section{Discussion}

Our conclusion from the previous two sections is that \jg\ bears some
resemblance to FU Ori stars, albeit with many unusual characteristics,
and that all other possible explanations for the large amplitude 
photometric increase and warm spectrum encounter insurmountable difficulties.
The evidence for pre-existing circumstellar dust, 
plus the composite absorption spectrum, the broad absorption lines, 
and the strong wind -- all sustained several years 
post-peak -- suggests an accretion-driven burst accompanied 
by an accelerated wind.  We thus consider it an acceptable hypothesis that \jg\ 
is just an unusually hot FU Ori disk. 

\begin{figure}
\includegraphics[scale=0.5]{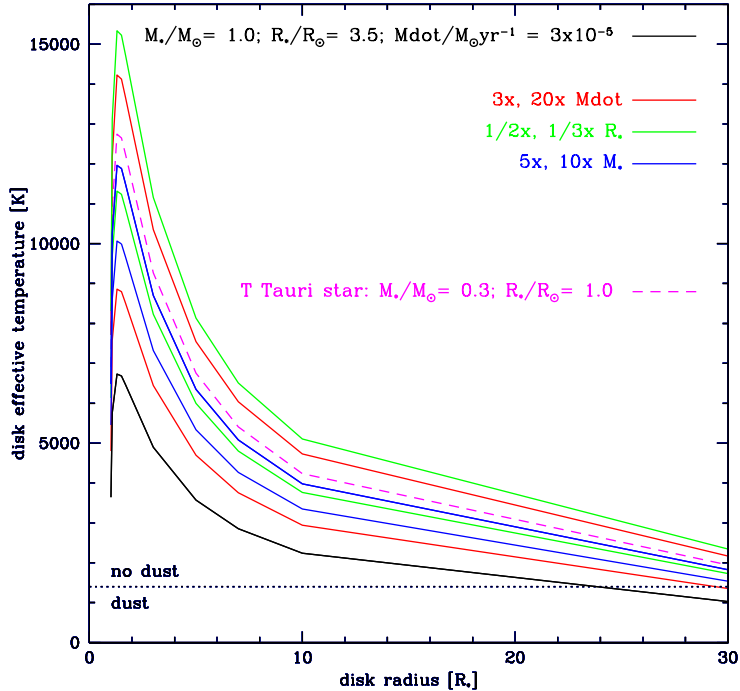}
\vskip -1truein
\caption{Classical accretion disk models showing the plausible parameters
that could produce a disk with maximum temperature in the range required by 
the broad-lined absorption spectrum of \jg.  A fiducial FU Ori disk 
that reaches a maximum temperature of close to 7000 K, is shown in black. 
Red lines indicate increased accretion rate, green lines
decreased stellar radius, and blue lines increased stellar mass.
The dashed magenta line shows the stellar parameters for a young pre-main sequence
star having the inferred pre-outburst luminosity of 0.1 \lsun, 
which in addition to the implied lower mass, has a smaller radius 
(i.e. older pre-main sequence age of 2-3 Myr) relative to 
the fiducial solar-mass case, 
in order to achieve the required higher disk temperature.
Other combinations are possible.
}
\label{fig:accdisk}
\end{figure}

In classical viscous accretion disk theory \citep{lbp1974}, 
the maximum temperature of the disk occurs at $\approx 1.4 R_*$,  and has a value of $0.49 (3G\dot{M}M_*/8\pi\sigma R_*^3)^{1/4}$
with temperature falling off as r$^{-3/4}$.  High disk temperature, such as
we observe, thus requires one or more of: high mass accretion rate, 
high stellar mass, or small stellar radius compared to a canonical 
FU Ori disk. As shown in Figure~\ref{fig:accdisk}, 
mass above 5 \msun\ or radius closer to 1 \rsun, 
or a very high accretion rate above $10^{-4}$ \msun/yr
(with the other parameters fixed), would be required
in order to raise the inner disk temperature above 10,000 K. 

Guided by the implied low luminosity in the pre-outburst phase,
we can consider a low mass star, in the range 0.1-0.5 \msun, 
with a 0.8-1.2 \rsun\ radius, that would give the observed disk temperature.  
This object (magenta line in the Figure)
would have an age of a few Myr instead of the $<$1 Myr age 
of the fiducial case.  It would thus be a late-stage FU Ori event.
We note that this argument is consistent with the inferred source 
luminosity.  Rewriting the above equation in terms of accretion luminosity
instead of accretion rate, we find
$T_{max} = 3700 (L_{disk}/L_\odot)^{1/4} (R/R\odot)^{-1/2}$, and
for $T_{max} = 12,000$K and $L_{disk} = L = 114 L_\odot$,
that $R = 0.95 R_\odot$ is implied.

If alternate merge-burst or other explosive scenarios are correct, 
then there would have been rapid evolution (on roughly a month long time scale)
in the lightcurve decline and in the spectrum from early to late type, 
as an expanding shell cooled.  Instead, the lightcurve of \jg\ was only slowly 
declining, and there was no evolution in the spectrum over several months. 

\jg\ experienced a faster photometric decay 
than most other known FU Ori type bursts. 
It is still 3.5-4 mag above quiescence, however. 
Its optical rise amplitude of 6.7 mag was also larger than most FU Ori bursts. 
And the spectrum is hotter.
Despite the differences relative to canonical FU Ori stars, 
if \jg\ is indeed a member of this class, it is one of $<$10 
young stars to have been caught in the act, that is, 
to have had its outburst phase fully captured and well-documented
(better than almost all, in fact).

Among young star bursts and outbursts,
the popular FU Ori and EX Lup classes differ in 
their amplitudes, durations, and duty cycles, as well as in their physical interpretation.  
The former are large-scale disk instabilities while the latter smaller-scale 
events have a probable origin in the magnetospheric region. 
However, other young star outburst types are beginning to be characterized
in the literature, and the landscape of burst amplitude and burst time scale is not yet fully appreciated.  
These new object types include those with EX Lup type amplitudes but even shorter duration (few months) 
outbursts such as ASAS-SN13db (5.4 mag, $<$150 days; Holoien et al. 2014), 
PTF 15afq (3.5 mag, 200 days; Miller et al. 2015), 
V899 Mon (3 mag, $>$10 years; Ninan et al. 2015),
and 
ASAS-15qi (3.5 mag, 200 days; Herczeg et al. 2016),
and the even lower amplitude and more frequent (few days to week) bursts 
associated with accreting stars as studied by \cite{cody2017} in Upper Sco 
and \citet{stauffer2014, stauffer2016} in NGC 2264.  
\jg\ may be an example of similar diversity among the larger amplitude FU Ori type events.

The phase space of amplitude-timescale for young star outbursts is still being mapped out.
Time domain surveys are now producing quality lightcurves of young stars
with roughly month-long (e.g. the YSOVAR surveys with Spitzer, the K2 mission,
the VVV survey) to years-long (e.g. ASAS, PTF, PanSTARRS, ZTF, ATLAS) duration. 
Our understanding of the range of burst and outburst behavior exhibited by 
young stars will thus continue to grow.  Worth remembering is  
the reminder by \cite{herczeg2016}: ``measuring the duration
of an outburst usually requires impatient people to wait".  

\section{Conclusions}

We have presented the outburst of \jg\ near the W4 HII region,
documenting the early history of its photometric and
spectroscopic behavior.

The rise time and the amplitude of the \jg\ outburst are 
roughly consistent with an FU Orionis classification, 
though at an extreme in both parameters relative to 
other members of the class.
The characteristic change in spectral type with wavelength 
that is a hallmark of the class is present, as is \ion{Li}{1} absorption.
However, the spectral signatures are overall hotter 
in the early stages, both in the rest velocity (disk) spectrum 
and the outflowing (wind) spectrum, compared to other FU Ori stars. 
As shown in Figure~\ref{fig:accdisk}, the required temperatures can be
generated by a disk around a 0.3 \msun, 1.0 \rsun\ star having an accretion rate 
of $3\times10^{-5} M_\odot$/yr.
This would be a 1-3 Myr old progenitor rather than the more typically considered
$<1$Myr old FU Ori source.

Our classification of \jg\ as a member of the FU Ori class is tentative.
We also considered a large body of alternate explanations for our suite of observations.
However, we find no clearly better, or even equally good, alternate 
to the FU Ori outburst interpretation.
Continued spectral study as the source cools from it peak brightness, 
especially at infrared wavelengths, should better illuminate its nature.

\section{Acknowledgements}
We have made extensive use of the SIMBAD and ADS on-line resources in
sleuthing for possible analogs of \jg.  We also spent a considerable amount
of quality time with the NIST atomic database in attemping line identification
in the HIRES data, and we consulted the MILES population synthesis spectral database 
when searching for spectral templates.  
We are grateful to Cathie Clarke and Giuseppe Lodato for locating 
electronic copies of photometric data on the ``classical" FU Ori outbursts 
that were published in their 2005 paper, and to Ariel Langer who
contributed to our analysis of these lightcurves.
We thank Mark Heyer for locating CO channel maps from his 1998 paper.  
We benefited from allocations by Tom Soifer as Director
of the Spitzer Space Telescope and Lee Mundy as director of CARMA,
of small amounts of DDT to obtain the observations reported here.  
We thank the spectroscopic observers listed in Table~\ref{tbl:spec} who are not
otherwise acknowledged with authorship, especially Yi Cao, Sumin Tang, Jacob Jencsen, and Anna Ho. 
DPKB thanks N.M. Ashok and V. Venkataraman for help with some of the Mt. Abu observations.
The research at PRL is supported by the Department of Space, Government of India.  
We thank the staff members at the various observational facilities used, and the numerous instrument builders. 
The Intermediate Palomar Transient Factory project is a scientific collaboration among the California Institute of Technology, Los Alamos National Laboratory, the University of Wisconsin, Milwaukee, the Oskar Klein Center, the Weizmann Institute of Science, the TANGO Program of the University System of Taiwan, and the Kavli Institute for the Physics and Mathematics of the Universe.
Finally, LAH is grateful for the tolerance of the many young star and nova pundits 
on whom various versions of this story have been tried out over the past several years. 
And we thank the referee for comments that provided a valuable opportunity for us to re-examine our presentation. 

\vskip0.25truein
{\bf Facilities:}
\facility{PO:1.2m:PTF, PO:1.5m:GRBcam, MIRO:1.2m, Hale:DBSP, Hale:TSPEC, Keck:I:HIRES, Keck:I:LRIS, Keck:I:DEIMOS, Keck:I:MOSFIRE, APO:DIS, APO:TSPEC, 2MASS, Spitzer, WISE, NEOWISE, Swift:UVOT, Swift:XRT, CARMA, IRSA }


\vskip0.25truein

\begin{thebibliography}

\bibitem[Abt \& Morrell(1995)]{abt1995} Abt, H.~A., \& Morrell, N.~I.\ 1995, \apjs, 99, 135 

\bibitem[Anandarao et al.(2008)]{anandarao2008} Anandarao, B., Richardson, E.~H., Chakraborty, A., \& Epps, H.\ 2008, \procspie, 7014, 70142Y 

\bibitem[Azevedo et al.(2006)]{azevedo2006} Azevedo, R., Calvet, N., Hartmann, L., et al.\ 2006, \aap, 456, 225 

\bibitem[Banerjee et al.(2012)]{banerjee12} Banerjee, D.P.K. and Ashok, N.M., 2012, BASI, 40, 243

\bibitem[Baraffe et al.(1998)]{baraffe1998} Baraffe, I., Chabrier, G., Allard, F. \& Hauschildt, P.H. 1998, \aap, 337, 403

\bibitem[Beckwith et al.(1990)]{beckwith1990} Beckwith, S.~V.~W., Sargent, A.~I., Chini, R.~S., \& Guesten, R.\ 1990, \aj, 99, 924 

\bibitem[Bertin \& Arnouts(1996)]{bertin1996} Bertin, E., \& Arnouts, S.\ 1996, \aaps, 117, 393 

\bibitem[Breeveld et al.(2010)]{breeveld2010} Breeveld, A.~A., Curran, P.~A., Hoversten, E.~A., et al.\ 2010, \mnras, 406, 1687 

\bibitem[Carpenter et al.(2000)]{2000ApJS..130..381C} Carpenter, J.~M., Heyer, M.~H., \& Snell, R.~L.\ 2000, \apjs, 130, 381 

\bibitem[{{Cenko} {et~al.}(2006){Cenko}, {Fox}, {Moon}, {Harrison}, {Kulkarni},
  {Henning}, {Guzman}, {Bonati}, {Smith}, {Thicksten}, {Doyle}, {Petrie},
  {Gal-Yam}, {Soderberg}, {Anagnostou}, \& {Laity}}]{cenko2006}
{Cenko}, S.~B. {et~al.} 2006, \pasp, 118, 1396

\bibitem[Churchwell et al.(2009)]{churchwell2009} Churchwell, E., Babler, B.~L., Meade, M.~R., et al.\ 2009, \pasp, 121, 213 


\bibitem[Clarke et al.(2005)]{clarke2005} Clarke, C., Lodato, G., Melnikov, S.~Y., \& Ibrahimov, M.~A.\ 2005, \mnras, 361, 942 

\bibitem[Cody et al.(2017)]{cody2017} Cody, A.~M., Hillenbrand, L.~A., David, T.~J., et al.\ 2017, \apj, 836, 41 

\bibitem[Cushing et al.(2005)]{cushing2005} Cushing, M.~C., Rayner, J.~T., \& Vacca, W.~D.\ 2005, \apj, 623, 1115 

\bibitem[Cutri et al.(2003)]{cutri2003} Cutri, R.~M., Skrutskie, M.~F., van Dyk, S., et al.\ 2003, VizieR Online Data Catalog, 2246.  

\bibitem[Cutri \& et al.(2012)]{cutri2012} Cutri, R.~M., et al.\ 2012, VizieR Online Data Catalog, 2311.  


\bibitem[da Rio et al.(2012)]{dario2012} da Rio, N., Robberto, 
M., Hillenbrand, L.~A., Henning, T., \& Stassun, K.~G.\ 2012, \apj, 748, 14 

\bibitem[Georgelin \& Georgelin(1970)]{1970A&A.....6..349G} Georgelin, Y.~P., \& Georgelin, Y.~M.\ 1970, \aap, 6, 349 

\bibitem[Gramajo et al.(2014)]{gramajo2014} Gramajo, L.~V., Rod{\'o}n, J.~A., \& G{\'o}mez, M.\ 2014, \aj, 147, 140 

\bibitem[Grinin et al.(2001)]{grinin2001} Grinin, V.~P., Kozlova, O.~V., Natta, A., et al.\ 2001, \aap, 379, 482 

\bibitem[Gullbring et al.(1998)]{gullbring1998} Gullbring, E.,
Hartmann, L., Briceno, C., \& Calvet, N.\ 1998, \apj, 492, 323

\bibitem[Hartigan et al.(2004)]{hartigan2004} Hartigan, P., Edwards, S., \& Pierson, R.\ 2004, \apj, 609, 261 

\bibitem[Hartmann \& Kenyon(1996)]{HK1996} Hartmann, L., \& Kenyon, S.~J.\ 1996, \araa, 34, 207 

\bibitem[Herbig(1977)]{herbig1977} Herbig, G.~H.\ 1977, \apj, 217, 693 

\bibitem[Herbig(2009)]{herbig2009} Herbig, G.~H.\ 2009, \aj, 138, 448 

\bibitem[Herczeg et al.(2016)]{herczeg2016} Herczeg, G.~J., Dong, S., Shappee, B.~J., et al.\ 2016, \apj, 831, 133 

\bibitem[Heyer et al.(1998)]{heyer1998} Heyer, M.~H., Brunt, C., Snell, R.~L., et al.\ 1998, \apjs, 115, 241 

\bibitem[Heyer et al.(2001)]{heyer2001} Heyer, M.~H., Carpenter, J.~M., \& Snell, R.~L.\ 2001, \apj, 551, 852 

\bibitem[Hillenbrand et al.(2018)]{hillenbrand2018} Hillenbrand, L.~A., Contreras Pe{\~n}a, C., Morrell, S., et al.\ 2018, \apj, 869, 146 

\bibitem[Hinkley et al.(2013)]{hinkley2013} Hinkley, S., Hillenbrand, L., Oppenheimer, B.~R., et al.\ 2013, \apjl, 763, L9 

\bibitem[Holoien et al.(2014)]{2014ApJ...785L..35H} Holoien, T.~W.-S., Prieto, J.~L., Stanek, K.~Z., et al.\ 2014, \apjl, 785, L35 

\bibitem[Hora et al.(2007)]{hora2007} Hora, J., Adams, J., Allen, L., et al.\ 2007, Spitzer Proposal ID 40184  


\bibitem[Kraus et al.(2016)]{kraus2016} Kraus, S., Caratti o Garatti, A., Garcia-Lopez, R., et al.\ 2016, \mnras, 462, L61 

\bibitem[Kudritzki et al.(1999)]{kudritzki1999} Kudritzki, R.~P., Puls, J., Lennon, D.~J., et al.\ 1999, \aap, 350, 970 

\bibitem[Kulkarni(2013)]{kulkarni2013} Kulkarni, S.~R.\ 2013, ATel, 4807 

\bibitem[K{\"o}nigl et al.(2011)]{konigl2011} K{\"o}nigl, A., Romanova, M.~M., \& Lovelace, R.~V.~E.\ 2011, \mnras, 416, 757 

\bibitem[Kurosawa \& Romanova(2012)]{kurosawa2012} Kurosawa, R., \& Romanova, M.~M.\ 2012, \mnras, 426, 2901 

\bibitem[Lagrois \& Joncas(2009)]{2009ApJ...691.1109L} Lagrois, D., \& Joncas, G.\ 2009, \apj, 691, 1109 

\bibitem[Laher et al.(2014)]{laher2014}
Laher, R.~R., Surace, J., Grillmair, C.~J., et al.\ 2014, \pasp, 126, 674 

\bibitem[Law et al.(2009)]{law2009} Law, N.~M., Kulkarni, S.~R., Dekany, R.~G., et al.\ 2009, \pasp, 121, 1395 

\bibitem[Lefever et al.(2010)]{lefever2010} Lefever, K., Puls, J., Morel, T., et al.\ 2010, \aap, 515, A74 

\bibitem[Leggett et al.(2006)]{leggett2006} Leggett, S.~K., Currie, M.~J., Varricatt, W.~P., et al.\ 2006, \mnras, 373, 781 

\bibitem[Li \& McCray(1993)]{1993ApJ...405..730L} Li, H., \& McCray, R.\ 1993, \apj, 405, 730 

\bibitem[Lynden-Bell \& Pringle(1974)]{lbp1974} Lynden-Bell, D., \& Pringle, J.~E.\ 1974, \mnras, 168, 603 

\bibitem[Mainzer et al.(2014)]{mainzer2014} Mainzer, A., Bauer, J., Cutri, R.~M., et al.\ 2014, \apj, 792, 30 

\bibitem[Masci et al.(2017)]{masci2017} Masci, F.~J., Laher, R.~R., Rebbapragada, U.~D., et al.\ 2017, \pasp, 129, 014002 

\bibitem[Megeath et al.(2008)]{megeath2008} Megeath, S.~T., Townsley, L.~K., Oey, M.~S., \& Tieftrunk, A.~R.\ 2008, Handbook of Star Forming Regions, Volume I, 4, 264 

\bibitem[Merrill(1943)]{merrill1943} Merrill, P.~W.\ 1943, \pasp, 55, 242 

\bibitem[Metzger et al.(2012)]{metzger2012} Metzger, B.~D., Giannios, D., \& Spiegel, D.~S.\ 2012, \mnras, 425, 2778 

\bibitem[Miller et al.(2011)]{miller2011} Miller, A.~A., 
Hillenbrand, L.~A., Covey, K.~R., et al.\ 2011, \apj, 730, 80 

\bibitem[Miller et al.(2015)]{2015ATel.7428....1M} Miller, A.~A., Hillenbrand, L.~A., Bilgi, P., et al.\ 2015, ATel, 7428  

\bibitem[Ninan et al.(2015)]{2015ApJ...815....4N} Ninan, J.~P., Ojha, D.~K., Baug, T., et al.\ 2015, \apj, 815, 4 

\bibitem[Nisini et al.(2005)]{nisini2005} Nisini, B., Bacciotti, F., Giannini, T., et al.\ 2005, \aap, 441, 159 

\bibitem[Ofek et al.(2012)]{ofek2012} Ofek, E.~O., Laher, R., Surace, J., et al.\ 2012, \pasp, 124, 854 

\bibitem[Rei et al.(2018)]{rei2018} Rei, A.~C.~S., Petrov, P.~P., \& Gameiro, J.~F.\ 2018, \aap, 610, A40.

\bibitem[Reipurth \& Aspin(2010)]{RA2010} Reipurth, B., \& Aspin, C.\ 2010, Evolution of Cosmic Objects through their Physical Activity, 19 

\bibitem[Roming et al.(2005)]{roming2005} Roming, P.~W.~A., Kennedy, T.~E., Mason, K.~O., et al.\ 2005, \ssr, 120, 95 

\bibitem[Semkov et al.(2010)]{semkov2010} Semkov, E.~H., Peneva, S.~P., Munari, U., Milani, A., \& Valisa, P.\ 2010, \aap, 523, LL3 

\bibitem[Semkov et al.(2017)]{semkov2017} Semkov, E.~H., Peneva, S.~P., \& Ibryamov, S.~I.\ 2017, Bulgarian Astronomical Journal, 26, 57 

\bibitem[Silva \& Cornell(1992)]{1992ApJS...81..865S} Silva, D.~R., \& Cornell, M.~E.\ 1992, \apjs, 81, 865 

\bibitem[Smith et al.(2011)]{smith2011} Smith, N., Li, W., Silverman, J.~M., Ganeshalingam, M., \& Filippenko, A.~V.\ 2011, \mnras, 415, 773 

\bibitem[Soubiran et al.(2008)]{soubiran2008} Soubiran, C., Bienaym{\'e}, O., Mishenina, T.~V., \& Kovtyukh, V.~V.\ 2008, \aap, 480, 91 

\bibitem[Stauffer et al.(2014)]{stauffer2014} Stauffer, J., Cody, A.~M., Baglin, A., et al.\ 2014, \aj, 147, 83 

\bibitem[Stauffer et al.(2016)]{stauffer2016} Stauffer, J., Cody, A.~M., Rebull, L., et al.\ 2016, \aj, 151, 60 

\bibitem[Strope et al.(2010)]{strope2010} Strope, R.~J., Schaefer, B.~E., \& Henden, A.~A.\ 2010, \aj, 140, 34 

\bibitem[Terebey et al.(2003)]{terebey2003} Terebey, S., Fich, M., Taylor, R., Cao, Y., \& Hancock, T.\ 2003, \apj, 590, 906 

\bibitem[Vazdekis et al.(2003)]{vazdekis2003} Vazdekis, A., Cenarro, A.~J., Gorgas, J., Cardiel, N., \& Peletier, R.~F.\ 2003, \mnras, 340, 1317 

\bibitem[Verdugo et al.(1999)]{verdugo1999} Verdugo, E., Talavera, A., \& G{\'o}mez de Castro, A.~I.\ 1999, \aap, 346, 819 

\bibitem[Xu et al.(2006)]{xu2006} Xu, Y., Reid, M.~J., Menten, K.~M., \& Zheng, X.~W.\ 2006, \apjs, 166, 526 


\bibitem[York et al.(2000)]{york2000} York, D.~G., Adelman, J., Anderson, J.~E., Jr., et al.\ 2000, \aj, 120, 1579 


\end{thebibliography}
\end{document}